\documentclass[11 pt]{article}
\usepackage{amssymb,latexsym,amsmath,amsfonts}
\usepackage{subfigure,color}
\usepackage{graphicx, epsf}
\usepackage[active]{srcltx}
\usepackage[active]{srcltx}
\numberwithin{equation}{section}

\newtheorem{theorem}{Theorem}

\begin{document}

\title{Smooth and non-smooth traveling wave solutions of some generalized Camassa-Holm equations}

\author{T. Rehman\footnote{University of Central Florida, Department of Mathematics,USA, taslima.rehman@knights.ucf.edu}$\;$
G. Gambino\footnote{Corresponding author. Department of Mathematics, University of Palermo, Italy, gaetana@math.unipa.it}$\;$
S. Roy Choudhury\footnote{University of Central Florida, Department of Mathematics,USA, choudhur@cs.ucf.edu}}

%
%
%

\maketitle

\begin{abstract}
In this paper we employ two recent analytical approaches to investigate the possible classes of traveling wave solutions of some members of a recently-derived integrable family of generalized Camassa-Holm (GCH) equations.
A recent, novel application of phase-plane analysis is employed to analyze the singular traveling wave equations of three of the GCH NLPDEs, i.e. the possible non-smooth peakon and cuspon solutions. One of the considered GCH equations supports both solitary (peakon) and periodic (cuspon) cusp waves in different parameter regimes. The second equation does not support singular traveling waves and the last one supports four-segmented, non-smooth $M$-wave solutions.

Moreover, smooth traveling waves of the three GCH equations are considered. Here, we use a recent technique to derive convergent multi-infinite series solutions for the homoclinic orbits of their traveling-wave equations, corresponding to pulse (kink or shock) solutions respectively of the original PDEs. We perform many numerical tests in different parameter regime to pinpoint real saddle equilibrium points of the corresponding GCH equations, as well as ensure simultaneous convergence and continuity of the multi-infinite series solutions for the homoclinic orbits anchored by these saddle points. Unlike the majority of unaccelerated convergent series, high accuracy is attained with relatively few terms. We also show the traveling wave nature of these pulse and front solutions to the GCH NLPDEs.
\end{abstract}



\section{Introduction}\label{Sec1}

The Camassa-Holm (CH) equation:

\begin{equation}
\label{CH}
u_t+2ku_x-u_{xxt}+3uu_x=2u_x u_{xx}+uu_{xxx},
\end{equation}

\noindent
where $u = u(x, t)$ and $k$ is a constant (and subscripts denote partial derivatives), came to prominence with the work of Camassa et al. in 1993 and 1994 \cite{CH93}, where it was argued that the equation could be taken as a model for the unidirectional propagation of waves in shallow water. This equation has attracted much research interest in recent years both from analytical and numerical point of view \cite{BC07,CGS12,CHH94,DIS10,DKST09,DLSS06,HR09,LSS05,M10,W06,XWZ12}.

For $k = 0$, Camassa and Holm showed that Eq. \eqref{CH} has peakons of the form $u(x,t)=ce^{-|x-ct| }$. In mathematics and physics, a soliton is a solitary wave packet or pulse that maintains its shape while traveling at constant speed. This type of wave has been the focus interest since solitons are thus stable, and do not disperse over time. Peakons are a type of non-smooth soliton, discovered by Camassa and Holm; these waves have a sharp peak where it has a discontinuous derivative. The wave profile is similar to the graph of the function $e^{-|x| }$.

Eq. \eqref{CH} can be rewritten in the following form:
\begin{equation}\label{CH2}
(1-D_x^2 ) u_t=-2ku_x-3uu_x+2u_x u_{xx}+uu_{xxx},\qquad    u=u(x,t),\quad    D_x=\frac{\partial}{\partial x}.
\end{equation}
It belongs to the class:
\begin{equation}\label{CHclass}
(1-D_x^2 ) u_t=F(u,u_x,u_{xx},u_{xxx}, \dots),
\end{equation}
which has attracted much interest, particularly the possible integrable member of equation \cite{J03,J02,DHH02}.

The Camassa-Holm equation is integrable by the inverse scattering transform. It possesses an infinite hierarchy of local conservation laws, bi-Hamiltonian structure and the various other remarkable properties of integrable equations. Despite its non-evolutionary form, the Camassa-Holm equation possesses an infinite hierarchy of local higher symmetries \cite{N09}.
Until 2002, the Camassa-Holm equation was the only known integrable example of the type of Eq. \eqref{CHclass}. Later, Degasperis and Procesi (see \cite{DHH02} and references therein) found another nonlinear PDE with similar properties, and this, the so-called DP equation, has been studied quite intensively.

More recently, Novikov \cite{N09} and Mikhailov and Novikov \cite{MN02} showed that there are other examples of NLPDEs in the class of Eq. \eqref{CHclass} which are integrable. Novikov presented a detailed summary of integrable and homogeneous polynomial generalizations of the Camassa-Holm type equation with quadratic and cubic nonlinearities.

In this paper, the dynamical behavior of the traveling wave solutions of some of these generalized Camassa-Holm equations is discussed. In particular, we consider the following three nonlinear PDEs (NLPDEs) from Novikov's list of $27$ integrable generalized Camassa-Holm equations \cite{N09}:
\begin{eqnarray}
\label{eq4}
(1-D_x^2 ) u_t&=&D_x (4-D_x^2 ) u^2,\\
\label{eq2}
(1-D_x^2 ) u_t&=&D_x (2+D_x ) [(2-D_x )u]^2,\\\label{eq3}
(1-D_x^2 ) u_t&=&D_x (u^2 u_{xx}-u_x^2 u_{xx}+uu_x^2-u^3).
\end{eqnarray}
Since these integrable generalized CH equations are new, and the properties of the solutions of only one of them has been considered in any sort of detail \cite{HW08}, we investigate the possible traveling wave solutions of Eqs. \eqref{eq4}--\eqref{eq3} in detail in this paper.
Two separate approaches are employed.

Since Eqs. \eqref{eq4}--\eqref{eq3} are integrable generalized CH equations, non-smooth solutions as in the CH equation (or in other GCH equations \cite{RTW10,ZTW10}) are a definite possibility. To investigate these, we employ a somewhat unusual variant of phase-plane analysis \cite{LD07} which has been recently applied to consider peakon and cuspon solutions of a wide variety of NLPDEs.

We also consider regular smooth traveling wave solutions of our system of Eqs. \eqref{eq4}--\eqref{eq3}. As it is well known, homoclinic and heteroclinic orbits of the traveling wave ODE (of any PDE) correspond to pulse and front (shock or kink) solutions of the governing PDE.
In particular, we apply a recently developed technique \cite{CG13,W09} to analytically compute convergent multi-infinite series solutions for the possible homoclinic and heteroclinic orbits of the traveling-wave ODEs of Eqs. \eqref{eq4}--\eqref{eq3}. They correspond to convergent series for pulse or front (shock) solutions of these generalized CH Eqs. \eqref{eq4}--\eqref{eq3}. Since the later terms in the series fall off exponentially, we show high accuracy may be obtained for the pulse  and front shapes using only small number of terms. The actual convergence of such series is analogous to the earlier treatments \cite{CG13,W09}, and it is omitted here.

The plan of the paper is the following: in Section 2, the recently developed theory for singular traveling-wave ODEs is reviewed \cite{LD07,DF10}. This is then applied to our generalized CH Eqs. \eqref{eq4}--\eqref{eq3} in Section \ref{Sec3}. Section \ref{Sec4} develops analytic pulse and front solutions of Eqs. \eqref{eq4}--\eqref{eq3}.

\section{Phase plane analysis of traveling wave equations having singularities}\label{Sec2}
\setcounter{figure}{0}
\setcounter{equation}{0}

In this section, we briefly review some background material on the phase plane analysis of dynamical systems having singularities \cite{LD07}.

\subsection{The dynamics of the first type of singular traveling waves}\label{firstType}

Let us consider the following first type of singular traveling systems:
%
%
\begin{equation}\label{gen_ld}
\frac{d \phi}{d z}=y,\qquad\qquad \frac{d y}{d z}=-\frac{G'(\phi)y^2+F(\phi)}{G(\phi)},
\end{equation}
where $F$ and $G$ are smooth nonlinear functions (at least $\mathcal{C}^2$ - functions in order to guarantee the existence and uniqueness of the solutions of the initial value problem). The system \eqref{gen_ld} can be thought as
the general form of the traveling wave system of some interesting physical models, see \cite{LD07}.
The system of Eq. \eqref{gen_ld} also admits a first integral:
\begin{equation}\label{1int_ld}
H(\phi,y)=y^2 G^2(\phi )+2\int G(\phi)F(\phi)d\phi=h,
\end{equation}
where $h$ is the integral constant.

Assume that:
\begin{itemize}
\item[$a)$]\noindent $\phi =\phi_s$ is the unique simple zero of $G(\phi)$ and $G'(\phi_s)\neq 0$;
\item[$b)$]\noindent $\phi =\phi_{1,2}$, $\phi_1<\phi_2$ are the only two simple zeros of $F(\phi)$ and $F'(\phi_1)>0$, $F'(\phi_2)<0$,
\end{itemize}
therefore $P_{1,2}\equiv (\phi_{1,2},0)$ are the two equilibria of the system \eqref{gen_ld} and the right hand side of the second equation of \eqref{gen_ld} is discontinuous in the straight line $\phi=\phi_s$ of the $(\phi ,y)$ phase plane. If  $Y=-\displaystyle\frac{F(\phi_s)}{G'(\phi_s)}>0$ there exist two critical points $S_{\pm} (\phi_s,\pm\sqrt{Y})$ on the \textit{singular straight line} $\phi=\phi_s$. We denote $h_{1,2}=H(\phi_{1,2},0)$ and $h_s=H (\phi_s,\pm\sqrt{Y})$.

The analysis of \eqref{gen_ld} is based on the following {three steps}:
\begin{itemize}
\item[$i)$]\noindent make an independent variable in such a way that the \textit{singular system} \eqref{gen_ld} becomes a \textit{regular system};
\item[$ii)$]\noindent discuss and analyze the dynamical behavior of the associated regular system;
\item[$iii)$]\noindent use the dynamical behavior of the regular system to obtain the wave profiles determined by all the bounded solutions of the singular system.
\end{itemize}

As the step $i)$, let us make the transformation $dz =G(\phi)d\zeta$, for $\phi\neq \phi_s$ and the Eq. \eqref{gen_ld} becomes:
\begin{equation}\label{gen_ld_reg}
\frac{d \phi}{d \zeta}=G(\phi)y,\qquad\qquad \frac{d y}{d \zeta}=-(G'(\phi)y^2+F(\phi)).
\end{equation}
Notice that, for the system \eqref{gen_ld_reg}, the straight line $\phi=\phi_s$ is an invariant straight line.

The systems of equations in \eqref{gen_ld} and \eqref{gen_ld_reg} have the same invariant curve solutions, the main difference between Eqs. \eqref{gen_ld} and \eqref{gen_ld_reg} is the parametric representation of the orbit: near $\phi=\phi_s$ Eq. \eqref{gen_ld_reg} uses the \textit{fast time variable} $\zeta$, while Eq. \eqref{gen_ld} uses the \textit{slow time variable} $z$. Hence, for the step $ii)$, we study the associated regular system of Eq. \eqref{gen_ld_reg} in order to get the phase portraits of Eq. \eqref{gen_ld}.

Via standard linear stability analysis for the Eq. \eqref{gen_ld_reg}, we obtain that the determinant of the Jacobian matrix computed at the critical points $P_{1,2} (\phi_{1,2},0)$ and $S_{1,2} (\phi_s,\pm\sqrt{Y})$ is respectively given by:
\begin{equation}\label{jac_reg}
J(\phi_{1,2},0)=F'(\phi_{1,2})G(\phi_{1,2}),\qquad\qquad J(\phi_s,\pm\sqrt{Y})=-2Y(G'(\phi_s))^2,
\end{equation}
thus $S_{1,2}$ are saddle points (being $J<0$). Moreover,
if $\phi_s\neq\phi_{1,2}$ and $G(\phi_1 )>0$, the equilibrium $(\phi_1,0)$ is a saddle, when $G(\phi_1 )<0$ it is a center (see the above assumption $b)$). Analogously, if $\phi_s\neq\phi_{1,2}$ and $G(\phi_2)>0$, the equilibrium $(\phi_2,0)$ is a center, when $G(\phi_2 )<0$ it is a saddle .
If $\phi_s$ coincide with $\phi_{1}$ or  $\phi_{2}$, the corresponding equilibria $(\phi_1,0)$ or $(\phi_2,0)$ are second - order critical points.

For the step $iii)$, we have to obtain the wave profiles for the singular system \eqref{gen_ld}. Even though Eq. \eqref{gen_ld} has the same invariant level curves as Eq. \eqref{gen_ld_reg}, the smooth property of orbits of Eq. \eqref{gen_ld} with respect to time variable $z$ should be investigated, as the line $\phi=\phi_s$ is not an orbit of Eq. \eqref{gen_ld}. Therefore the singular straight line $\phi=\phi_s$ of Eq. \eqref{gen_ld} has to be geometrically distinguished from the straight line solution $\phi=\phi_s$ of Eq. \eqref{gen_ld_reg}, as detailed in the following three main theorems (whose proofs may be found in \cite{LD07}).

\subsubsection{Main theorems to identify the profiles of waves}

In this section we review three theorems, introduced in \cite{LD07}, in order to identify the profiles of traveling wave solutions of Eqs. \eqref{gen_ld} determined by different phase orbits of Eq. \eqref{gen_ld_reg}.
\begin{figure}
\begin{center}
\subfigure[] {\epsfxsize=2 in \epsfbox{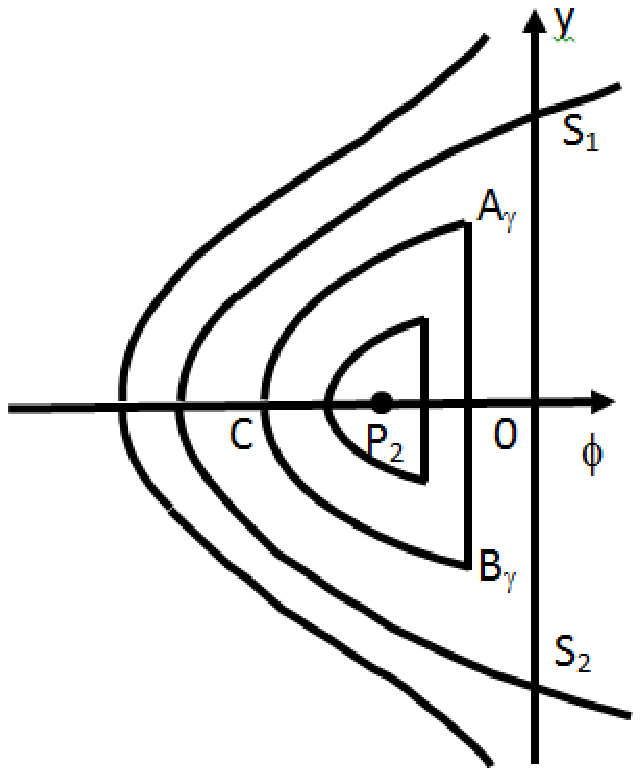}}
\subfigure[] {\epsfxsize=2 in \epsfbox{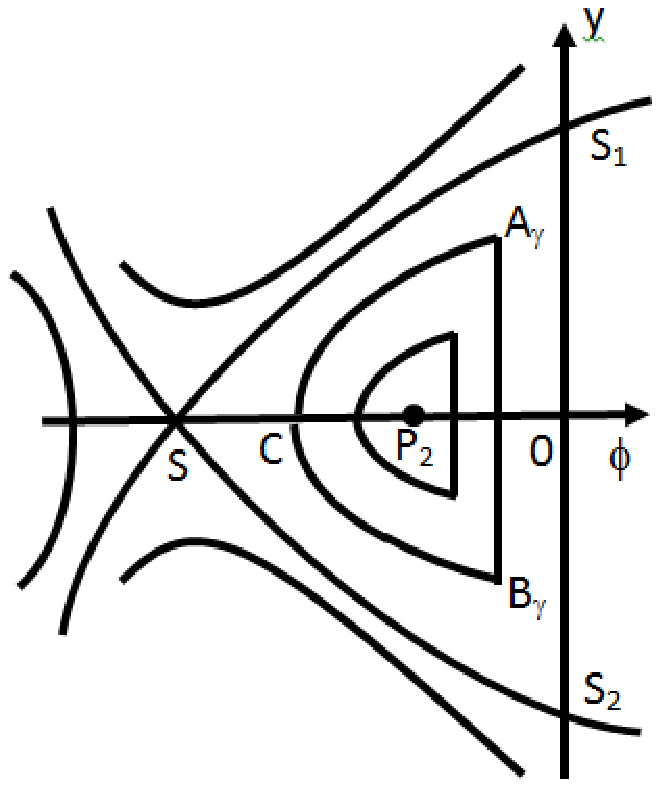}}
\end{center}
\caption{\label{pp1} Two types of phase portraits of Eq.\eqref{gen_ld_reg}, see \cite{LD07}.}
\end{figure}
We consider two possible types of phase portraits of Eq. \eqref{gen_ld_reg}, drawn in Fig.\ref{pp1}. The point $P_2 (\phi_2,0)$, placed on the left of the straight line $\phi=\phi_s=0$, is a center, whose boundary curves consist of the segment $S_1 S_2$ and the arc $\widehat{S_1 S_2}$ defined by the level curve $H(\phi ,y)=h_s$, as given in the Eq. \eqref{1int_ld}. These boundary curves are heteroclinic orbits for system \eqref{gen_ld_reg} and along both these orbits the phase points $(\phi(\zeta),y(\zeta))$ of Eq. \eqref{gen_ld_reg} tend to the equilibria $S_1$ and $S_2$ when $\zeta\rightarrow  \pm\infty$; but the boundary segment $S_1 S_2$ lies on the singular straight line of Eq. \eqref{gen_ld}. An orbit $\gamma$ of the family of periodic orbits in the periodic annulus of $P_2$ is a closed branch of the invariant curves $H(\phi,y)=h_{\gamma}$, where $h_{\gamma}\in(H(\phi_2,0),h_s)$.

\begin{figure}
\begin{center}
\subfigure[] {\epsfxsize=2 in \epsfbox{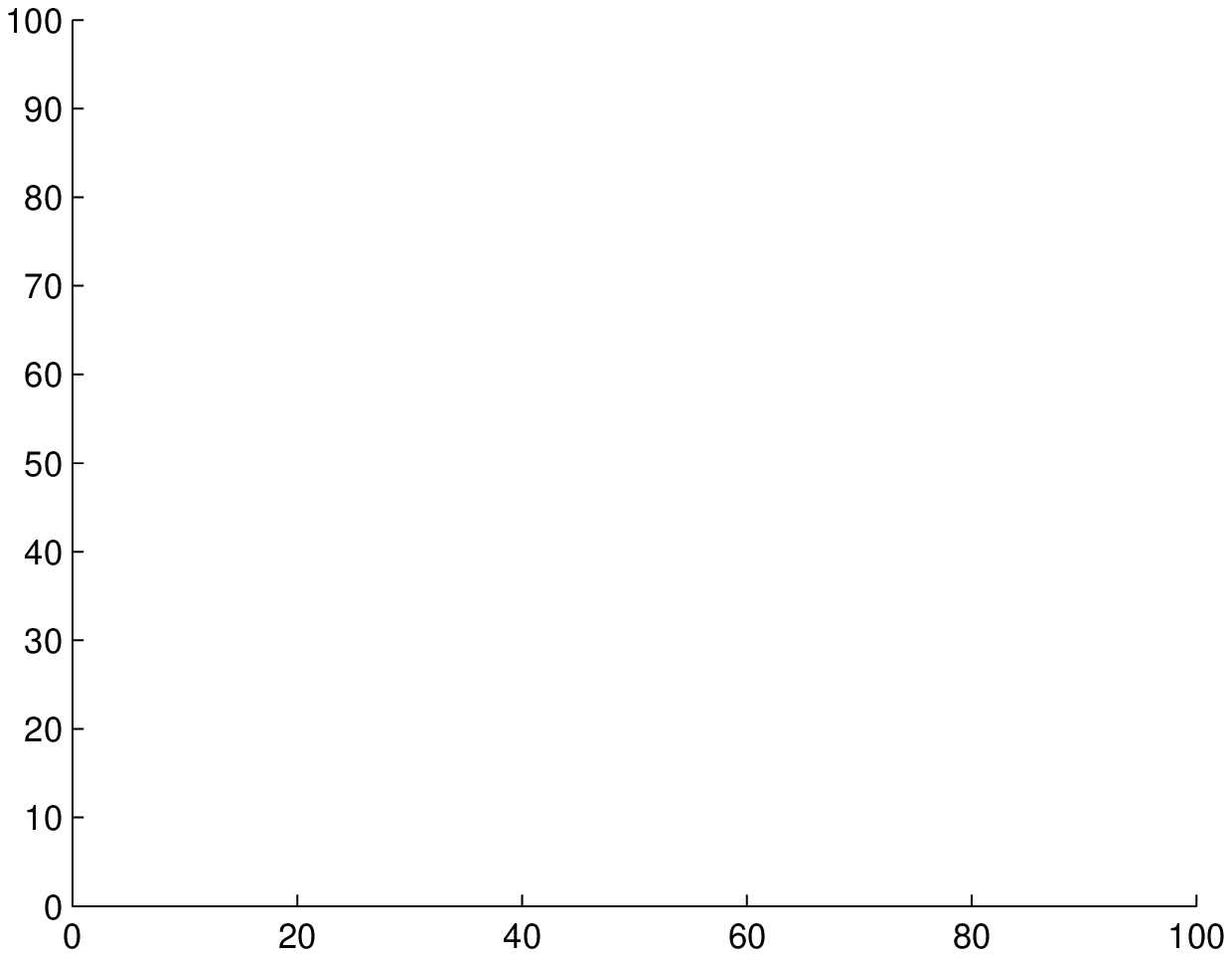}}
\subfigure[] {\epsfxsize=2 in \epsfbox{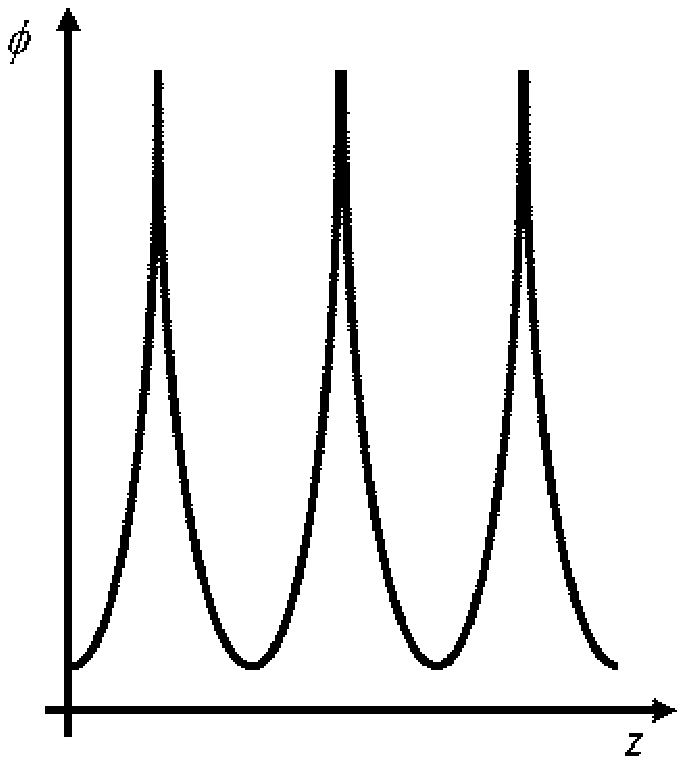}}
\end{center}
\caption{\label{pp2} (a) The rapid jump of $\phi'(z)$ near the singular line. (b) The periodic cuspons $\phi(z)$ corresponding to the phase plot given in Fig.\ref{pp1}(a).}
\end{figure}

\begin{figure}
\begin{center}
\subfigure[] {\epsfxsize=2 in \epsfbox{vuoto.eps}}
\subfigure[] {\epsfxsize=2 in \epsfbox{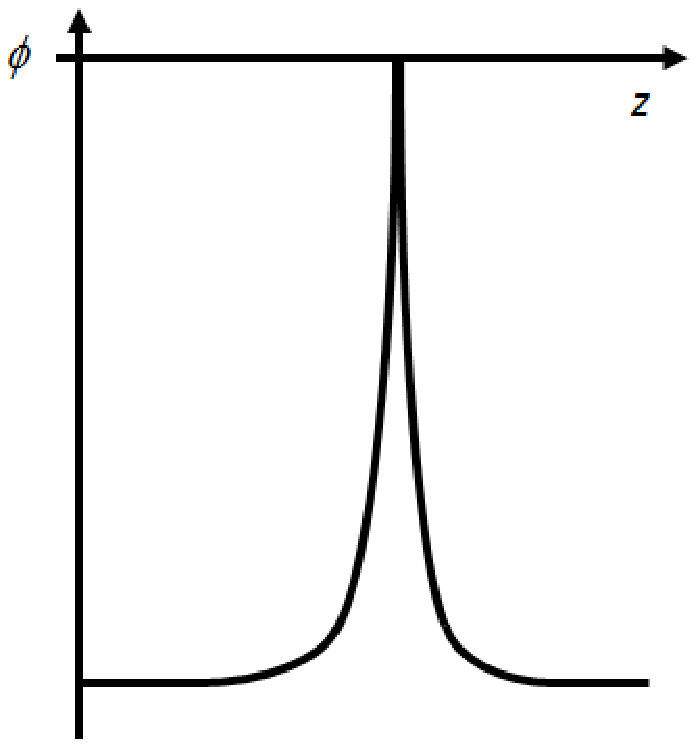}}
\subfigure[] {\epsfxsize=2 in \epsfbox{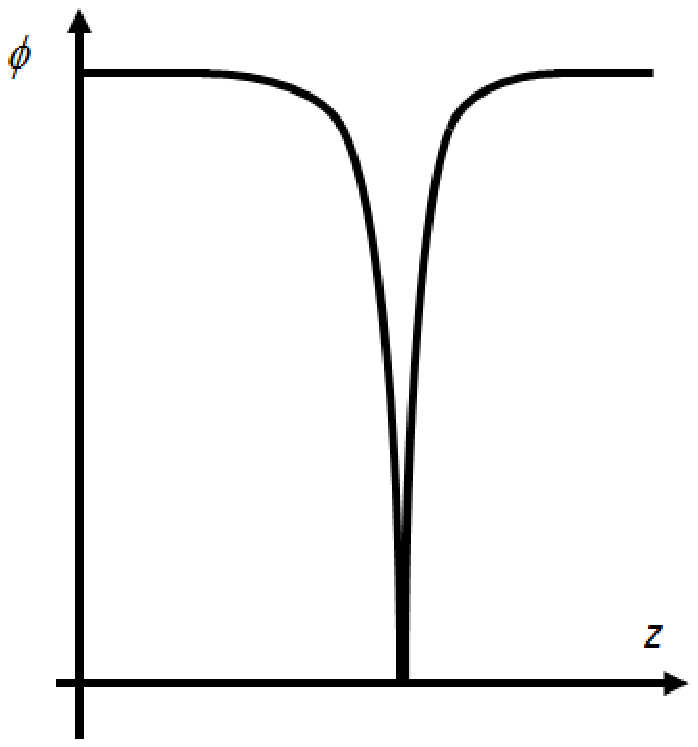}}
\end{center}
\caption{\label{pp3} (a) The rapid jump of $\phi'(z)$ near the singular line. (b),(c) The peakon $\phi(z)$ (solitary waves of two types) corresponding to the phase plot given in Fig.\ref{pp1}(b).}
\end{figure}

The following theorems now apply.
\begin{theorem}\label{theo1}
\textbf{\textsc{(The rapid jump property of $\displaystyle \frac{d\phi}{dz}=y$ near the singular straight line):}}
When $h\rightarrow h_s$, the periodic orbits of the periodic annulus surrounding $P_2$ approach the boundary curves.
Let $\left(\phi,\displaystyle\frac{d\phi}{dz}=y\right)$ be a point on the periodic orbit $\gamma$  of Eq. \eqref{gen_ld}. Then, along the line segment $A_{\gamma\varepsilon}B_{\gamma\varepsilon}$ near the straight line $\phi=\phi_s$, in a very short time interval of $z, \displaystyle \frac{d\phi}{dz}=y$ jumps down rapidly.
\end{theorem}
\begin{theorem}\label{theo2}
\textbf{\textsc{(Existence of the finite time intervals of solutions with respect to $z$ in the positive or negative directions):}}
Let $\left(\phi,\displaystyle\frac{d\phi}{dz}=y\right)$ be the parametric representation of an orbit $\gamma$ of system of Eq. \eqref{gen_ld} and $(\phi_s,\pm\sqrt{Y})$ be two points on the singular straight line
$\phi = \phi_s$. Suppose that one of the following three conditions holds:
\begin{itemize}
\item[a)]\noindent $Y>0$ and, along the orbit $\gamma$, as $z$ increases or decreases, the phase point $(\phi(z), y(z))$ tends to the points $(\phi_s,\pm\sqrt{Y})$, respectively.
\item[b)]\noindent $Y=0$ and, along the orbit $\gamma$, as $z$ increases or decreases, the phase point $(\phi(z), y(z))$ tends to the point $(\phi_s,0)$ and $\gamma$ is in contact with the $y$-axis at the point $(\phi_s,0)$.
\item[b)]\noindent Along the orbit $\gamma$, as $z$ increases or decreases, the phase point $(\phi(z), y(z))$ approaches the straight line $\phi = \phi_s$ in the positive direction or negative direction respectively, and $\lim_{\phi\rightarrow \phi_s}|y|=\infty$.
\end{itemize}
Then, there exists a finite value $z =\tilde{z}$ such that $\lim_{z\rightarrow \tilde{z}}\,\phi(z)=\phi_s$.
\end{theorem}
From Theorems \ref{theo1} and \ref{theo2} and some qualitative considerations on the two phase plots given in Fig.\ref{pp1}, it follows the Theorem \ref{theo3} (see the detailed discussion in \cite{LD07}):
\begin{theorem}\label{theo3}
$\ $
\begin{enumerate}
\item\noindent The arch curves in Fig.\ref{pp1}(a) defined by $H(\phi,y)=h_s$ gives rise to a periodic cusp wave of the peak type, called \textit{cuspons} as in Fig.\ref{pp2}(b).
\item\noindent The curve triangle in Fig.\ref{pp1}(b) defined by $H(\phi,y)=h_s$ gives rise to a solitary cusp wave solutions of the peak type, called \textit{peakon} as in Fig.\ref{pp3}(b).
    \end{enumerate}
    \end{theorem}


Having summarized the theorems relevant to the first class of traveling wave equations, we next consider a second class of traveling wave equations.

\subsection{The dynamics of the second type of singular traveling waves}\label{secondType}

Let us consider the following second class of singular traveling wave systems:
\begin{equation}\label{gen_df}
\frac{d \phi}{d z}=y,\qquad\qquad \frac{d y}{d z}=\frac{Q(\phi, y)}{f(\phi,y)},
\end{equation}
as given in \cite{DF10}. Assume that $f(\phi,y)$ and $Q(\phi, y)$ are sufficiently regular functions satisfying the following condition:
\begin{equation}\label{gen_df_cond}
y\frac{\partial f(\phi,y)}{\partial \phi}+\frac{\partial Q(\phi,y)}{\partial y}\equiv 0,
\end{equation}
which implies there exists a first integral of Eq. \eqref{gen_df}. Notice that $\displaystyle\frac{d y}{d z}=\frac{Q(\phi, y)}{f(\phi,y)}$  is not defined on the set of real planar curves $f(\phi ,y)=0$ and when the phase point $(\phi ,y)$ passes through every branch of $f(\phi ,y)=0$, the quantity $\displaystyle\frac{d y}{d z}$ changes sign \cite{DF10}. Similarly to the step $i)$ given in Section \ref{firstType}, we introduce the new variable $\zeta$ to obtain the following regular system, associated to Eq. \eqref{gen_df}:
\begin{equation}\label{gen_df_reg}
\frac{d \phi}{d \zeta}=yf(\phi,y),\qquad\qquad \frac{d y}{d \zeta}={Q(\phi, y)},
\end{equation}
where ${d z}=f(\phi,y)d\zeta$, for $f(\phi,y)\neq 0$.

For the second type of singular traveling system of Eq. \eqref{gen_df}, the existence of the singular curve $f(\phi,y)= 0$ implies that there may exist a breaking wave solution $\phi(z)$ of the corresponding nonlinear wave equation on the singular curve $f(\phi,y)= 0$, even though the associated regular system of Eq. \eqref{gen_df_reg} has a family of smooth periodic solutions and homoclinic or heteroclinic orbits.

Having laid out the basic background theory of singular traveling waves, we apply it next to the traveling wave equations of the three generalized Camassa Holm (GCH) Eqs. \eqref{eq4}--\eqref{eq3}.

\section{Phase plane analysis of generalized Camassa-Holm equations: possible singular solutions}\label{Sec3}

\subsection{Phase portrait/possible solutions of Eq. \eqref{eq4}}\label{subsec3.4}

Eq. \eqref{eq4} can be simplified and written as:
\begin{equation}\label{eq4_pp}
u_t-u_{xxt}=\frac{\partial}{\partial x} \left(4u^2-2u_x^2-2uu_{xx}\right).
\end{equation}
Substituting $u(x,t)=\phi(x-ct)=\phi(z)$, where $z=x-ct$ and $c$ is the wave speed, into Eq. \eqref{eq4_pp} we obtain:

\begin{equation}\label{eq4_trav1}
-c\frac{d\phi}{dz}+c\frac{d^3\phi}{dz^3}=\frac{d}{dz}\left(4\phi^2-2\phi\frac{d^2\phi}{dz^2}-2
\left(\frac{d\phi}{dz}\right)^2\right).
\end{equation}

Integrating the above equation \eqref{eq4_trav1} once with respect to $z$, one gets:
\begin{equation}\label{eq4_trav}
-c\phi+c\frac{d^2\phi}{dz^2}=4\phi^2-2\phi\frac{d^2\phi}{dz^2}-2
\left(\frac{d\phi}{dz}\right)^2+g,
\end{equation}
where $g$ is the constant of integration.

Eq. \eqref{eq4_trav} is equivalent to the following $2$-dimensional system:
\begin{equation}\label{eq4_sd}
\begin{array}{ll}
\displaystyle\frac{d\phi}{dz}=y,\\
\,\\
\displaystyle\frac{dy}{dz}=\frac{4\phi^2-2y^2+c\phi+g}{c+2\phi},
\end{array}
\end{equation}
which is the traveling wave system for \eqref{eq4_pp}. The system \eqref{eq4_sd} belongs to the first type of singular traveling wave system \eqref{gen_ld}.

The second equation of \eqref{eq4_sd} is discontinuous along the singular straight line $\phi=\phi_s=-\displaystyle\frac{c}{2}$ of the $(\phi, y)$ phase plane.
Following the step $i)$ in Section \ref{firstType}, we make the transformation $dz=(c+2\phi)d\zeta$, obtaining the following regular system associated to Eq. \eqref{eq4_sd}:
\begin{equation}\label{eq4_sd_reg}
\begin{array}{ll}
\displaystyle\frac{d\phi}{d\zeta}=y(c+2\phi),\\
\,\\
\displaystyle\frac{dy}{d\zeta}={4\phi^2-2y^2+c\phi+g}.
\end{array}
\end{equation}
Since the first integral of both Eqs. \eqref{eq4_sd} and \eqref{eq4_sd_reg} are the same, thus both of them have the same phase orbits, except on the straight line $\phi=-\displaystyle\frac{c}{2}$.

The system of Eq. \eqref{eq4_sd_reg} has the following equilibrium points:
\begin{equation}\label{eq4_sd_equi}
\begin{split}
z_1&\,\equiv \left(\frac{-c-\sqrt{c^2-16g}}{8},0\right),\qquad z_2\equiv \left(\frac{-c-\sqrt{c^2-16g}}{8},0\right), \\
z_3&\,\equiv \left(\frac{c}{2},-\frac{\sqrt{c^2+2g}}{2}\right),\qquad \qquad z_4\equiv \left(\frac{c}{2}, \frac{\sqrt{c^2+2g}}{2}\right).
\end{split}
\end{equation}
Here $z_1$ and $z_2$ are the regular equilibrium points (corresponding to the critical points $P_{1,2}$ given in Section \ref{firstType}), while $z_3$ and $z_4$ are the singular equilibrium points (corresponding to $S_{1,2}$ given in Section \ref{firstType}).
We study the stability of these equilibrium points using the linearized system of Eq. \eqref{eq4_sd_reg}.

\begin{figure}
\begin{center}
\subfigure[] {\epsfxsize=2 in \epsfbox{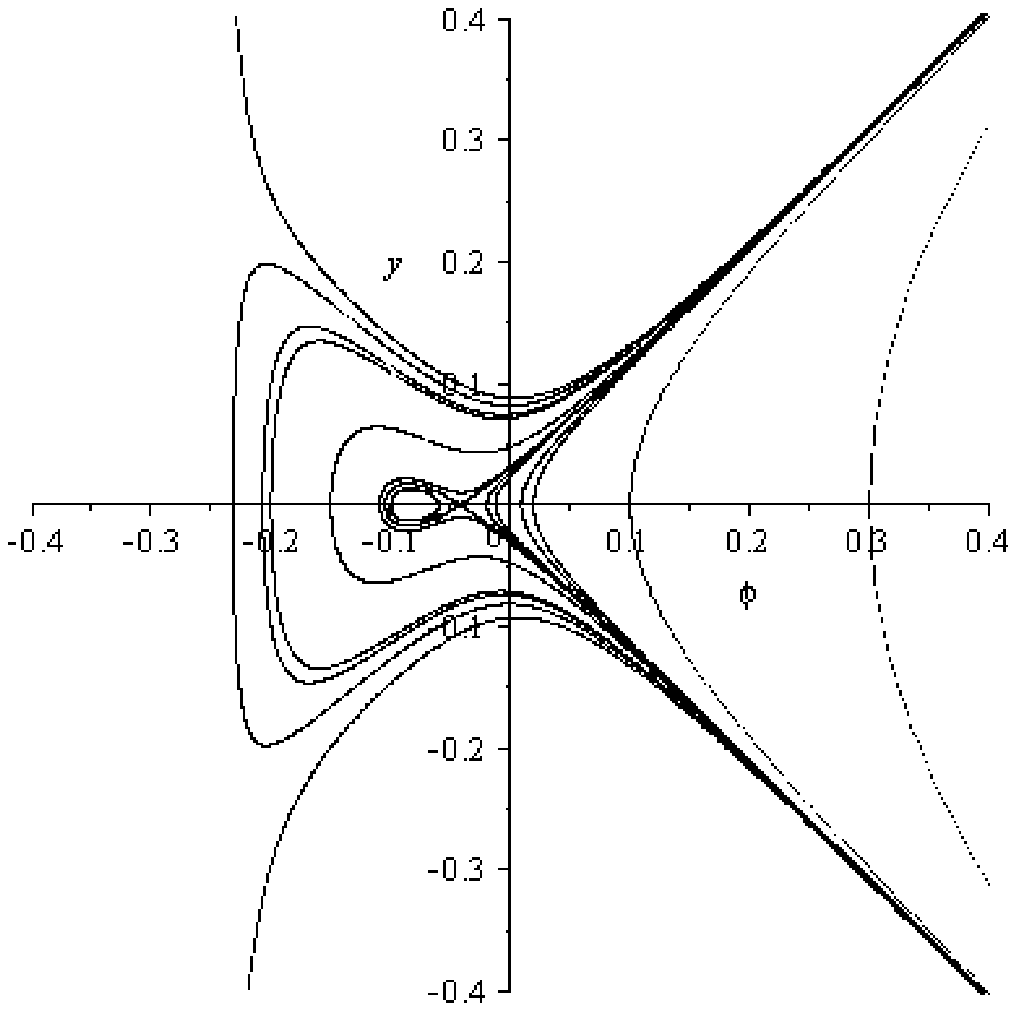}}
\subfigure[] {\epsfxsize=2 in \epsfbox{vuoto.eps}}
\subfigure[] {\epsfxsize=2 in \epsfbox{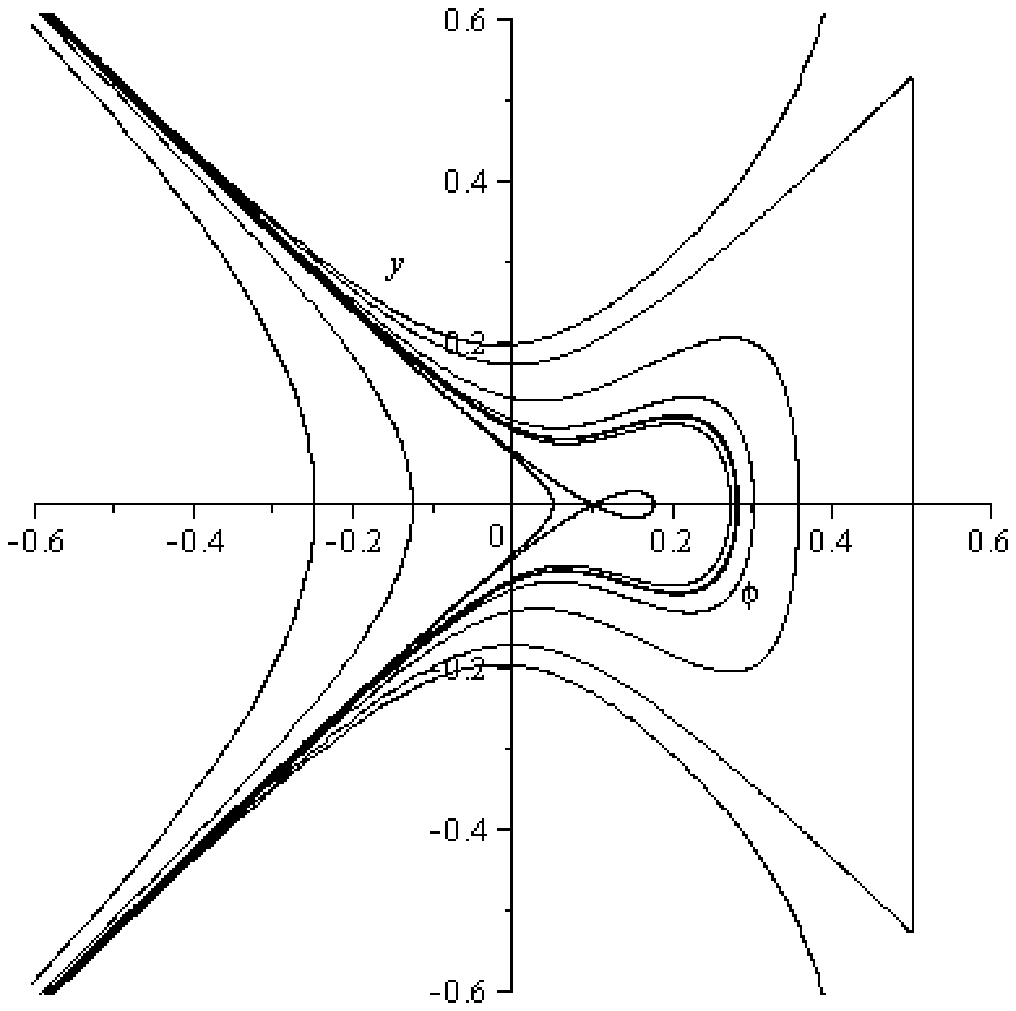}}
\subfigure[] {\epsfxsize=2 in \epsfbox{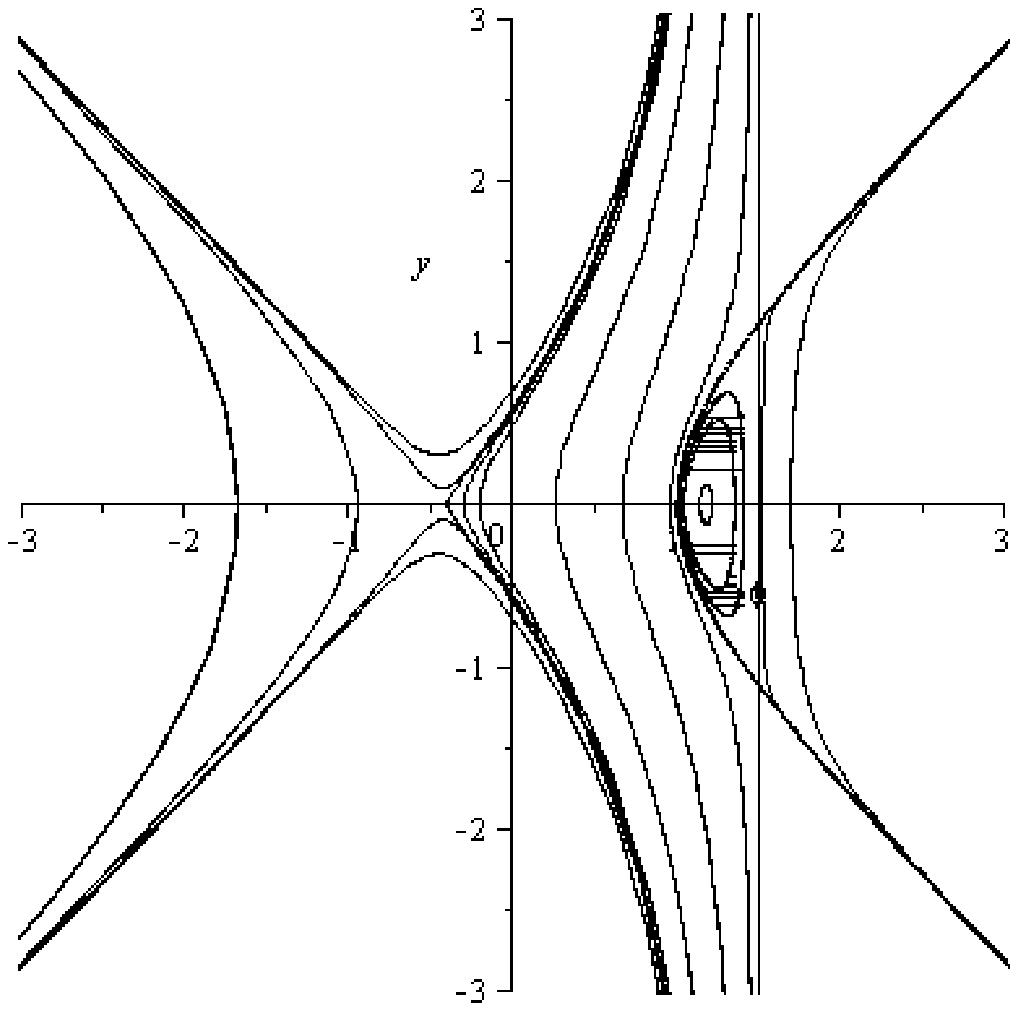}}
\subfigure[] {\epsfxsize=2 in \epsfbox{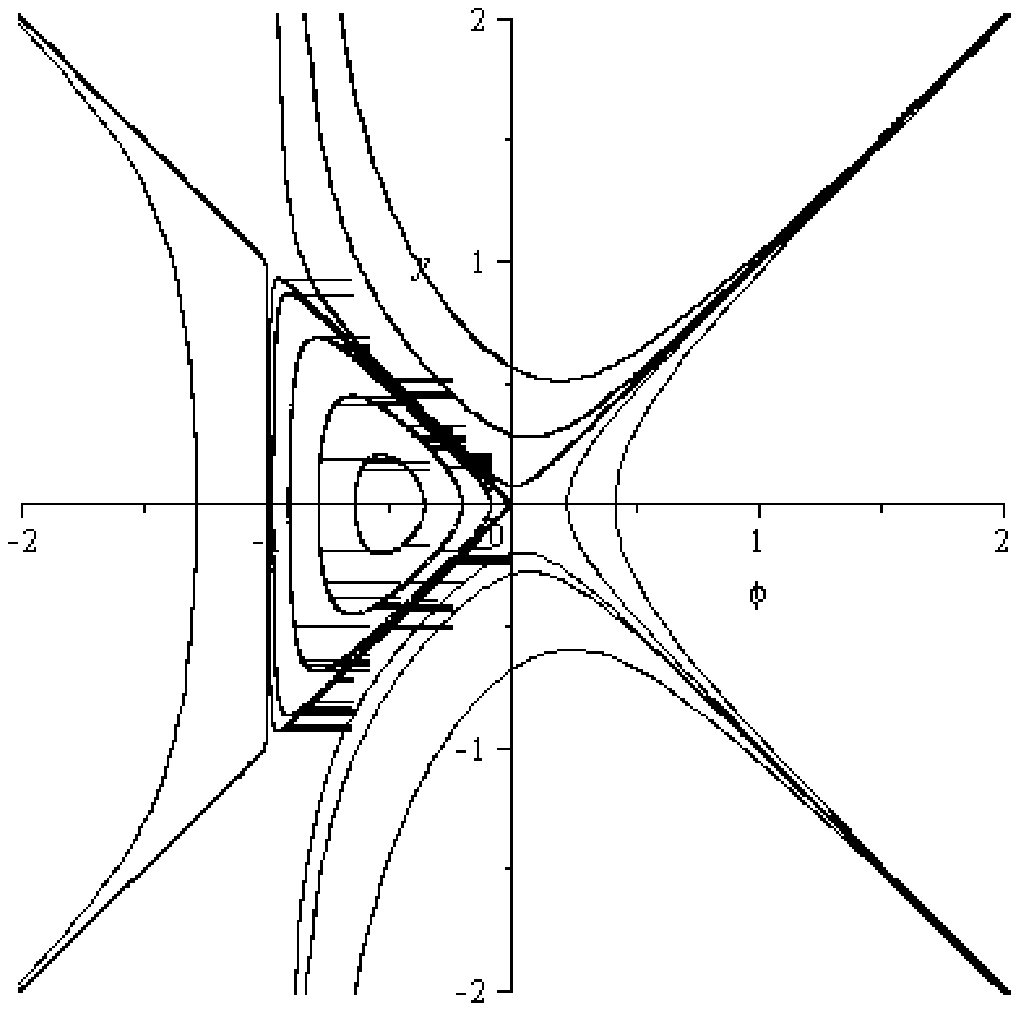}}
\subfigure[] {\epsfxsize=2 in \epsfbox{vuoto.eps}}
\end{center}
\caption{\label{eq4_Fig_phase} The phase portraits ($y$ plotted against $\phi$) of system \eqref{eq4_sd_reg}. (a) $c > 0$ and
$0 <g < \displaystyle\frac{c^2}{32}$. (b) $c < 0$ and  $0<g<\displaystyle\frac{c^2}{32}$. (c) $c > 0$ and $g>-\displaystyle\frac{c^2}{32}$. (d) $c < 0$  and $-\displaystyle\frac{c^2}{32}<g<0$.
(e) $c > 0$ and $g = 0$. (f) $c < 0$ and $g = 0$.}
\end{figure}

When $c>0$ and $0<g<\displaystyle\frac{c^2}{16}$, the equilibrium point $z_1$ is a center, while the equilibrium points $z_2, z_3$ and $z_4 $ are saddle points. The phase portrait is drawn in Fig.\ref{eq4_Fig_phase}(a).

When $c>0$, $g<0$ and $|g|<\displaystyle\frac{c^2}{2}$, the equilibrium point $z_1$ is a center, while the equilibrium points $z_2, z_3$  and $z_4$ are saddle points. The phase portrait is drawn in Fig.\ref{eq4_Fig_phase}(b).

When $c<0$ and $0<g<\displaystyle\frac{c^2}{16}$, the equilibrium points  $z_1, z_3$ and $z_4$ are saddle points, while $z_2$ is a center. The phase portrait is drawn in Fig.\ref{eq4_Fig_phase}(c).

When $c<0$, $g<0$ and $|g|<\displaystyle\frac{c^2}{2}$, the equilibrium points $z_1, z_3$ and $z_4$ are saddle points, while $z_2$ is a center. The phase portrait is drawn in Fig.\ref{eq4_Fig_phase}(d).

When the constant of integration $g$ of Eq. \eqref{eq4_trav} is zero, the equilibrium points in \eqref{eq4_sd_equi} reduces to:
\begin{equation}\label{eq4_sd_equi0}
z_1\equiv \left(0, 0\right),\qquad  z_2\equiv \left(-\frac{c}{4}, 0\right),\qquad
z_3\equiv \left(-\frac{c}{2},-\frac{c}{2}\right),\qquad z_4\equiv \left(-\frac{c}{2},\frac{c}{2}\right),
\end{equation}
and $\forall c\neq 0$ the singular equilibrium points $z_1, z_2$ and the regular equilibrium $z_1$ are saddles; the regular equilibrium $z_2$ is a center. 	
The phase portrait of Eq. \eqref{eq4_sd_reg} for $g = 0$ are drawn in Fig.\ref{eq4_Fig_phase}(e) and (f) for $c > 0$ and $c < 0$ respectively.

Now, Theorem \ref{theo2} can be seen (following the details in \cite{LD07}) to apply to the closed orbits adjacent to the singular straight line $\phi =\phi_s=-\displaystyle\frac{c}{2}$ in Fig.\ref{eq4_Fig_phase}(b),(d),(e) and (f).
In particular, in Fig.\ref{eq4_Fig_phase}(b) and (d), we notice that the closed orbits adjacent to the singular straight line have the arched curve form seen earlier in Fig.\ref{pp1}(a). Thus, from Theorem \ref{theo3}(1), we conclude that this arch curve gives rise to periodic cusp waves of the peak type, called cuspons. Similarly, in Fig.\ref{eq4_Fig_phase}(e) and (f), the closed orbits adjacent to the singular straight line have the curve triangular form seen in Fig.\ref{pp1}(b), and they give rise to solitary cusp waves of the peak type, called peakons,
as follows from Theorem \ref{theo3}(2).

\subsection{Phase portrait/possible solutions of Eq. \eqref{eq2}}\label{subsec3.2}
Let us rewrite Eq. \eqref{eq2} in the following form:
\begin{equation}\label{eq2_pp}
u_t-u_{xxt}=\frac{\partial}{\partial x} \left(8u^2-4uu_{xx}-2u_x^2+2u_x u_{xx}\right).
\end{equation}
By the usual substitution $u(x,t)=\phi(x-ct)=\phi(z)$ into Eq. \eqref{eq2_pp} and integrating one time the resulting equation with respect to $z$, we obtain the following traveling wave equation:
\begin{equation}\label{eq2_trav}
-c\phi +c \frac{d^2\phi}{dz^2}=8\phi^2-4\phi\frac{d^2\phi}{dz^2}-2\left(\frac{d\phi}{dz}\right)^2+2\frac{d\phi}{dz}\frac{d^2\phi}{dz^2}+g,
\end{equation}
where $g$ is the constant of integration.
Eq. \eqref{eq2_trav} is equivalent to the following $2$-dimensional dynamical system:
\begin{equation}\label{eq2_sd}
\begin{array}{ll}
\displaystyle\frac{d\phi}{dz}=y, \\
\,\\
\displaystyle\frac{dy}{dz}=\frac{8\phi^2-2y^2+c\phi+g}{c+4\phi-2y},
\end{array}
\end{equation}
belonging to the second class of singular traveling wave system given in Eq. \eqref{gen_df}. Here $c+4\phi-2y=0$ defines the set of real planar curves along which the second equation of the system \eqref{eq2_sd} is discontinuous. Moreover, the quantity $\displaystyle\frac{dy}{dz}$ changes its sign as the phase point $(\phi ,y)$ passes through every branch of $c+4\phi-2y=0$.

Following the procedure described in Section \eqref{secondType}, we make the coordinate transformation $dz= (c+4\phi -2y)d\zeta$, for $c+4\phi -2y\neq 0$, to obtain the following regular system associated to the system \eqref{eq2_sd}:
\begin{equation}\label{eq2_sd_reg}
\begin{array}{ll}
\displaystyle\frac{d\phi}{d\zeta}=y(c+4\phi-2y), \\
\,\\
\displaystyle\frac{dy}{d\zeta}={8\phi^2-2y^2+c\phi+g}.
\end{array}
\end{equation}
Let  $L(\phi)=8\phi^2+c\phi+g$ and $L'(\phi)=16\phi +c$; for a fixed $c>0$, it is straightforward show that  the following hold:
\begin{itemize}
\item\noindent $g>0$. When $g>\displaystyle\frac{c^2}{32}$, $L(\phi)$ has no real zero; when $g=\displaystyle\frac{c^2}{32}$, $L(\phi)$ has a double zero $z_{12}=-\displaystyle\frac{c}{16}$; when $g<\displaystyle\frac{c^2}{32}$, $L(\phi)$ has two simple zeros $z_1<-\displaystyle\frac{c}{16}<z_2$.
\item\noindent $g<0$. When $|g|>\displaystyle\frac{c^2}{32}$, $L(\phi)$ has two simple zeros $z_1<0<z_2$; when $|g|=\displaystyle\frac{c^2}{32}$, $L(\phi)$ has two simple zeros $z_1<-g<0<z_2$; when $|g|<\displaystyle\frac{c^2}{32}$, $L(\phi)$ has two simple zeros $z_1<-g<0<z_2$.
    \end{itemize}
The stationary states of system \eqref{eq2_sd_reg} are the points $z_1\equiv \left(\displaystyle\frac{-c+\sqrt{c^2-32g}}{16},0\right)$,
$z_2\equiv \left(\displaystyle\frac{-c\sqrt{c^2-32g}}{16},0\right)$ and $z_3\equiv \left(\displaystyle\frac{-c^2+2g}{6c},\frac{-c^2+4g}{6c}\right)$. Here $z_1$ and $z_2$ are the regular equilibrium points, while $z_3$ is a singular equilibrium point.  By linear stability analysis of Eq. \eqref{eq2_sd_reg} we classify the qualitative behaviour of these equilibrium points.

For  $ c>0$ and $0<g<\displaystyle\frac{c^2}{32}$, the equilibrium points $z_1$ and $z_3$ are saddle, while the equilibrium point $z_2$ is a center. The phase portrait is drawn in Figure 5(a).

For $g>0, c > 0$ and $g=\displaystyle\frac{c^2}{32}$, the equilibrium point $z_1=z_2\equiv\left(-\displaystyle\frac{c}{16},0\right)$ is a center, while the equilibrium point $z_3$ is a saddle.

\begin{figure}
\begin{center}
\subfigure[] {\epsfxsize=2 in \epsfbox{vuoto.eps}}
\subfigure[] {\epsfxsize=2 in \epsfbox{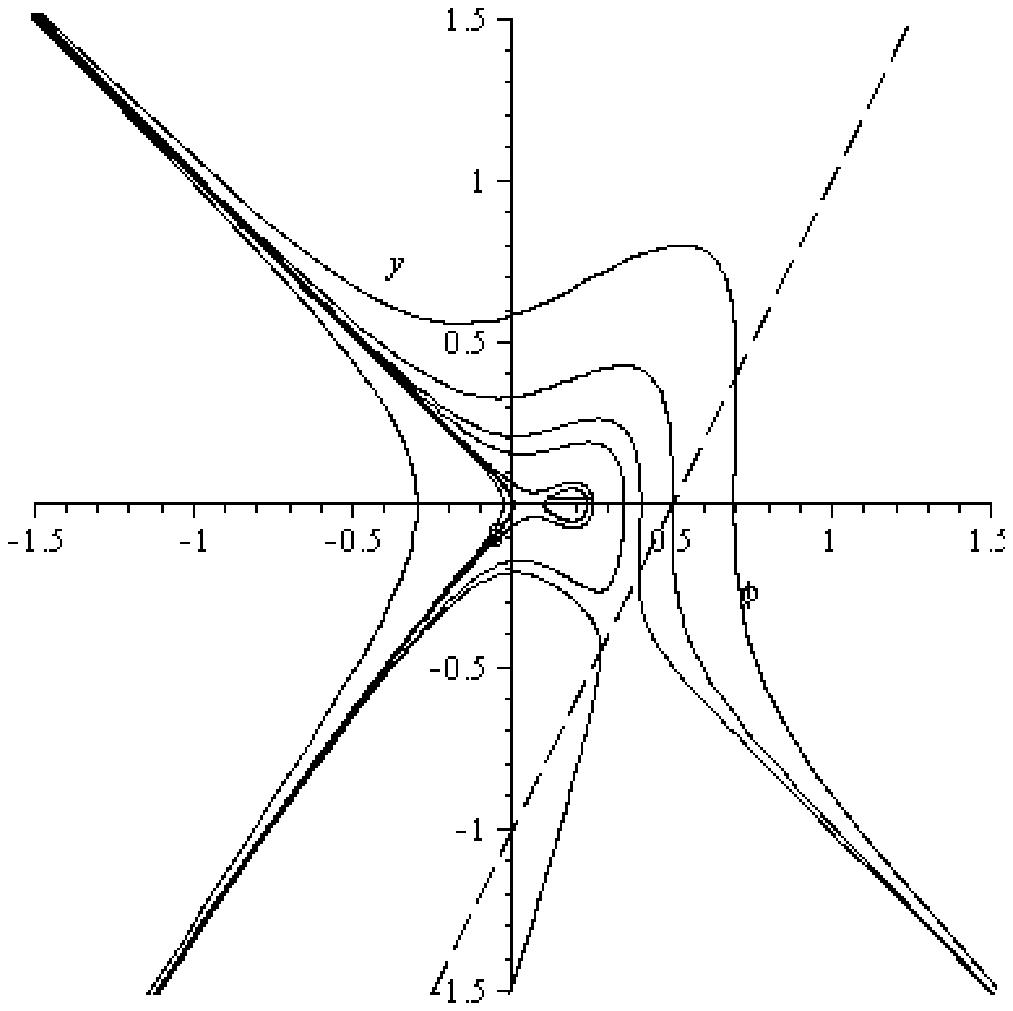}}
\subfigure[] {\epsfxsize=2 in \epsfbox{vuoto.eps}}
\subfigure[] {\epsfxsize=2 in \epsfbox{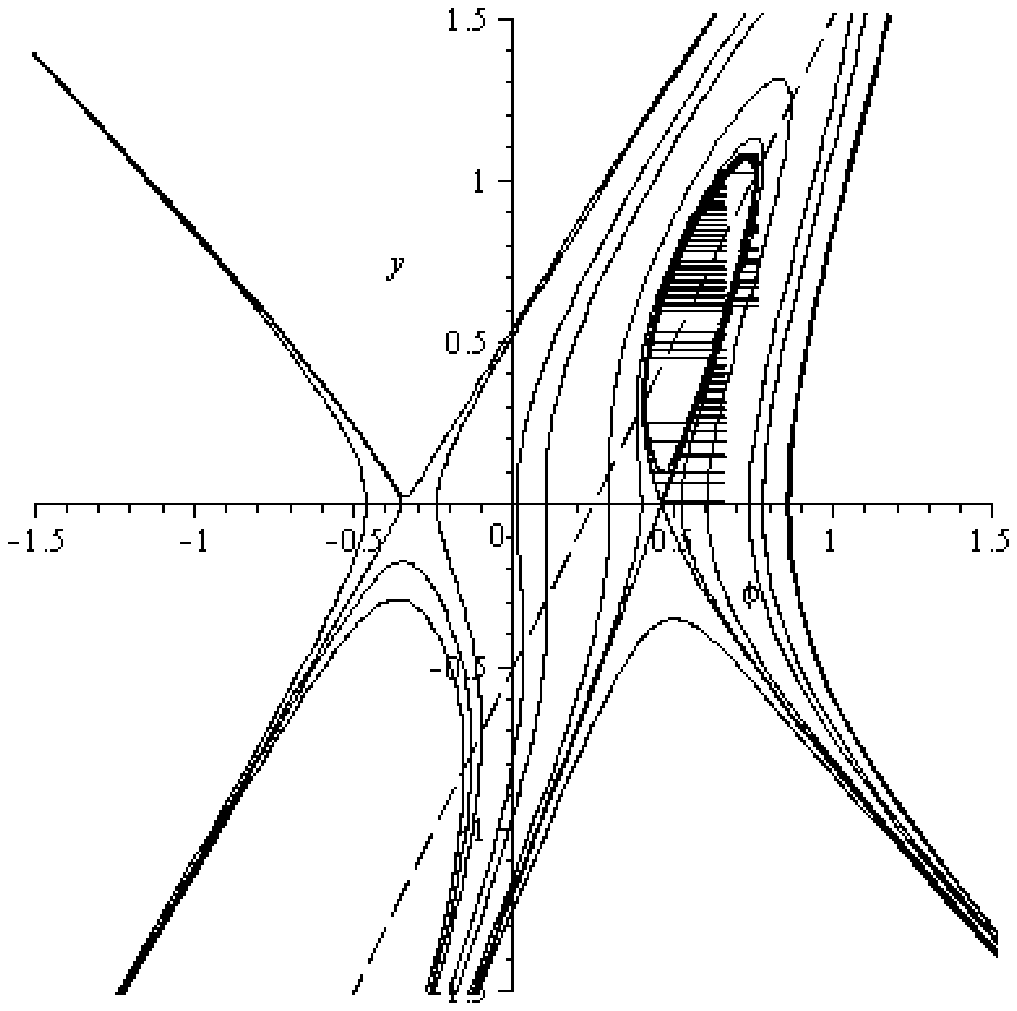}}
\subfigure[] {\epsfxsize=2 in \epsfbox{vuoto.eps}}
\subfigure[] {\epsfxsize=2 in \epsfbox{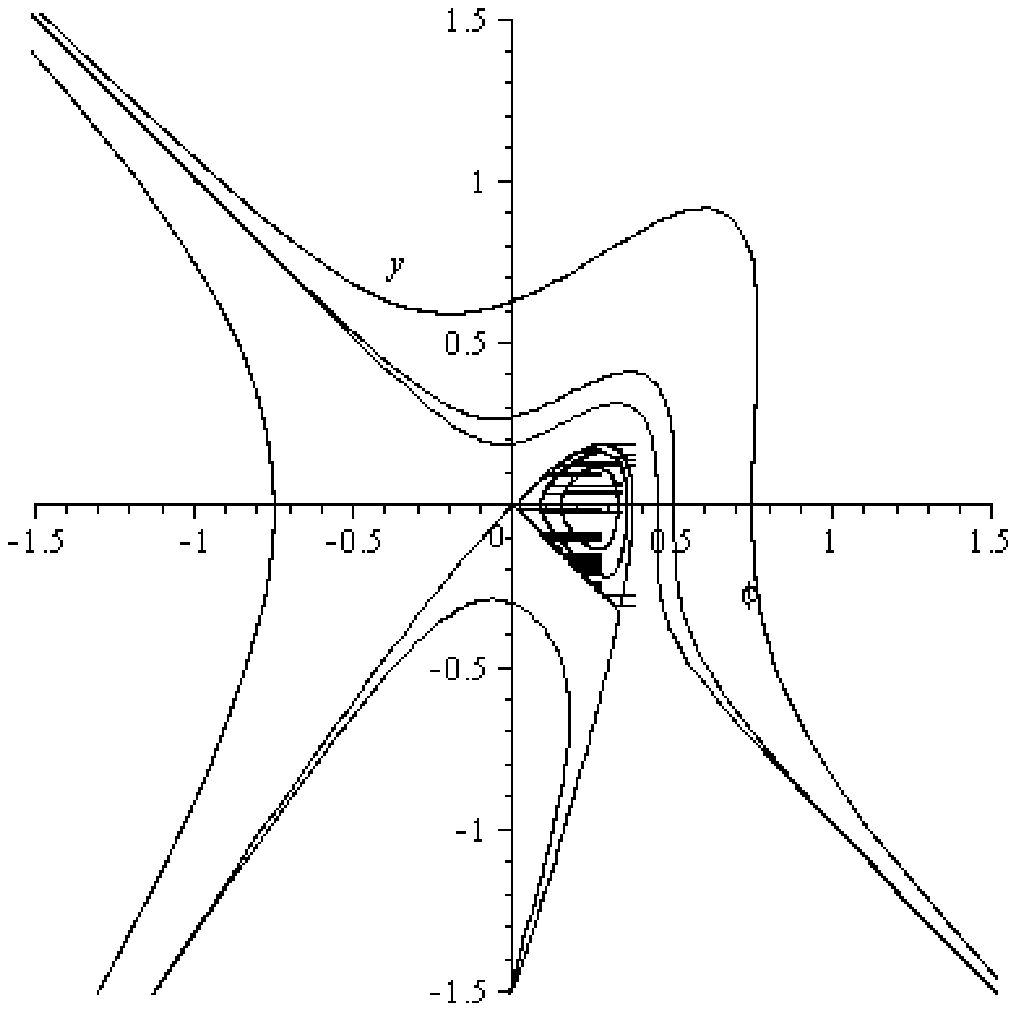}}
\end{center}
\caption{\label{eq2_Fig_phase} The phase portraits ($y$ plotted against $\phi$) of system \eqref{eq2_sd_reg}. The dotted line is the singular line $c+4\phi-2y=0$. (a) $c > 0$ and
$0 <g < \displaystyle\frac{c^2}{32}$. (b) $c < 0$ and  $0<g<\displaystyle\frac{c^2}{32}$. (c) $c > 0$ and $g>-\displaystyle\frac{c^2}{32}$. (d) $c < 0$  and $-\displaystyle\frac{c^2}{32}<g<0$.
(e) $c > 0$ and $g = 0$. (f) $c < 0$ and $g = 0$.}
\end{figure}

For $c < 0$ and  $0<g<\displaystyle\frac{c^2}{32}$, the equilibrium points $z_2$ and $z_3$ are saddle, while the equilibrium point $z_1$ is a center. The phase portrait is drawn in Fig.\ref{eq2_Fig_phase}(b).

For $g < 0, c > 0$ and $g>-\displaystyle\frac{c^2}{32}$, the equilibrium points $z_1$ and $z_3$ are saddle, while the equilibrium point $z_2$ is a center. The phase portrait is drawn in Fig.\ref{eq2_Fig_phase}(c).

For $c < 0$  and $-\displaystyle\frac{c^2}{32}<g<0$, the equilibrium points $z_2$ and $z_3$ are saddle, while the equilibrium point $z_1$ is a center. The phase portrait is drawn in Fig.\ref{eq2_Fig_phase}(d).
 	
For $g=0$, the system \eqref{eq2_sd_reg} has the regular equilibria $z_1\equiv (0,0)$, $z_2\equiv\left(-\displaystyle\frac{c}{8},0\right)$ and the singular equilibrium $z_3\equiv\left(-\displaystyle\frac{c}{6},-\displaystyle\frac{c}{6}\right)$. Of these $z_1$ and $z_3$ are saddle points, whereas $z_2$ is a center. The phase portraits of Eq. \eqref{eq2_sd_reg} for $g=0$ are drawn in Fig.\ref{eq2_Fig_phase}(e) and Fig.\ref{eq2_Fig_phase}(f) for $c > 0$ and $c < 0$ respectively.

From Fig.\ref{eq2_Fig_phase} we see that there are no closed orbits adjacent to the singular straight line $c+4\phi-2y=0$ which, as they limit to the singular straight line $c+4\phi-2y=0$, could give us singular solutions. And from the discussion and theorems given in Section \ref{Sec2}, in the absence of either closed arched curves or curved triangles, neither singular peakons nor cuspons are possible.

\subsection{Phase portrait/possible solutions of Eq. \eqref{eq3}}\label{subsec3.3}
Let us rewrite Eq. \eqref{eq3} in the following useful form:
\begin{equation}\label{eq3_pp}
u_t-u_{xxt}=D_x \left(u^2 u_{xx}-u_x^2 u_{xx}+uu_x^2-u^3\right).
\end{equation}
By substituting $u(x,t)=\phi(x-ct)=\phi(z)$ in \eqref{eq3_pp} and integrating with respect to $z$, we obtain:
%
%
\begin{equation}\label{eq3_trav}
-c\phi+c\frac{d^2\phi}{dz^2}= \phi^2\frac{d^2\phi}{dz^2}+\phi
\left(\frac{d\phi}{dz}\right)^2-\left(\frac{d\phi}{dz}\right)^2\frac{d^2\phi}{dz^2}-\phi ^3-g,
\end{equation}
where $g$ is the constant of integration. Eq. \eqref{eq3_trav} is the traveling wave equation of \eqref{eq3_pp} and it is equivalent to the following planar dynamical system:
\begin{equation}\label{eq3_sd}
\begin{array}{ll}
\displaystyle\frac{d\phi}{dz}=y,\\
   \,\\
\displaystyle\frac{dy}{dz}=\frac{\phi y^2+c\phi-g-\phi^3}{c-\phi^2+y^2}.
\end{array}
\end{equation}
Eq. \eqref{eq3_sd} belongs to the second type of singular traveling wave system described in Section \ref{secondType}.
The second equation in \eqref{eq3_sd} is discontinuous on the planar curves $c-\phi^2+y^2=0$ and $\displaystyle\frac{dy}{dz}$ changes its sign as the phase point $(\phi, y)$ passes through every branch of $c-\phi^2+y^2=0$.
In order to obtain the regular system associated to Eq. \eqref{eq3_sd}, we make the transformation $dz=(c-\phi^2+y^2)d\zeta$, for $c-\phi^2+y^2\neq 0$ and we get:
\begin{equation}\label{eq3_sd_reg}
\begin{array}{ll}
\displaystyle\frac{d\phi}{d\zeta}=y(c-\phi^2+y^2),\\
   \,\\
\displaystyle\frac{dy}{d\zeta}=-g+\phi(y^2-\phi^2+c).
\end{array}
\end{equation}
Let $P(\phi)=\phi^3-c\phi+g$ and $P'(\phi)=3\phi^2-c$. For a fixed $c>0$ the following hold:
\begin{itemize}
\item\noindent	$g>0$. When $g>\displaystyle\frac{2c}{3}\sqrt{\frac{c}{3}}$, $P(\phi)$  has only a negative zero $z_3<-\displaystyle\sqrt{\frac{c}{3}}$; when $g=\displaystyle\frac{2c}{3}\sqrt{\frac{c}{3}}$, $P(\phi)$ has one simple zero $z_3$ and a double zero $z_{2,1}=\displaystyle\sqrt{\frac{c}{3}}$; when $g<\displaystyle\frac{2c}{3}\sqrt{\frac{c}{3}}$, $P(\phi)$ has three simple zeros $z_3<0<z_2<\displaystyle\sqrt{\frac{c}{3}}<z_1<\sqrt{c}$.
\item\noindent	$g<0$. When $|g|>\displaystyle\frac{2c}{3}\sqrt{\frac{c}{3}}$, $P(\phi)$ has only a positive zero $z_1>\displaystyle\sqrt{\frac{c}{3}}$; when $|g|=\displaystyle\frac{2c}{3}\sqrt{\frac{c}{3}}$, $P(\phi)$ has one simple zero $z_1$ and a double zero $z_{23}=-\displaystyle\sqrt{\frac{c}{3}}$; when $|g|<\displaystyle\frac{2c}{3}\sqrt{\frac{c}{3}}$, $P(\phi)$ has three simple zeros $-\sqrt{c}<z_3<-\displaystyle\sqrt{\frac{c}{3}}<z_2<0<\displaystyle\sqrt{\frac{c}{3}}<z_1$.
\end{itemize}
Suppose $\mathcal{M}(z_i,0)$ be the coefficient matrix of the linearized system \eqref{eq3_sd_reg} at the critical point $S_i(z_i,0)$. Then, the jacobian can be computed as follows:
\begin{equation}\label{eq3_jac}
J(z_i,0)={\rm det}\mathcal{M}(z_i,0)=(c-z_i^2)(3z_i^2-c).
\end{equation}
Let us define $h_i=H(z_i,0),\ i=1,2,3$, where $H$ is the first integral of the system \eqref{eq3_sd_reg}. It is clear from Eq. \eqref{eq3_sd_reg} that when $g>\displaystyle\frac{2c}{3}\sqrt{\frac{c}{3}}$ (or $|g|>\displaystyle\frac{2c}{3}\sqrt{\frac{c}{3}}$, $g<0$), the unique critical point $S_3 (z_3,0)$ of \eqref{eq3_jac} is a saddle point; when $g^*<g<\displaystyle\frac{2c}{3}\sqrt{\frac{c}{3}}$ the point $S_1 (z_1,0)$ is a center and $S_2 (z_2,0)$ is a saddle point, where the value $g^*$ is defined such that the homoclinic orbit of Eq. \eqref{eq3_sd_reg}, determined by the first integral $H(\phi ,y)=h_2$ to the saddle point $S_2 (z_2,0)$, passes through the point $S_s (\sqrt{c},0)$, with $h_2=\displaystyle\frac{1}{4} g^*\sqrt{c}$.

If $g=0$, for both $c>0$ and $c<0$ the equilibrium point $(0,0)$ is a saddle point.

\begin{figure}
\begin{center}
\subfigure[] {\epsfxsize=2 in \epsfbox{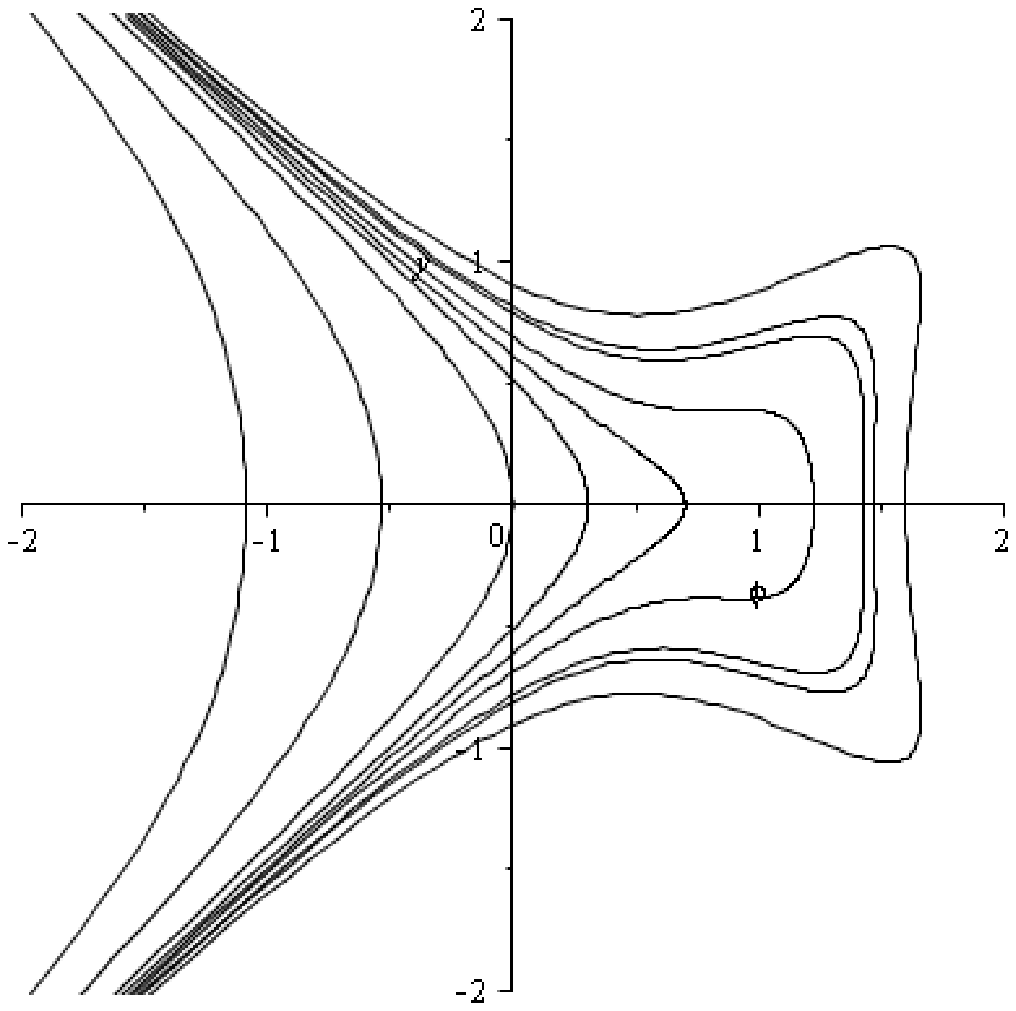}}
\subfigure[] {\epsfxsize=2 in \epsfbox{vuoto.eps}}
\subfigure[] {\epsfxsize=2 in \epsfbox{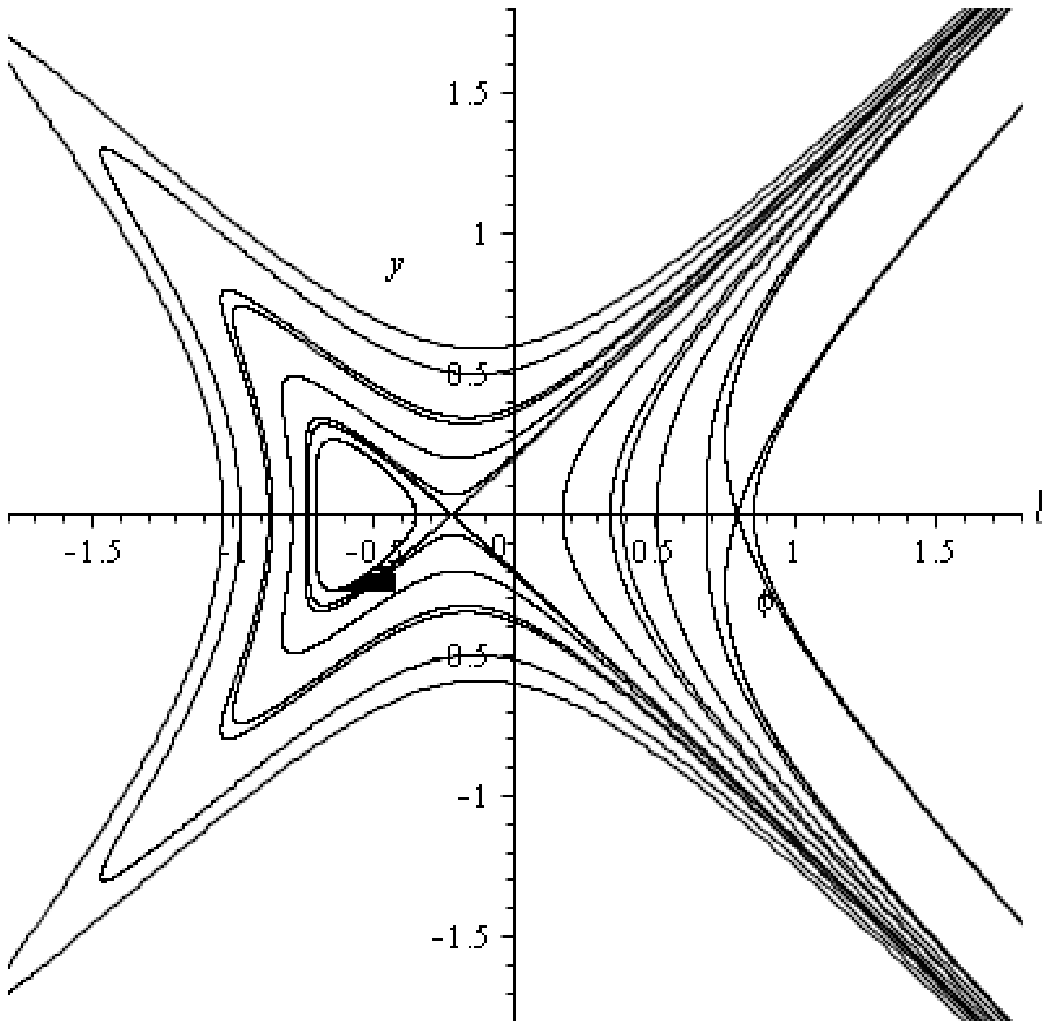}}
\subfigure[] {\epsfxsize=2 in \epsfbox{vuoto.eps}}
\subfigure[] {\epsfxsize=2 in \epsfbox{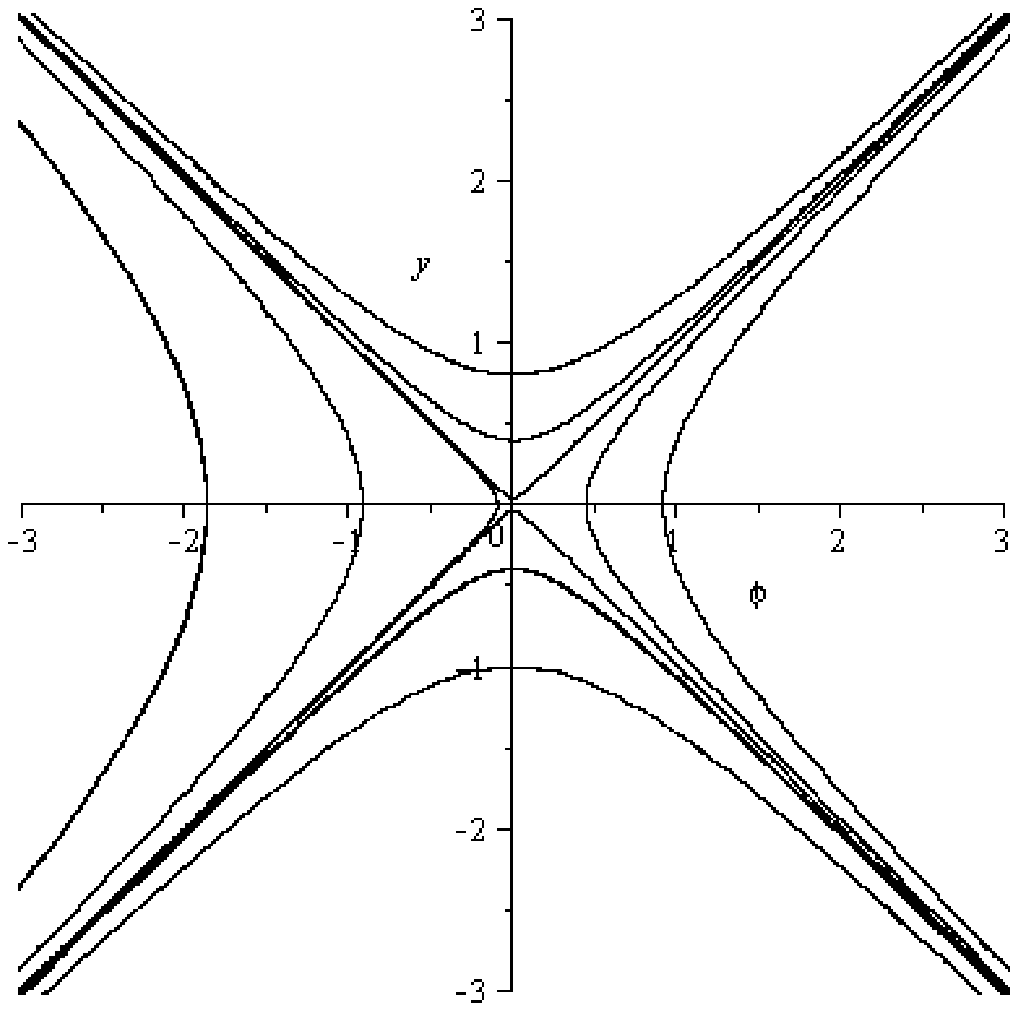}}
\subfigure[] {\epsfxsize=2 in \epsfbox{vuoto.eps}}
\end{center}
\caption{\label{eq3_Fig_phase} The phase portraits ($y$ plotted against $\phi$) of system \eqref{eq3_sd_reg}. (a) $c > 0$ and
$g > 0, g > \displaystyle\frac{2c}{3}\sqrt{\frac{c}{3}}$. (b) $c > 0$ and
$0<g < \displaystyle\frac{2c}{3}\sqrt{\frac{c}{3}}$. (c) $c > 0$ and
$g < 0, |g| < \displaystyle\frac{2c}{3}\sqrt{\frac{c}{3}}$. (d) $c > 0$ and
$g < 0, |g| > \displaystyle\frac{2c}{3}\sqrt{\frac{c}{3}}$.
(e) $c > 0$ and $g = 0$. (f) $c < 0$ and $g = 0$.}
\end{figure}

By using the above results we can draw the phase portrait of \eqref{eq3_sd_reg}, depending on the parameters $(g,c)$. The phase portraits of \eqref{eq3_sd_reg} are shown in Fig.\ref{eq3_Fig_phase} where the graph of the hyperbola $\phi^2-y^2=c$ is also shown Fig.\ref{eq3_Fig_phase}(b). We note that, unlike the case of having singular straight line as in system \eqref{gen_ld}, for the system \eqref{eq3_sd_reg} the hyperbola $\phi^2-y^2=c$ is not a solution. For every fixed $c>0$, when $0<g<g^*$, system \eqref{eq3_sd_reg} has got a family of periodic orbits defined by the first integral $H(\phi ,y)=h$. There exists homoclinic orbit of system \eqref{eq3_sd_reg} defined by $H(\phi ,y)=h_2$, which transversely intersect the hyperbola $\phi^2-y^2=c$ at two points, $P^\pm (\phi_s,\pm y_s)$, where $\phi_s=\displaystyle\frac{h_2}{g}$ and $y_s=\sqrt{c-\left(\displaystyle\frac{h_2}{g}\right)^2}$.

Singular solutions of Eq. \eqref{eq3} have been considered earlier, see \cite{DF10}. In particular, there are
four-segmented M-waves. Hence, using these phase-plots, we will consider the regular solutions of \eqref{eq3} subsequently in Section \ref{subsec4_3}.

\section{Regular pulse and front solutions of generalized Camassa-Holm equations: analytic solutions for homoclinic and heteroclinic orbits }\label{Sec4}

In this section, we change gears and consider regular pulse and front solutions of the generalized Camassa-Holm Eqs. \eqref{eq4}--\eqref{eq3}
by calculating convergent, multi-infinite, series solutions for the possible homoclinic and heteroclinic orbits of their traveling wave equations.

Homoclinic orbits of dynamical systems have been widely treated in recent years by a variety of approaches. For instance, an early review integrating bifurcation theoretical and numerical approaches was given in \cite{C98}. Homoclinic orbits are important in applications for a variety of reasons. In the context of ODE systems, they are often anchors for the local dynamics in their vicinity. Under certain conditions, their existence may indicate the existence of chaos in their neighborhood \cite{K95,G94}. In a totally different setting, if the governing dynamical system is the traveling-wave ODE for a partial differential equation or equations, its homoclinic orbits correspond to the solitary wave or pulse solutions of the PDEs, which have many important uses and applications in nonlinear wave propagation theory, nonlinear optics, and in various other settings \cite{C98}.

We employ a recently developed approach \cite{CG13,W09}, using the method of undetermined coefficients to derive heteroclinic and homoclinic orbits of Eqs. \eqref{eq4_trav}, \eqref{eq2_trav} and \eqref{eq3_trav}. Convergent analytic series for these orbits corresponding to pulse/front solutions of the generalized CH Eqs. \eqref{eq4}--\eqref{eq3} are derived and investigated here.

\subsection{Infinite Series for homoclinic orbits of Eq. \eqref{eq4_trav}}\label{subsec4_4}

We rewrite Eq. \eqref{eq4_trav} in the following useful form:
\begin{equation}\label{eq4_trav_s}
-c\phi+c\phi^{''}=4\phi^2-2\phi\phi^{''}-2\phi^{'2}+g,
\end{equation}
where the prime $'$ indicates the derivative with respect to $z$.
Let $\bar{z}$ an equilibrium of Eq.\eqref{eq4_trav_s}, corresponding to a regular equilibrium point $(x_0,0)$ of the Eq. \eqref{eq4_sd}. We assume to choose the parameters $c$ and $g$ in such a way that $\bar{z}$ is a saddle point and a homoclinic orbit to this equilibrium is given (see as an example Fig.\ref{eq4_Fig_phase}(a) and (c)).	
Let us now proceed to construct the homoclinic orbit of Eq. \eqref{eq4_trav_s}.

We look for a solution of the following form:
\begin{equation}\label{eq4_hom_orbit}
\phi(z)=\left\{\begin{array}{lll}
\phi^+(z)\qquad z>0\\
0\qquad\qquad z=0\\
\phi^-(z)\qquad z<0
\end{array}\right.
\end{equation}
where:
\begin{equation}\label{eq4_hom_spsm}
\phi^+(z)=x_0+\sum_{k=\,1}^{\infty} a_k e^{k\alpha z}, \qquad \phi^-(z)=x_0+\sum_{k=\,1}^{\infty} b_k e^{k\beta z},
\end{equation}
$\alpha<0$ and $\beta>0$ are undetermined constants and $a_k, b_k$, with $k\geq 1$, are, at the outset, arbitrary coefficients. Substituting the series \eqref{eq4_hom_spsm} for $\phi^+(z)$ we obtain the following expressions for each term of \eqref{eq4_trav_s}:
\begin{eqnarray}\label{eq4_terms1}
\phi^{''}&=&\sum_{k=\,1}^{\infty}a_k(k\alpha)^2e^{k\alpha z},\\\label{eq4_terms2}
\phi^{2}&=&\sum_{k=\,2}^{\infty}\sum_{i=\,1}^{k-1}a_{k-i}a_ie^{k\alpha z}
+2x_0\sum_{k=\,1}^{\infty}a_k e^{k\alpha z}+x_0^2,\\\label{eq4_terms3}
\phi\phi^{''}&=&\sum_{k=\,2}^{\infty}\sum_{i=\,1}^{k-1}(k-i)^2\alpha^2a_{k-i}a_ie^{k\alpha z}
+x_0\sum_{k=\,1}^{\infty}(k\alpha)^2a_k e^{k\alpha z},\\\label{eq4_terms4}
\phi^{'2}&=&\sum_{k=\,2}^{\infty}\sum_{i=\,1}^{k-1}i\alpha^2a_{k-i}a_ie^{k\alpha z}.
\end{eqnarray}
Using \eqref{eq4_terms1}-\eqref{eq4_terms4} into the Eq. \eqref{eq4_trav_s} we have:
\begin{equation}\label{eq4_series}
\begin{split}
-cx_0-4x_0^2&\,-g+
\sum_{k=1}^{\infty}((k\alpha)^2 (c+2x_0)-(8x_0+c))a_k e^{k\alpha z}\\
&\,=\sum_{k=2}^{\infty}\sum_{i=1}^{k-1} (2(k-j)i\alpha^2+2(k-i)^2\alpha^2-3)a_{k-i}a_i e^{k\alpha z}=0.
\end{split}
\end{equation}
As $x_0$ is an equilibrium of Eq. \eqref{eq4_trav_s},  $-cx_0-4x_0^2-g=0$. Comparing the coefficients of $e^{k\alpha z}$ for each $k$, one has for $k=1$:
\begin{equation}\label{eq4_series1}
(\alpha^2 (c+2x_0)-8x_0-c)a_1=0.
\end{equation}
Assuming $a_1\neq0$ (otherwise $a_k=0$ for all $k>1$ by induction), results in the two possible values of $\alpha$:
\begin{equation}\label{4_series_eig}
\alpha_{1}= \sqrt{\frac{8x_0+c}{c+2x_0}},\qquad\qquad \alpha_{2}=-\sqrt{\frac{8x_0+c}{c+2x_0}}.
\end{equation}
We are dealing with the case when the equilibrium $x_0$ is a saddle. In this case, as our series solution \eqref{eq4_hom_orbit} needs to converge for $z >0$, we pick the negative root $\alpha=\alpha_2$ (here we skip all the details on how choosing $c$ and $g$ in such a way that the eigenvalues $\alpha_i$ are real and opposite, as they are given in Section \ref{subsec3.4}). Thus we have:
\begin{equation}\label{eq4_series_2}
F(2\alpha_2)a_2=4(1-\alpha^2_2)a_1^2,
\end{equation}
where  $F(k\alpha_2)=(k\alpha)^2 (c+2x_0)-(8x_0+c)$ and the coefficient $a_2$ is easily obtained in terms of $a_1$ as follows:
\begin{equation}\label{eq4_series_coeff2}
a_2=\frac{4(1-\alpha^2_2)}{F(2\alpha_2)}a_1^2.
\end{equation}
For $k=3$ we obtain:
\begin{equation}\label{eq4_series_coeff3}
a_3=\frac{2(-9\alpha_2^2+4)a_1a_2}{F(3\alpha_2)}.
\end{equation}
Once substituted the formula \eqref{eq4_series_coeff2} into the Eq. \eqref{eq4_series_coeff3}, one obtains $a_3$ in terms of $a_1$. For $k>2$ one has:
\begin{equation}\label{eq4_series_coeffk_a}
a_k=\sum_{i=1}^{k-1}\frac{(2(k-i)i\alpha_2^2+2(k-i)^2\alpha_2^2-3)a_{k-i}a_i}{F(k\alpha_2)}.
\end{equation}
Therefore for all $k$ the series coefficients $a_k$ can be iteratively computed in terms of $a_1$:
\begin{equation}\label{eq4_series_coeffk}
a_k=\varphi_k a_1^k,                                                                                  \end{equation}
where $\varphi_k,\ k>1$ are functions which can be obtained using Eq. \eqref{eq4_series_coeff2}-\eqref{eq4_series_coeffk_a}. They depend on $\alpha_2$ and the constant coefficients of the Eq. \eqref{eq4_trav_s}.
The first part of the homoclinic orbit corresponding to $z>0$ has thus been determined in terms of $a_1$:         		 %
\begin{equation}\label{eq4_series_p}
\phi^+ (z)=x_0+a_1 e^{\alpha_2 z}+\sum_{k=2}^\infty \varphi_k a_1^k e^{k\alpha_2 z}.                                                 \end{equation}
Notice that the Eq. \eqref{eq4_trav_s} is reversible under the standard reversibility of classical mechanical systems:
\begin{equation}\label{eq4_norev}
z\rightarrow -z,\qquad   (\phi, \phi', \phi^{''})\rightarrow (\phi, -\phi', \phi^{''}).
\end{equation}
Mathematically, this property would translate to solutions having odd parity in $z$.
Therefore the series solution for $z < 0$ can be easily obtained based on the intrinsic symmetry property of the equation, i.e.:
\begin{equation}\label{eq4_series_m}
\phi^- (z)=x_0-a_1 e^{\alpha_2 z}-\sum_{k=2}^\infty \varphi_k a_1^k e^{k\alpha_2 z}.                                                 \end{equation}
We want to construct a solution continuous at $z=0$, therefore we impose:
\begin{equation}\label{eq4_cont}
x_0+a_1+\sum_{k=2}^\infty \varphi_k a_1^k=0.
\end{equation}
Hence we choose $a_1$ as the nontrivial solutions of the above polynomial equation \eqref{eq4_cont}.
In practice the Eq. \eqref{eq4_cont} is numerically solved and the corresponding series solutions are not unique.

\begin{figure}
\begin{center}
{\epsfxsize=3 in \epsfbox{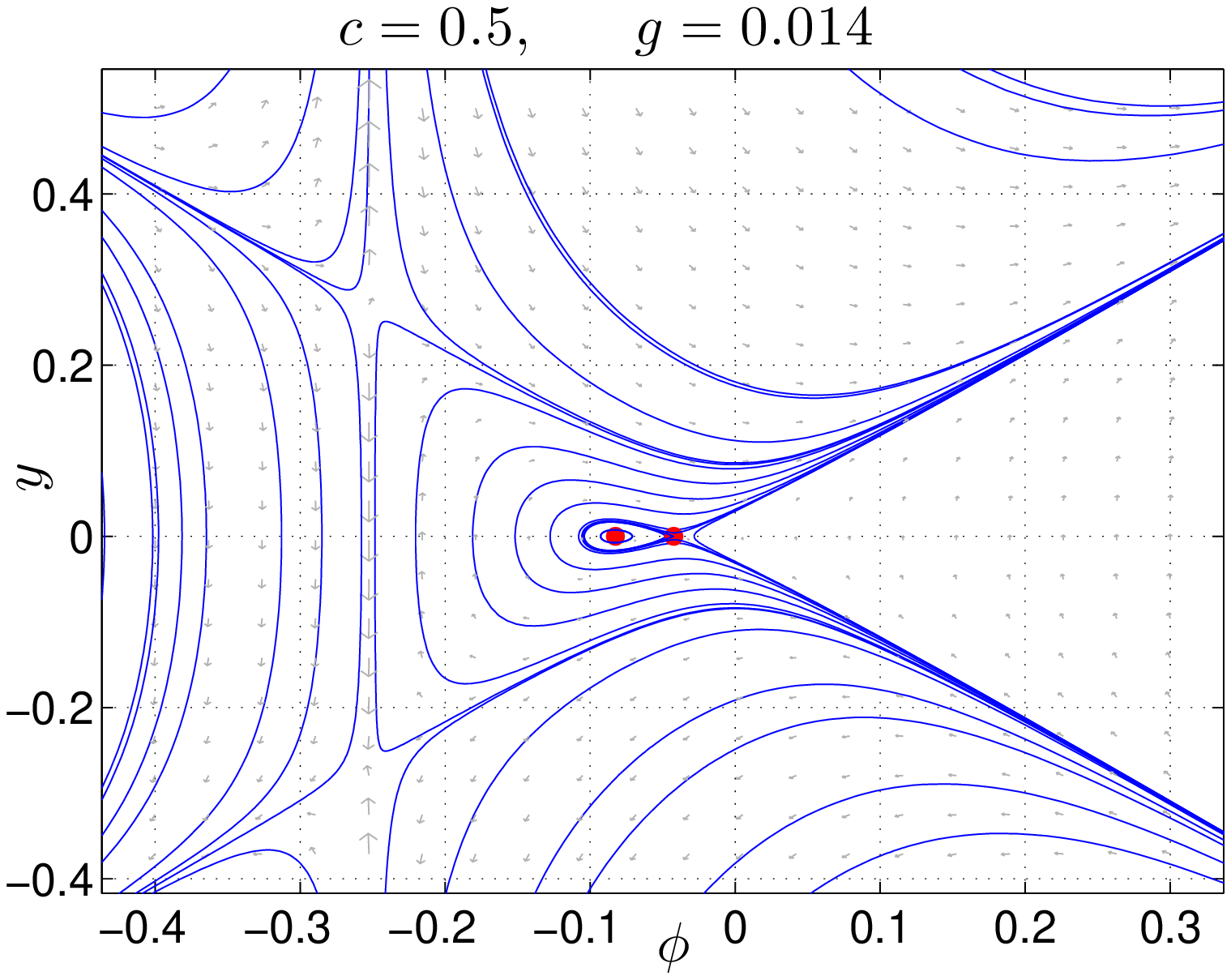}}
\end{center}
\caption{\label{Fig_eq4Hom1} Phase plane orbits of Eq. \eqref{eq4_sd_reg} when $c = 0.5$ and $g = 0.014$.}
\end{figure}

\begin{figure}
\begin{center}
\subfigure[] {\epsfxsize=3 in\epsfbox{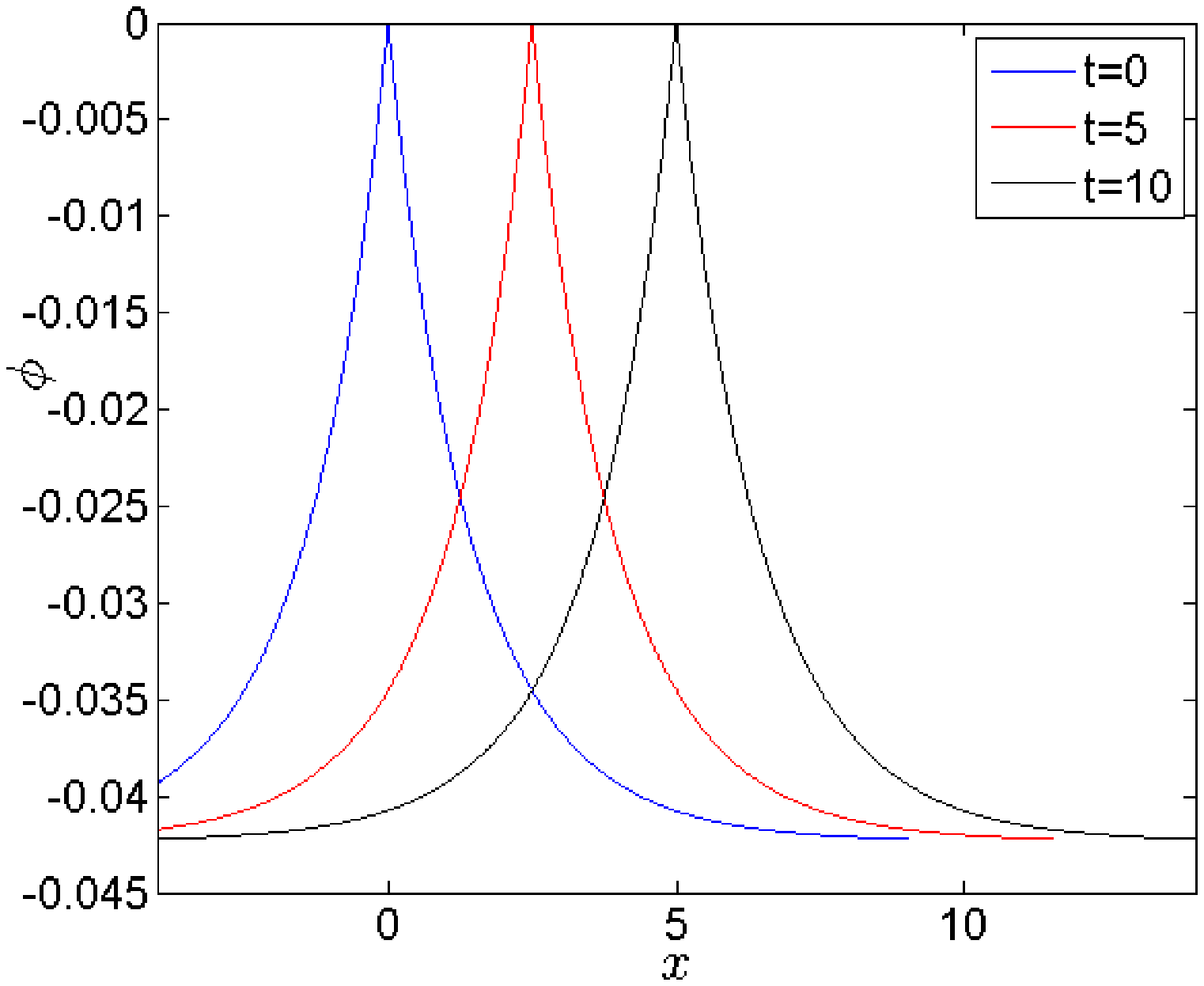}}
\subfigure[] {\epsfxsize=3 in \epsfbox{vuoto.eps}}
\end{center}
\caption{\label{eq4_Fig_hom1} The parameters are chosen as $c = 0.5$ and $g = 0.014$. (a) The series solution $\phi(z)$ in \eqref{eq4_hom_orbit} for the homoclinic orbit to the saddle point $(x_0=-0.0423,0)$ plotted as a function of $x$ for different values of $t$, showing traveling wave nature of the solution of Eq. \eqref{eq4_trav_s}. Here $a_1 = 0.0357$ is the only solution of the continuity equation \eqref{eq4_cont} truncated at $M=10$. (b) Plot of $a_k$ in \eqref{eq4_series_coeffk_a} versus $k$ shows the series coefficients are converging.}
\end{figure}

Let us now choose $c = 0.5$ and $g = 0.014$. In this case the equation \eqref{eq4_trav} admits two real equilibria: $\bar{z}_1=-0.0827$  and $\bar{z}_2=-0.0423$. In this parameter regime $\bar{z}_1$ is a center and $\bar{z}_2$ is a saddle.
Let us build the homoclinic orbit to the saddle point $(x_0,0)$, where $x_0=\bar{z}_2$, which is shown in Fig.\ref{Fig_eq4Hom1}.
We find the continuous solution for the homoclinic orbit shown in Fig.\ref{eq4_Fig_hom1}.
Notice that Fig.\ref{eq4_Fig_hom1} also shows the traveling nature of the solution. Here $a_1 = 0.0357$. For this choice of the parameters the series solution converges, as shown in  Fig.\ref{eq4_Fig_hom1}(b) where the $a_k$ rapidly goes to zero.

\begin{figure}
\begin{center}
{\epsfxsize=3 in \epsfbox{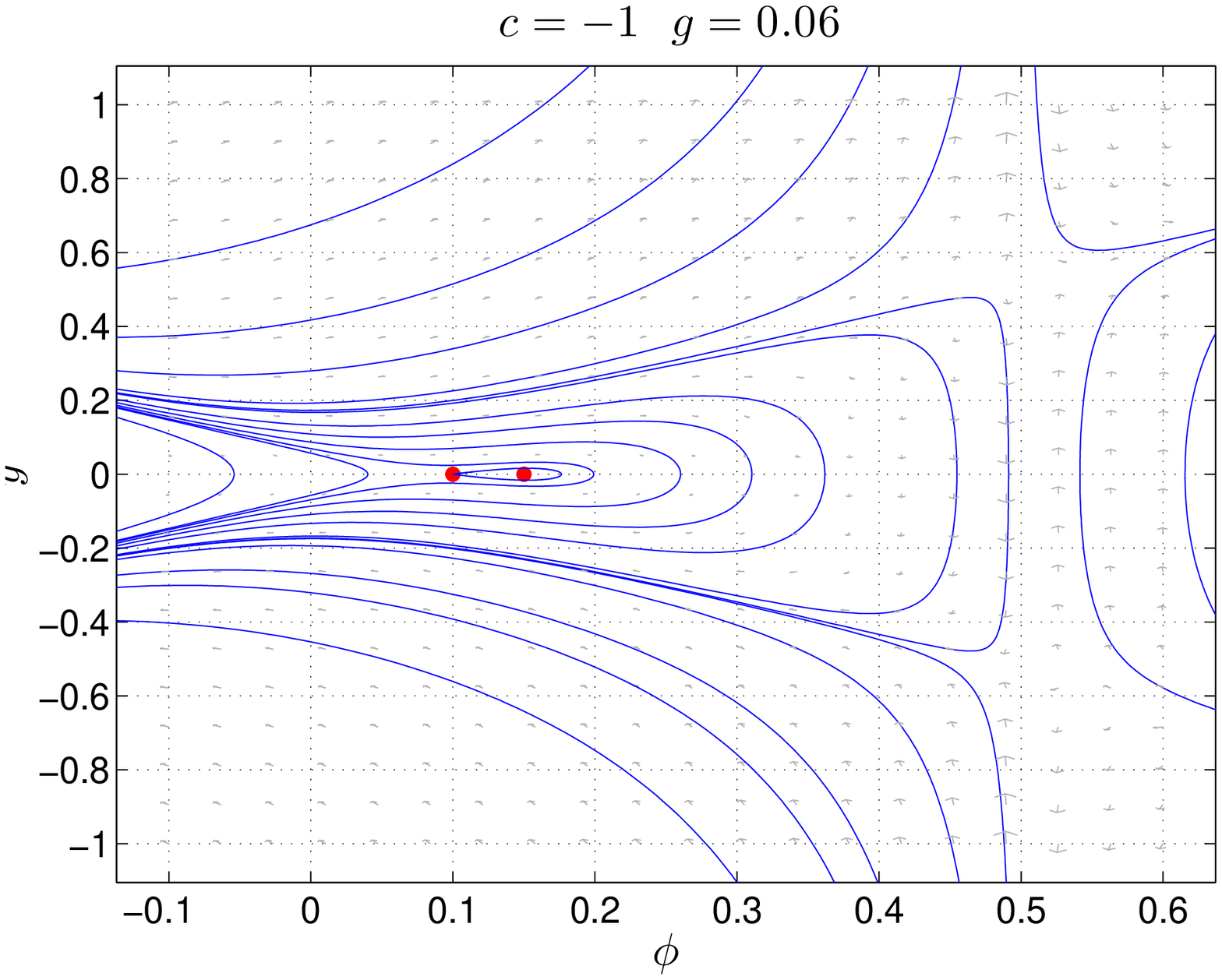}}
\end{center}
\caption{\label{Fig_eq4Hom2} Phase plane orbits of Eq. \eqref{eq4_sd_reg} when $c = -1$ and $g = 0.06$.}
\end{figure}

Let us consider another numerical example. We choose $c = -1$ and $g = 0.06$. In this case the Eq. \eqref{eq4_trav_s} admits two real equilibria, the two equilibrium points are $\bar{z}_1=0.1$ and $\bar{z}_2=0.15$. In this parameter regime $\bar{z}_1$ is a saddle (with eigenvalues  $\alpha_{1,2}=\pm 0.5$) and $\bar{z}_2$ is a center.
Let us build the homoclinic orbit to the saddle $(x_0,0)$, where $x_0=\bar{z}_1$ (see the phase portrait in Fig.\ref{Fig_eq4Hom2}).

\begin{figure}
\begin{center}
\subfigure[] {\epsfxsize=2.7 in \epsfbox{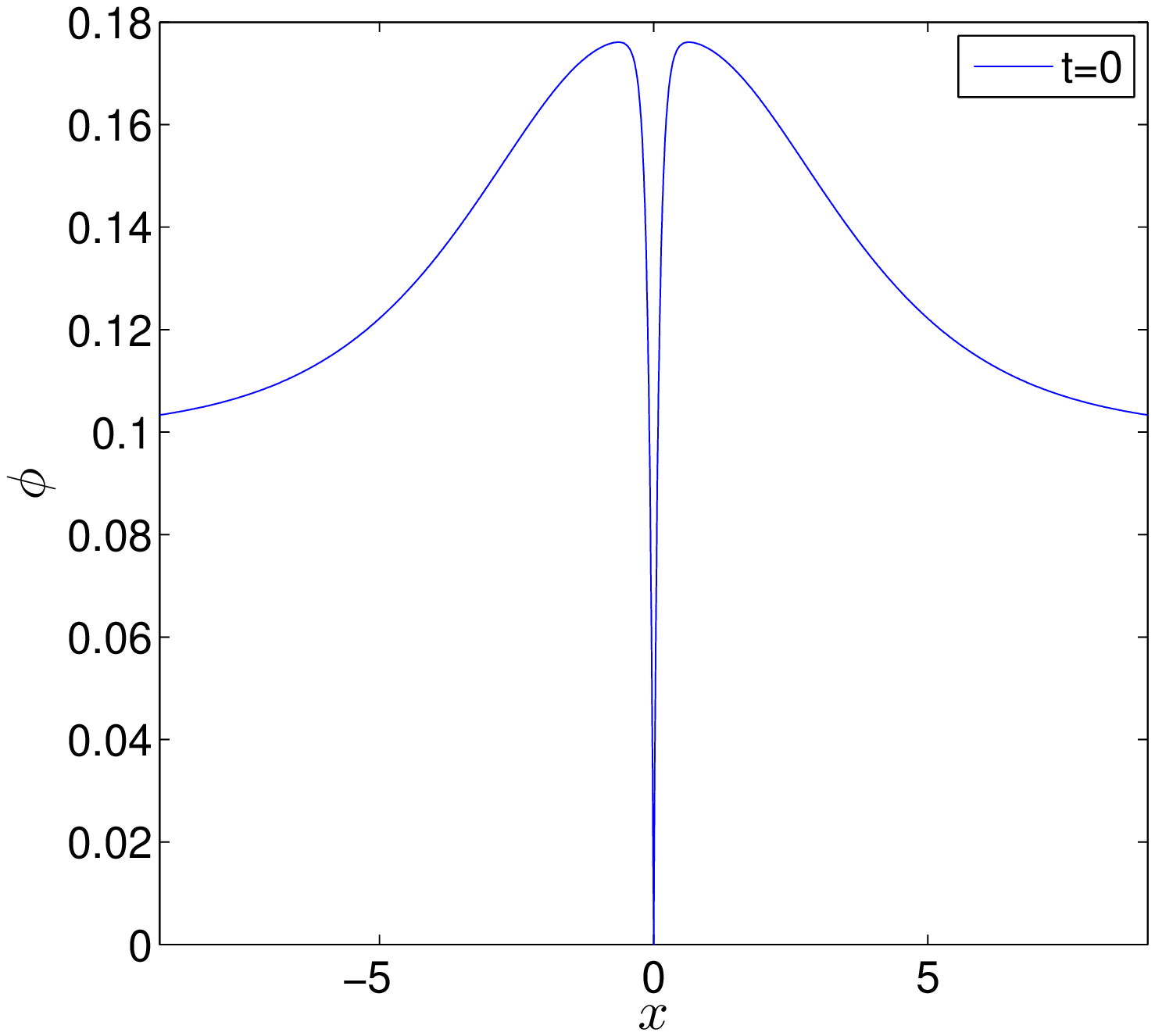}}
\subfigure[] {\epsfxsize=3 in \epsfbox{vuoto.eps}}
\end{center}
\caption{\label{eq4_Fig_hom2} The parameters are chosen as $c = -1$ and $g = 0.06$. The continuity condition \eqref{eq4_cont} truncated at $M=10$ does not admit a unique solution. Here we choose the solution $a_1= 0.3066$. (a) The series solution $\phi(z)$ in \eqref{eq4_hom_orbit} for the homoclinic orbit  to the saddle point $(x_0=0.1,0)$ plotted as a function of $x$ at $t=0$. (b) Plot of $a_k$ in \eqref{eq4_series_coeffk_a} versus $k$ shows the series coefficients do not converge.}
\end{figure}

The solution is not unique as the continuity condition admits more than one solution. Choosing $a_1= 0.3066$ the series solution has been shown in Fig. \ref{eq4_Fig_hom2}(a), but the series coefficients $a_k$ do not converge.

\begin{figure}
\begin{center}
\subfigure[] {\epsfxsize=3 in\epsfbox{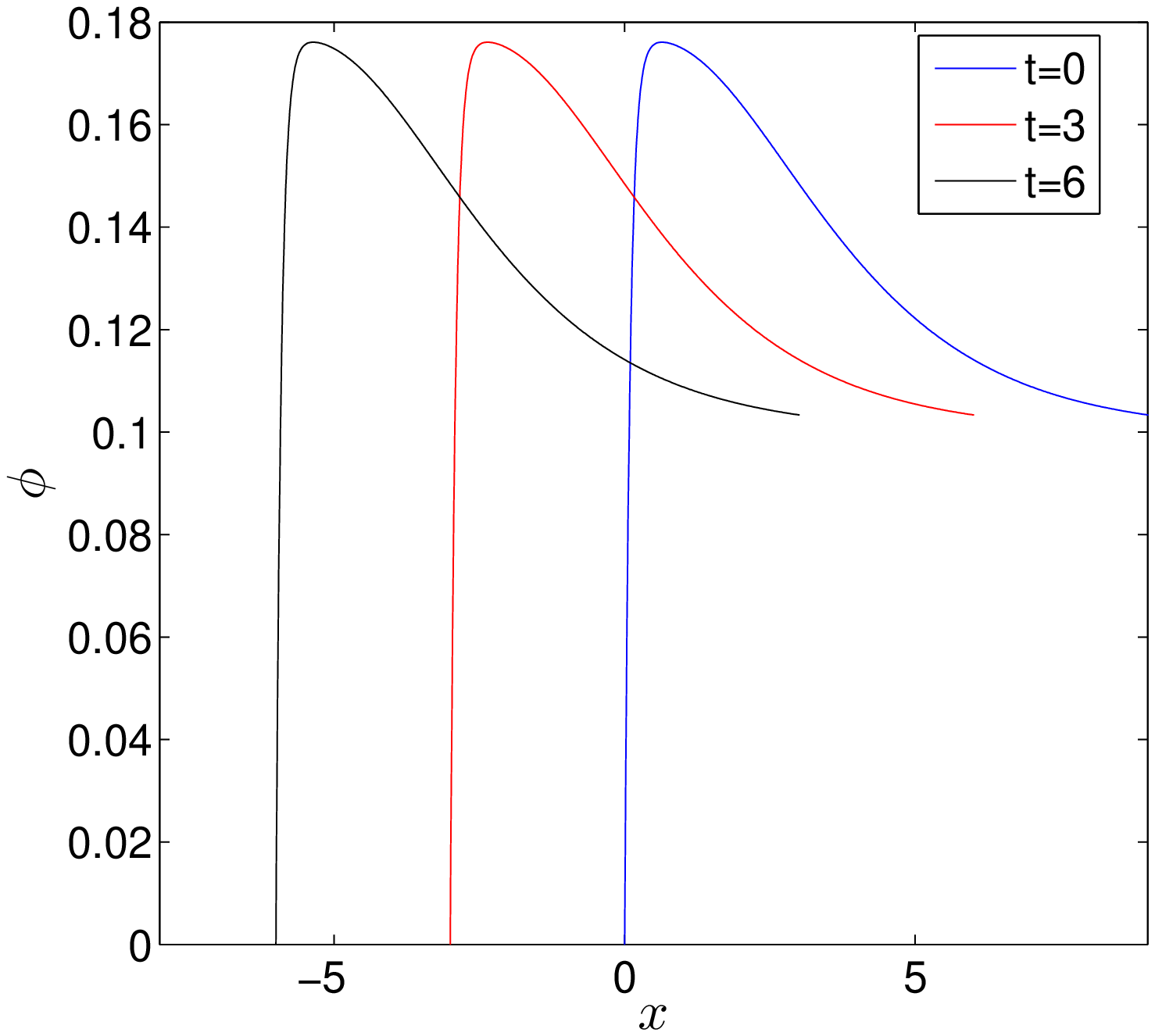}}
\end{center}
\caption{\label{eq4_Fig_hom2_trav} The parameters are chosen as in Fig.\ref{eq4_Fig_hom2}. The series solution $\phi(z)$ for the homoclinic orbit \eqref{eq4_hom_orbit} to the saddle point $(x_0=0.1,0)$ plotted as a function of $x$ for different values of $t$, showing traveling wave nature of the solution of Eq. \eqref{eq4_trav_s}.}
\end{figure}

\begin{figure}
\begin{center}
\subfigure[] {\epsfxsize=3 in\epsfbox{vuoto.eps}}
\subfigure[] {\epsfxsize=3 in \epsfbox{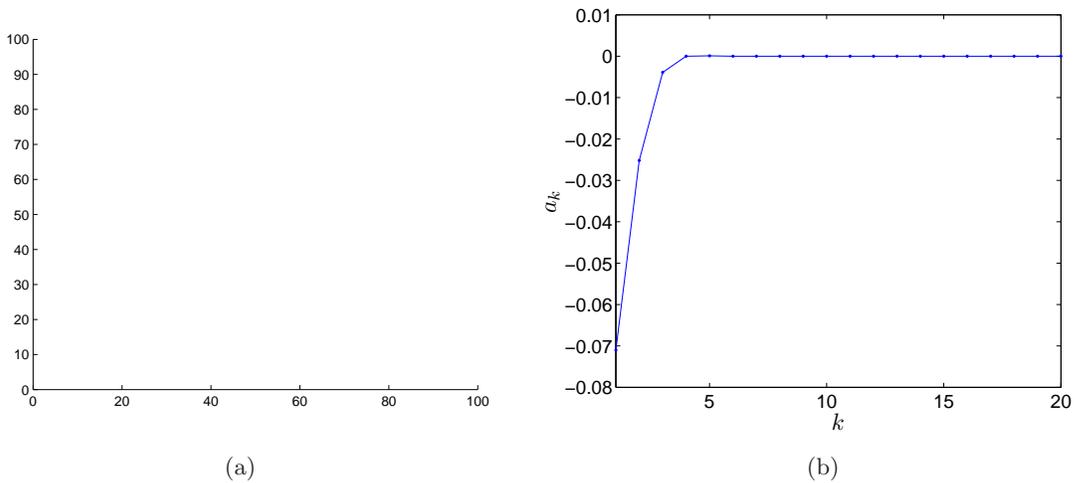}}
\end{center}
\caption{\label{eq4_Fig_hom2c} The system parameters $c$ and $g$ are chosen as in Fig.\ref{eq4_Fig_hom2}. Here we choose $a_1=-0.0710$ as solution of the continuity condition \eqref{eq4_cont} truncated at $M=20$.  (a) The series solution $\phi(z)$ in \eqref{eq4_hom_orbit} for the homoclinic orbit to the saddle point $(x_0=0.1,0)$ plotted as a function of $x$ for different values of $t$, showing traveling wave nature of the solution of Eq. \eqref{eq4_trav_s}. (b) Plot of $a_k$ in \eqref{eq4_series_coeffk_a} versus $k$ shows the series coefficients converge.}
\end{figure}

To show the traveling nature of the solution (when $c = -1$ and
$g = 0.06$) we plot just half part of the solution, for $z > 0$ to make the figure clearer, see in Fig.\ref{eq4_Fig_hom2_trav}.
Choosing $a_1 = -0.0710$, both the convergence of the series coefficients and the continuity at the origin have been obtained, as shown in Fig.\ref{eq4_Fig_hom2c}.

\subsection{Infinite Series for homoclinic orbits of Eq. \eqref{eq2_trav}}\label{subsec4_2}

Leu us use the same notation as in Section \ref{subsec4_4} to write Eq. \eqref{eq2_trav} as follows:
\begin{equation}\label{eq2_trav_s}
-c\phi+c\phi^{''}=8\phi^2-4\phi\phi^{''}-2\phi^{'2}+2\phi'\phi^{''}+g,
\end{equation}
where the prime $'$ indicates the derivative with respect to $z$.

As shown in Section \ref{subsec3.2}, the Eq. \eqref{eq2_trav} admits saddle regular critical points for a suitable choice of $c$ and $g$ and homoclinic orbits to such points have been observed (see Figs.\ref{eq2_Fig_phase}).
Let us now proceed to construct, at first formally, the series solution for the homoclinic orbit to such a point, here indicated with $(x_0, 0)$ (corresponding to an equilibrium $\bar{z}$ of Eq.\eqref{eq2_trav_s}, with the same qualitative nature).

We look for a solution of the same form given in \eqref{eq4_hom_orbit} and \eqref{eq4_hom_spsm},
with $\alpha<0$ and $\beta>0$ undetermined constants and $a_k, b_k$, with $k\geq 1$, arbitrary coefficients at the outset. From the expression for $\phi^+(z)$ it follows that the terms into the Eq. \eqref{eq2_trav_s} can be written as:
\begin{eqnarray}\label{eq2_terms1}
\phi^{''}&=&\sum_{k=1}^{\infty}(k\alpha)^2 a_k e^{k\alpha z},\\\label{eq2_terms2}
\phi^2&=&x_0^2+2x_0\sum_{k=1}^{\infty} a_k e^{k\alpha z}+\sum_{k=2}^{\infty}\sum_{i=1}^{k-1} a_{k-i}a_i e^{k\alpha z},\\\label{eq2_terms3}
\phi'^2&=&\sum_{k=2}^{\infty}\sum_{i=1}^{k-1}(k-i)i\alpha^2 a_{k-i}a_i e^{k\alpha z},\\\label{eq2_terms4}
\phi^{''}(2\phi-\phi')&=&2x_0\sum_{k=1}^{\infty}(k\alpha)^2 a_k e^{k\alpha z}
+\sum_{k=2}^{\infty}\sum_{i=1}^{k-1} (k-i)^2(2-i\alpha)\alpha^2 a_{k-i}a_i e^{k\alpha z}.
\end{eqnarray}
Using \eqref{eq2_terms1}-\eqref{eq2_terms4} in the Eq. \eqref{eq2_trav_s} we have:
\begin{equation}\label{eq2_series}
\begin{split}
\sum_{k=1}^{\infty}((k\alpha)^2 (c+4x_0)-(c+16x_0))a_k e^{k\alpha z}=\quad&\,\quad\\
cx_0+8x_0^2+g-\sum_{k=2}^{\infty}\sum_{i=1}^{k-1} (8-2(k-i)^2&\,(2-i\alpha)\alpha^2-2(k-i)i\alpha^2)a_{k-i}a_i e^{k\alpha z},
\end{split}
\end{equation}
where the quantity $cx_0+8x_0^2+g=0$ is identically zero, being $x_0$ an equilibrium. Now we compare the coefficients of $e^{k\alpha z}$ for each $k$ and for $k=1$ we obtain:
\begin{equation}\label{eq2_series1}
(\alpha^2 (c+4x_0)-(c+16x_0))a_1=0.
\end{equation}
Assuming $a_1\neq0$ (otherwise $a_k=0$ for all $k>1$ by induction), results in the two possible values of $\alpha$:
\begin{equation}\label{eq2_series_eig}
\alpha_{1}= \sqrt{\frac{c+16x_0}{c+4x_0}},\qquad\qquad \alpha_{2}=-\sqrt{\frac{c+16x_0}{c+4x_0}}.
\end{equation}
As $x_0$ is a saddle point (here we skip all the details about the possible choice of $c$ and $g$, as it has been done in Section \ref{subsec3.2}), the quantities $\alpha_{1,2}$ have opposite real parts.
As our solution given in \eqref{eq4_hom_orbit} needs to converge for $z >0$, we pick the negative root $\alpha=\alpha_1$. Thus we have:
\begin{equation}\label{eq2_series_2}
F(k\alpha_2)a_2=(8-6\alpha_2^2+2\alpha_2^3)a_1^2,
\end{equation}
where  $F(k\alpha_2)=(k\alpha_2)^2 (c+4x_0)-(c+16x_0)$. Thus, we obtain:
\begin{equation}\label{eq2_series_coeff2}
a_2=\frac{(8-6\alpha_2^2+2\alpha_2^3)a_1^2}{F(2\alpha_2)}.
\end{equation}
Similarly for $k=3$:
\begin{equation}\label{eq2_series_coeff3}
a_3=\sum_{i=1}^2\frac{(8-2(3-i)\alpha_2^2(-i-3i\alpha_2+i^2\alpha_2+6))a_{3-i} a_i}{F(3\alpha_2)}.
\end{equation}
In general, one has for $k>2$:
\begin{equation}\label{eq2_series_coeffk}
a_k=\sum_{i=1}^2\frac{(8-2(k-i)i\alpha_2^2-2(k-i)^2(2-i\alpha_2)\alpha_2^2)a_{k-i} a_i}{F(k\alpha_2)}.
\end{equation}
By iterative substitutions of the coefficients $a_k, k>1$ already obtained at the previous steps, all the coefficients of Eq. \eqref{eq2_series_coeffk} can be written in terms of the coefficient $a_1$ as follows:
\begin{equation}\label{eq2_series_coeffk_a}
a_k=\varphi_k a_1^k,                                                                                  \end{equation}
where $\varphi_k, k>1$ are functions of $\alpha_2$ and the coefficients of the Eq. \eqref{eq2_trav_s}.
The first part of the homoclinic orbit corresponding to $z>0$ has thus been determined in terms of $a_1$:         		 %
\begin{equation}\label{eq2_series_p}
\phi^+ (z)=x_0+a_1 e^{\alpha_2 z}+\sum_{k=2}^\infty \varphi_k a_1^k e^{k\alpha_2 z}.                                                 \end{equation}
We shall now construct the second part of the homoclinic orbit corresponding to $z<0$. Since the Eq. \eqref{eq2_trav_s} is not reversible we do not have any symmetric property for the solution, therefore we impose $\phi^-(z)$ has the form given as in \eqref{eq4_hom_spsm}, where the real part of $\beta>0$, because the solution $\phi^- (z)$ needs to converge for $z<0$. Working as for $z>0$, we obtain for $k=1$ the following equation:
\begin{equation}\label{eq2_series1b}
(\beta^2 (c+4x_0)-(c+16x_0))b_1=0.
\end{equation}
Assuming $b_1\neq0$ (otherwise $b_k=0$ for all $k>1$ by induction), the Eq. \eqref{eq2_series1b} has the same solutions as \eqref{eq2_series1}. Here we choose $\beta=\alpha_1>0$, therefore for $k>1$ we obtain:
\begin{equation}\label{eq2_series_coeffk_b}
b_k=\sum_{i=1}^2\frac{(8-2(k-i)i\alpha_1^2-2(k-i)^2(2-i\alpha_1)\alpha_1^2)b_{k-i} b_i}{F(k\alpha_1)},
\end{equation}
where the polynomial $F(k\alpha_1)=(k\alpha_1)^2 (c+4x_0)-(c+16x_0)$ is nonzero for $k>1$.
Therefore the series coefficients $b_k, k>1,$ can be easily obtained from Eq. \eqref{eq2_series_coeffk_b} as follows:
\begin{equation}\label{eq2_series_coeffk_b}
b_k=\psi_k b_1^k,                                                                                  \end{equation}
where $\psi_k, k>1$ are given in terms of $b_1$ and the coefficients of Eq. \eqref{eq2_trav_s}.

As we want to construct a solution continuous at $z=0$, we impose:
\begin{eqnarray}\label{eq2_cont_a}
\phi^+ (0)&=&x_0+a_1+\sum_{k=2}^\infty \varphi_k a_1^k=0,\\\label{eq2_cont_b}
\phi^- (0)&=&x_0+b_1+\sum_{k=2}^\infty \psi_k b_1^k=0,
\end{eqnarray}
hence we choose $a_1$ and $b_1$ as the nontrivial solutions of the above polynomial Eqs. \eqref{eq2_cont_a} and \eqref{eq2_cont_b}.

\begin{figure}
\begin{center}
{\epsfxsize=3.2 in\epsfbox{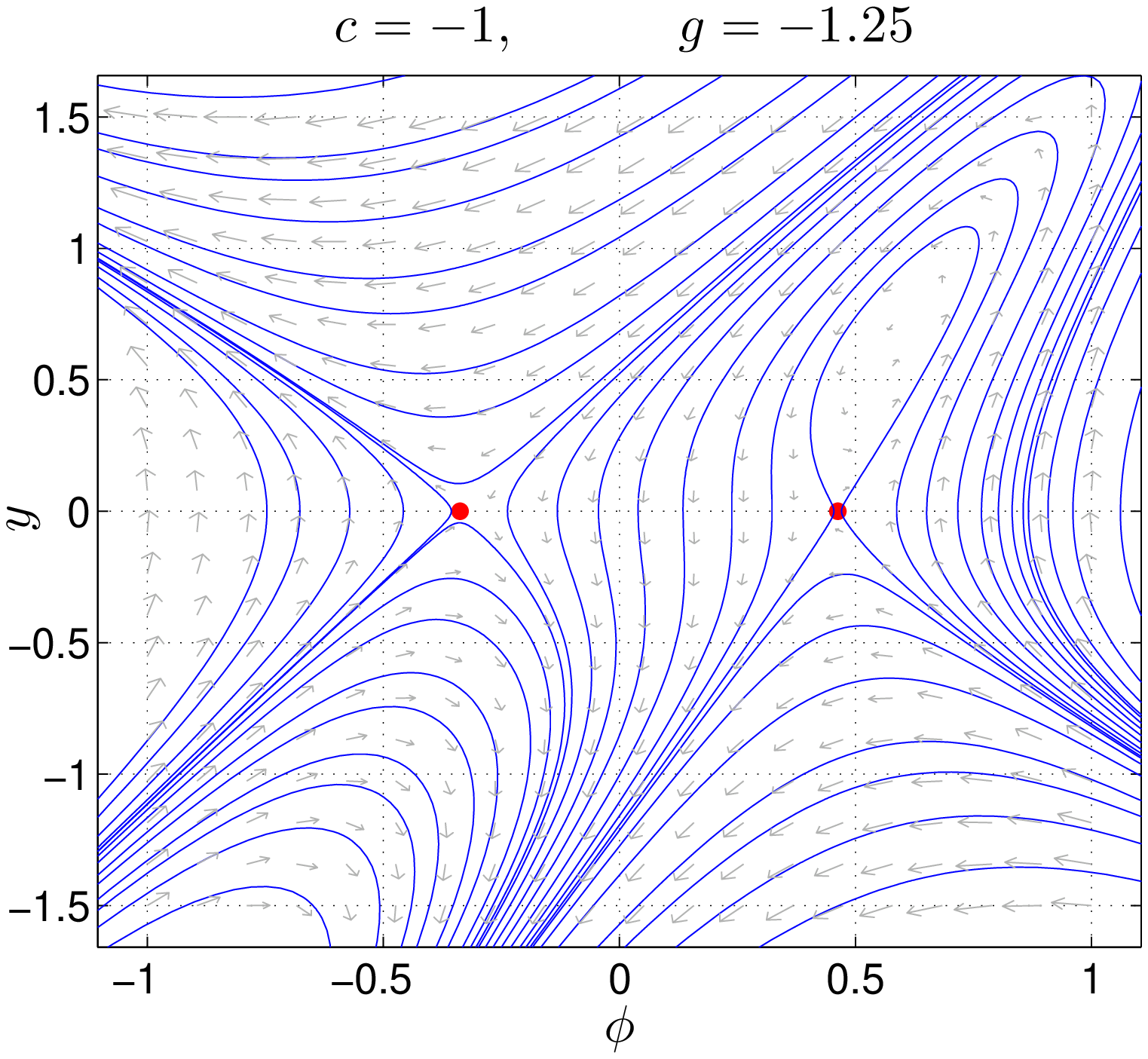}}
\end{center}
\caption{\label{Fig_eq2Hom1} Phase plane orbits of Eq. \eqref{eq2_sd_reg} when $c = -1$ and $g = -1.25$.}
\end{figure}

Let us consider the first numerical test. We choose $c = -1$ and $g = -1.25$. With this choice both the equilibria $z_1 = 0.4627$ and $z_2 =-0.3377$ are saddles, but only a homoclinic orbit to the point $z_1$ has been observed. Here we construct this orbit (see Fig.\ref{Fig_eq2Hom1}). The eigenvalues relative to the point $z_1$ are $\alpha_1 = 2.7434$ and $\alpha_2 = -2.7434$.

\begin{figure}
\begin{center}
\subfigure[] {\epsfxsize=2 in\epsfbox{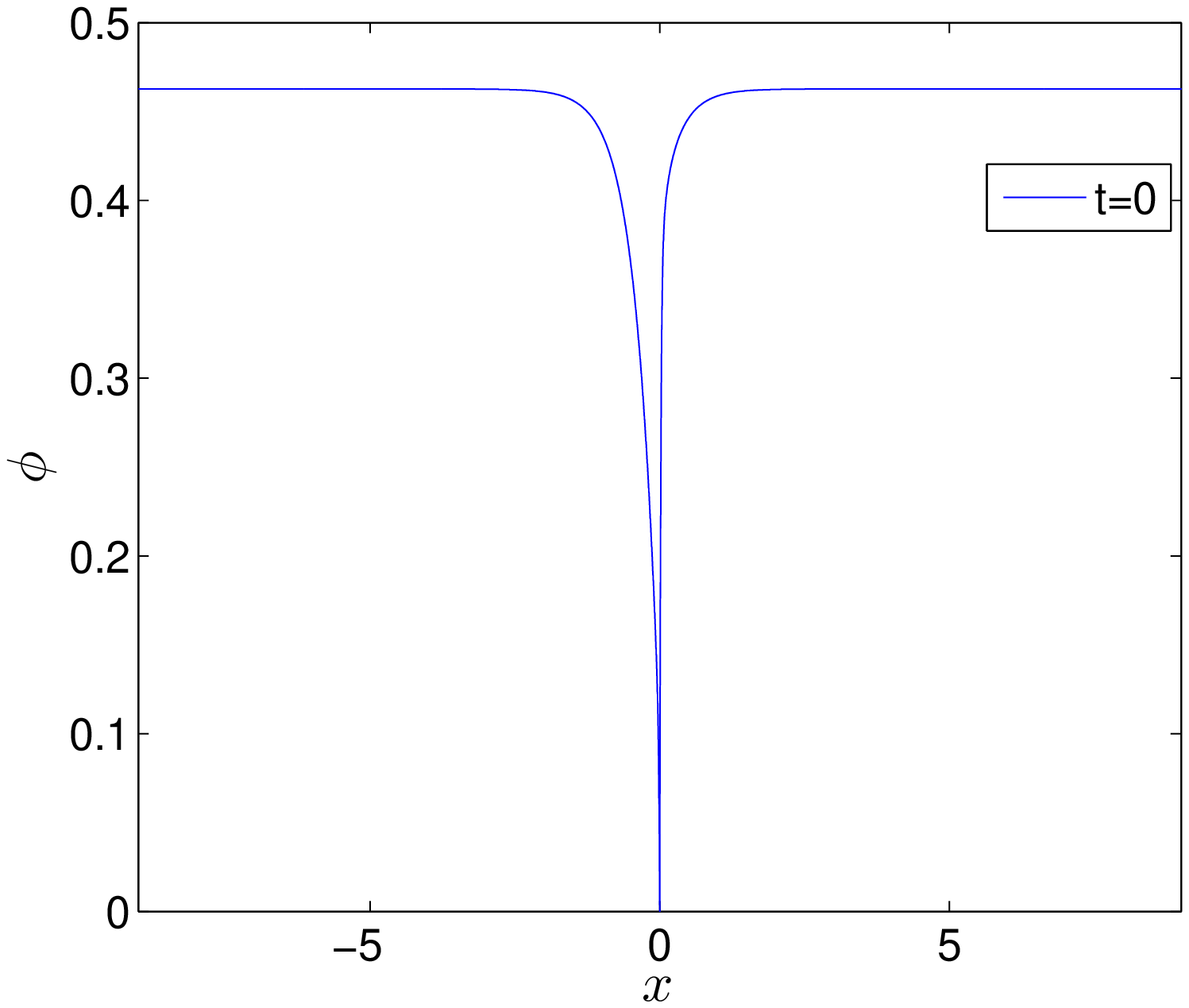}}
\subfigure[] {\epsfxsize=2 in \epsfbox{vuoto.eps}}
\subfigure[] {\epsfxsize=2 in \epsfbox{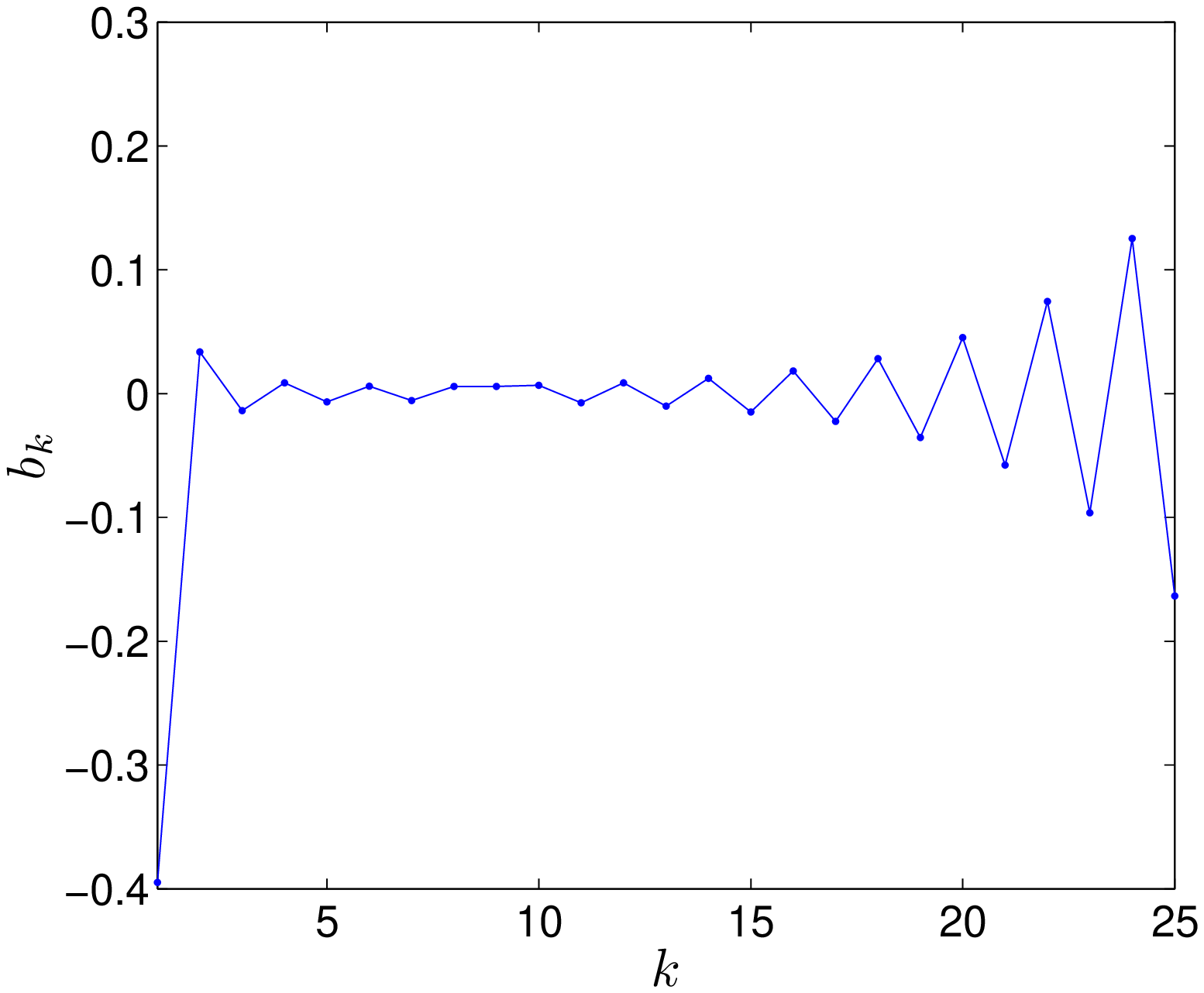}}
\end{center}
\caption{\label{eq2_Fig_hom1} The parameters are chosen as $c = -1$ and $g = -1.25$. We also choose $a_1 = -0.06$ and $b_1 = -0.3948$ as solutions respectively of the continuity conditions \eqref{eq2_cont_a} and \eqref{eq2_cont_b} truncated at $M=25$. (a) The series solution $\phi(z)$ (as in \eqref{eq4_hom_orbit}, with $a_k$ and $b_k$ respectively given in \eqref{eq2_series_coeffk} and \eqref{eq2_series_coeffk_b}) for the homoclinic orbit to the point $(x_0=0.4627,0)$ plotted as a function of $x$ at $t=0$. (b) Plot of $a_k$ in \eqref{eq2_series_coeffk} versus $k$ shows the series coefficients do not converge. (c) Plot of $b_k$ in \eqref{eq2_series_coeffk_b} versus $k$ shows the series coefficients do not converge.}
\end{figure}

If we truncate the series solution at $M=25$, the continuity Eq. \eqref{eq2_cont_a} has the only solution $a_1 = -0.06$ and the continuity Eq. \eqref{eq2_cont_b} has the only solution $b_1 = -0.3948$.	
Choosing these values for $a_1$ and $b_1$ we obtain the continuous solution as in Fig.\ref{eq2_Fig_hom1}(a). Nevertheless, the corresponding series coefficients $a_k$ and $b_k$ both diverge, as shown in Figs.\ref{eq2_Fig_hom1}(b)-(c).

\begin{figure}
\begin{center}
\subfigure[] {\epsfxsize=2 in\epsfysize=1.8 in\epsfbox{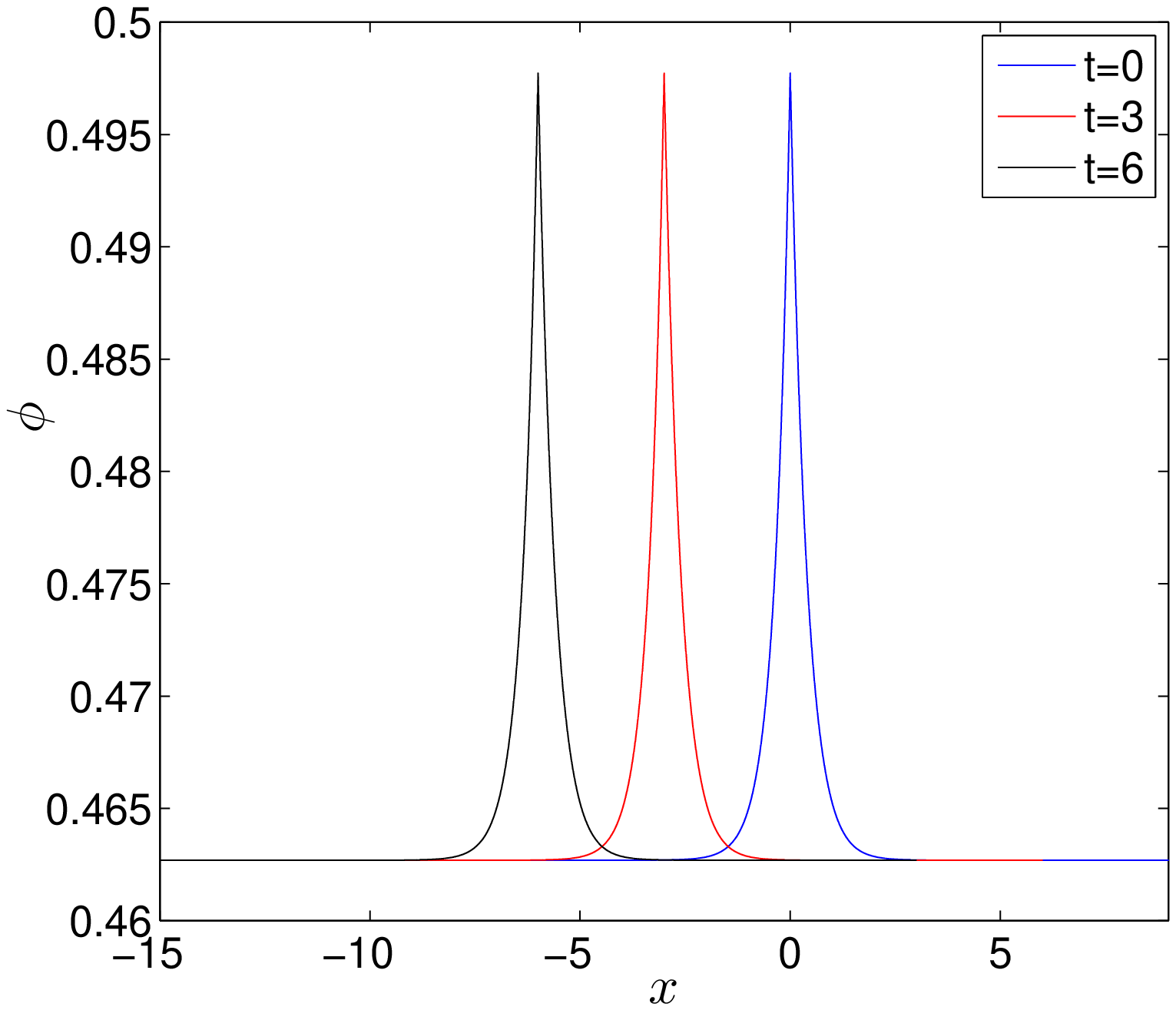}}
\subfigure[] {\epsfxsize=2 in \epsfysize=1.8 in\epsfbox{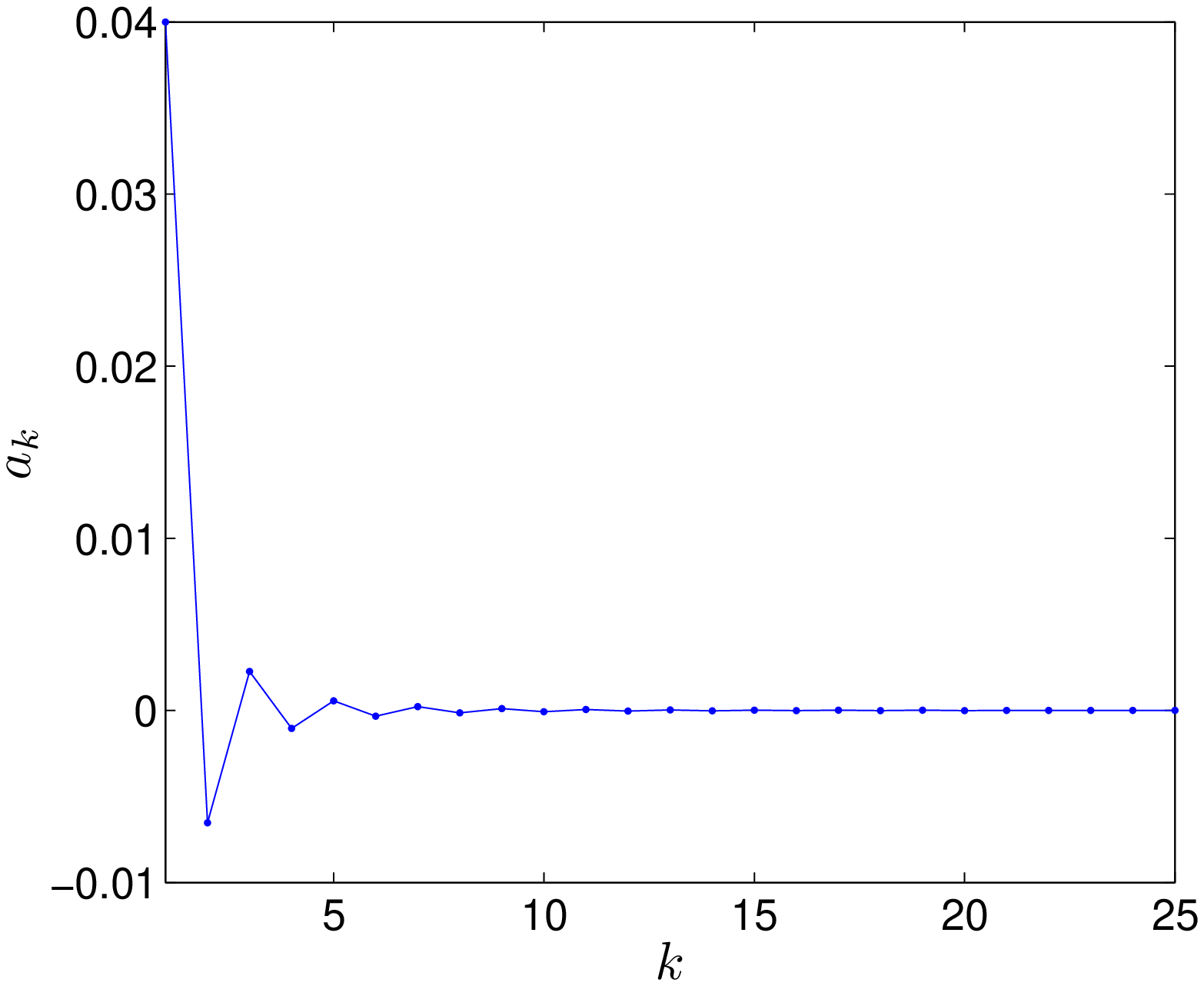}}
\subfigure[] {\epsfxsize=2 in \epsfysize=1.8 in\epsfbox{vuoto.eps}}
\end{center}
\caption{\label{eq2_Fig_hom1_con} The system parameters $c$ and $g$ are chosen as in Fig.\ref{eq2_Fig_hom1}. Here $a_1=0.04$ and $b_1=0.0348$ to have continuity at the origin $\phi^-(0)=\phi^+ (0)=0.4977$ for the series solution \eqref{eq4_hom_orbit}. (a) The series solution $\phi(z)$ (as in \eqref{eq4_hom_orbit}, with $a_k$ and $b_k$ respectively given in \eqref{eq2_series_coeffk} and \eqref{eq2_series_coeffk_b}) for the homoclinic orbit to the point $(x_0=0.4627,0)$ plotted as a function of $x$ at different instants $t$, showing the traveling nature of the solution. (b) Plot of $a_k$ in \eqref{eq2_series_coeffk} versus $k$ shows the series coefficients converge. (c) Plot of $b_k$ in \eqref{eq2_series_coeffk_b} versus $k$ shows the series coefficients converge.}
\end{figure}

To obtain the convergence of the series coefficients $a_k$ we arbitrarily choose $a_1=0.04$, see Fig.\ref{eq2_Fig_hom1_con}(b). With this choice of $a_1$ the solution, given as in \eqref{eq4_hom_orbit}, becomes discontinuous at the origin. Therefore, we choose $b_1=0.0348$ in such a way that
$\phi^-(0)=\phi^+ (0)=0.4977$ and the solution is still continuous into the origin. This value for $b_1$ also preserves the convergence of the series coefficients $b_k$, see Fig.\ref{eq2_Fig_hom1_con}(c). The traveling wave nature of the obtained continuous solution has been shown in Fig.\ref{eq2_Fig_hom1_con}(a). As it is obvious the solution is similar to a symmetric peakon having $\alpha_1 = 2.7434$ and $\alpha_2 = -2.7434$.

\begin{figure}
\begin{center}
{\epsfxsize=3.2 in\epsfbox{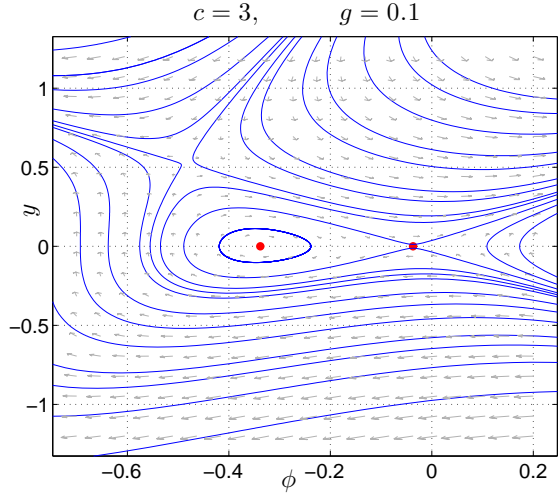}}
\end{center}
\caption{\label{Fig_eq2Hom2} Phase plane orbits of Eq. \eqref{eq2_sd_reg} when $c = 3$ and $g = 0.1$.}
\end{figure}

Let us consider a second numerical test choosing $c = 3$ and $g = 0.1$. In this case the equilibrium $z_1 = -0.0370$ is a saddle and $z_2 = -0.3380$ is a center. Here we construct the homoclinic orbit to the point $z_1$ as observed in Fig.\ref{Fig_eq2Hom2}.

\begin{figure}
\begin{center}
\subfigure[] {\epsfxsize=2 in\epsfbox{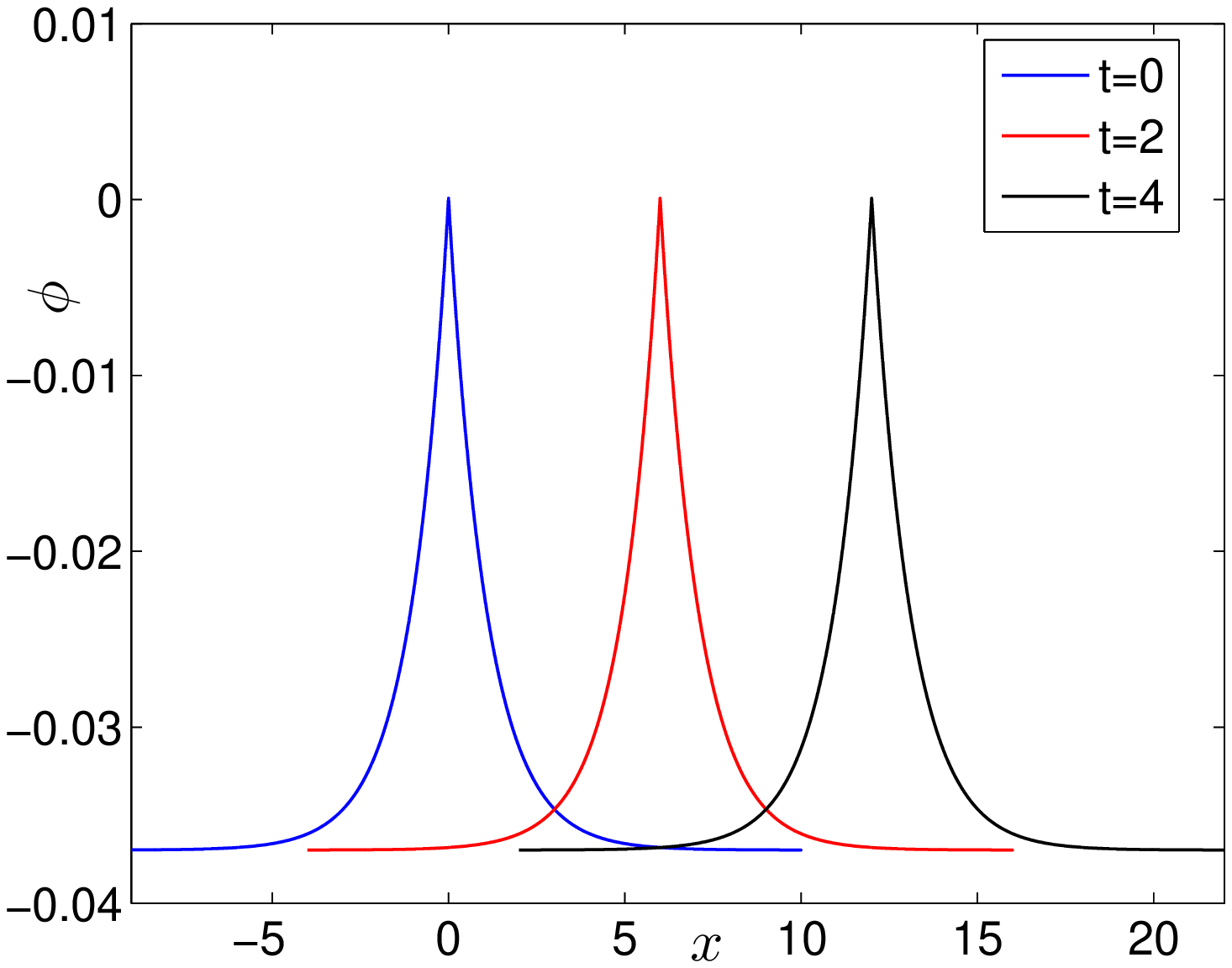}}
\subfigure[] {\epsfxsize=2 in \epsfbox{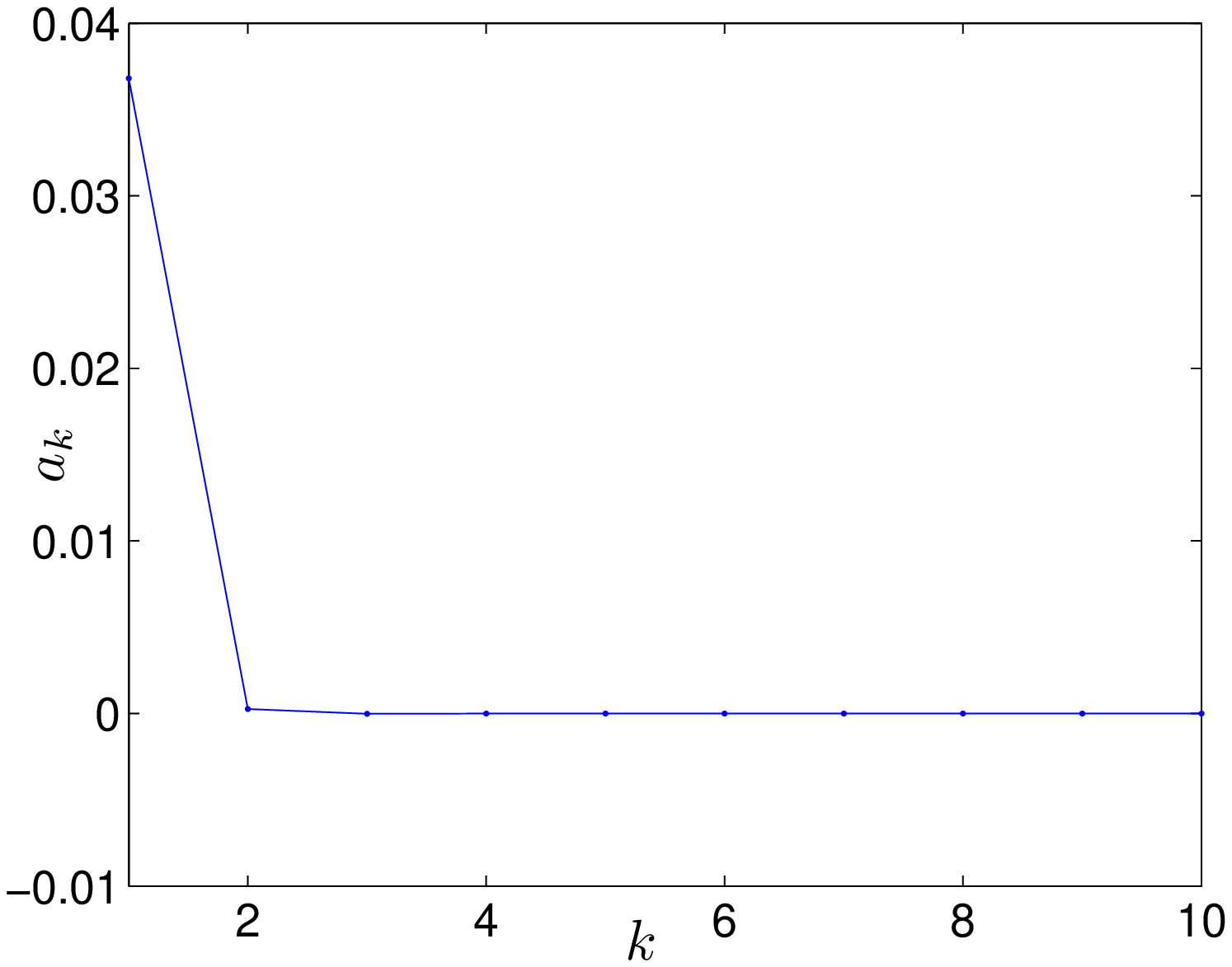}}
\subfigure[] {\epsfxsize=2 in \epsfbox{vuoto.eps}}
\end{center}
\caption{\label{eq2_Fig_hom2} The parameters are chosen as $c = 3$ and $g = 0.1$. (a) The series solution $\phi(z)$ (as in \eqref{eq4_hom_orbit}, with $a_k$ and $b_k$ respectively given in \eqref{eq2_series_coeffk} and \eqref{eq2_series_coeffk_b}) for the homoclinic orbit to the point $(x_0=-0.0370,0)$. Here $a_1=0.04$ and $b_1=0.0348$ are respectively the solutions of the continuity conditions \eqref{eq2_cont_a} and \eqref{eq2_cont_b} truncated at $M=10$. (b) Plot of $a_k$ in \eqref{eq2_series_coeffk} versus $k$ shows the series coefficients rapidly converge. (c) Plot of $b_k$ in \eqref{eq2_series_coeffk_b} versus $k$ shows the series coefficients rapidly converge.}
\end{figure}

The eigenvalues relative to the point $z_1$ are $\alpha_1 = 0.9189$ and $\alpha_2 = -0.9189$. We truncate the series solution at $M = 10$, the continuity Eq. \eqref{eq2_cont_a} has the only solution $a_1= -0.036$ and the continuity Eq. \eqref{eq2_cont_b} has the only solution $b_1 = -0.0362$. Choosing theses values for $a_1$ and $b_1$ we obtain the continuous solution as in Fig.\ref{eq2_Fig_hom2}(a). Moreover, the corresponding series coefficients $a_k$ and $b_k$ both converge, as shown in Figs.\ref{eq2_Fig_hom2}(b)-(c).

As it is obvious the solution is similar to a symmetric peakon having $\alpha_1 = 0.9189$ and $\alpha_2 =- 0.9189$. The traveling wave nature of the solution is shown in Fig.\ref{eq2_Fig_hom2}(a) for different time instants.

\begin{figure}
\begin{center}
{\epsfxsize=3.2 in\epsfbox{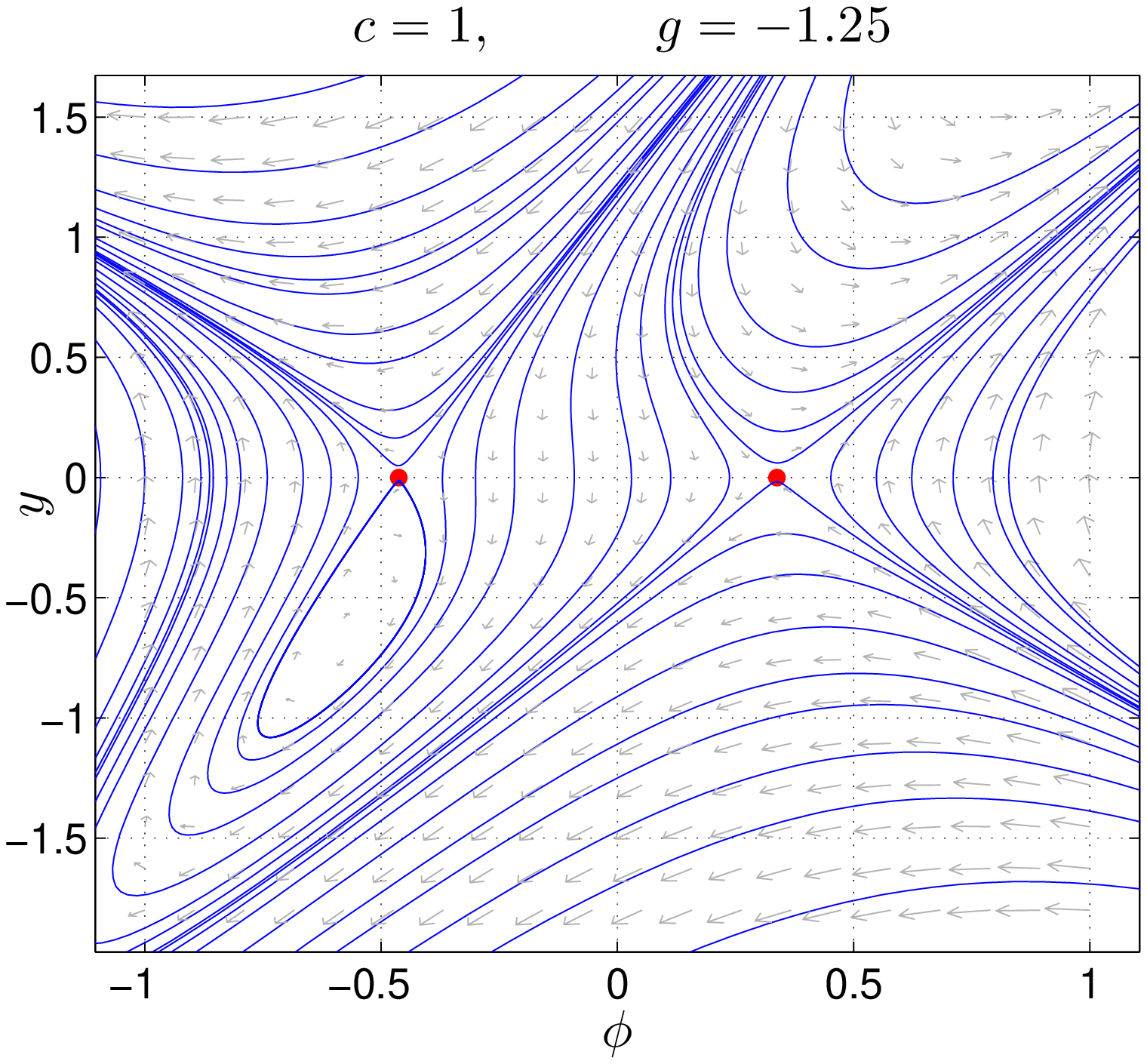}}
\end{center}
\caption{\label{Fig_eq2Hom3} Phase plane orbits of Eq. \eqref{eq2_sd_reg} when $c = 1$ and $g = -1.25$.}
\end{figure}

Notice the "symmetry" properties of the solution just changing the sign of $c$, as the equilibria $z_1$ and $z_2$ exchange their role. Choosing $c = 1$ and $g = -1.25$, the equilibria $z_1 =0.3377$ and $z_2 = -0.4627$ are both saddles, but only a homoclinic orbit to the point $z_2$ has been observed, see Fig.\ref{Fig_eq2Hom3}.

\begin{figure}
\begin{center}
\subfigure[] {\epsfxsize=2 in  \epsfysize=1.8 in\epsfbox{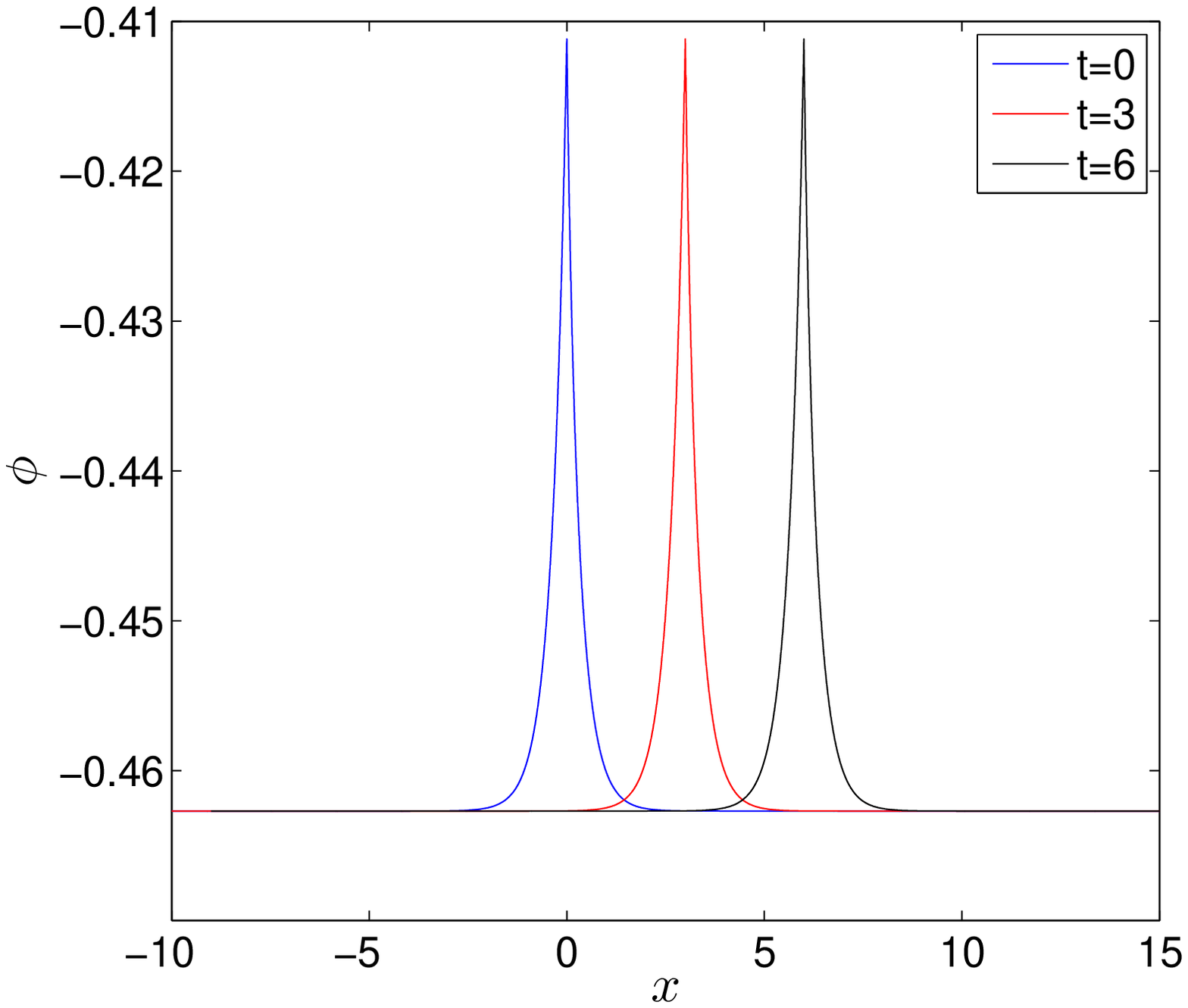}}
\subfigure[] {\epsfxsize=2 in \epsfysize=1.8 in\epsfbox{vuoto.eps}}
\subfigure[] {\epsfxsize=2 in \epsfysize=1.8 in\epsfbox{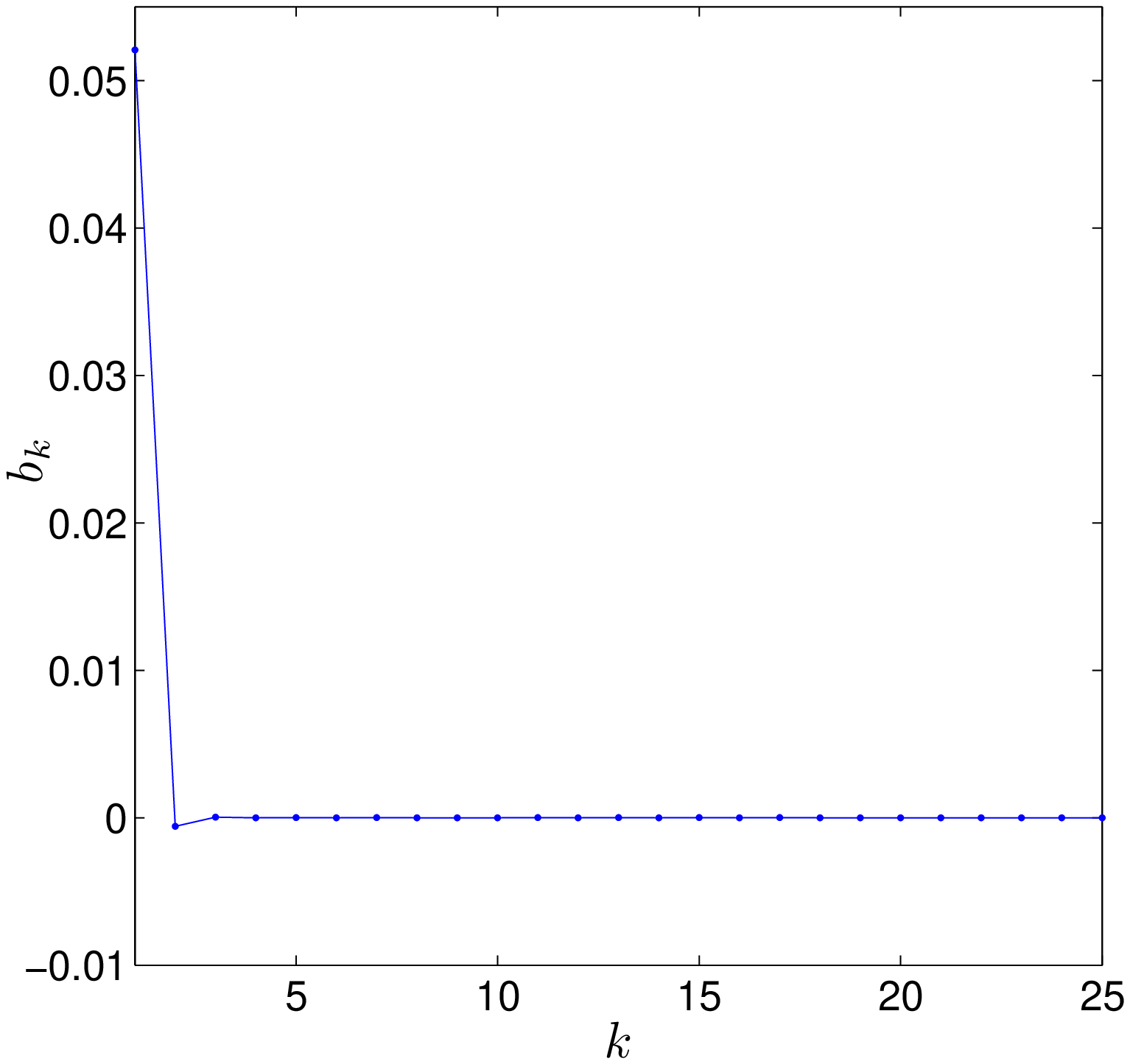}}
\end{center}
\caption{\label{eq2_Fig_hom3} The parameters are chosen as $c = 1$ and $g = -1.25$. (a) The series solution $\phi(z)$ (as in \eqref{eq4_hom_orbit}, with $a_k$ and $b_k$ respectively given in \eqref{eq2_series_coeffk} and \eqref{eq2_series_coeffk_b}) for the homoclinic orbit to the point $(x_0=-0.4627,0)$, plotted as a function of $x$ at different instants $t$, showing the traveling nature of the solutions. Here $a_1=0.04$ is arbitrarily chosen and $b_1=0.0521$  to have continuity at the origin. (b) Plot of $a_k$ in \eqref{eq2_series_coeffk} versus $k$ shows the series coefficients converge. (c) Plot of $b_k$ in \eqref{eq2_series_coeffk_b} versus $k$ shows the series coefficients converge.}
\end{figure}

The eigenvalues relative to the point $z_2$ are $\alpha_1 = 2.7434$ and $\alpha_2 = -2.7434$. If we truncate the series solution at $M=25$, the continuity Eq. \eqref{eq2_cont_a} has the only solution $a_1 =0.06$ and the continuity Eq. \eqref{eq2_cont_b} has the only solution $b_1 = 0.3948$. Choosing these values for $a_1$ and $b_1$ we obtain a continuous solution, but the corresponding series coefficients $a_k$ and $b_k$ both diverge, as already found in the "symmetric" case shown in Fig.\ref{eq2_Fig_hom1}.	
To obtain the convergence of the series coefficients $a_k$ we arbitrarily choose $a_1=0.04$. With this choice of $a_1$ the solution as in \eqref{eq4_hom_orbit} becomes discontinuous at the origin. Therefore, we choose $b_1=0.0521$ in such a way that $\phi^-(0)=\phi^+(0)=-0.4112$ and the solution is still continuous into the origin. This value for $b_1$ also preserves the convergence of the series coefficients $b_k$, see Fig.\ref{eq2_Fig_hom3}(c). The traveling nature of the obtained continuous solution has been shown in Fig.\ref{eq2_Fig_hom3}(a).
As it is obvious, the solution is similar to a symmetric peakon having $\alpha_1 = 2.7434$ and $\alpha_2 = -2.7434$.

\begin{figure}
\begin{center}
\subfigure[] {\epsfxsize=3 in \epsfysize=2.4 in\epsfbox{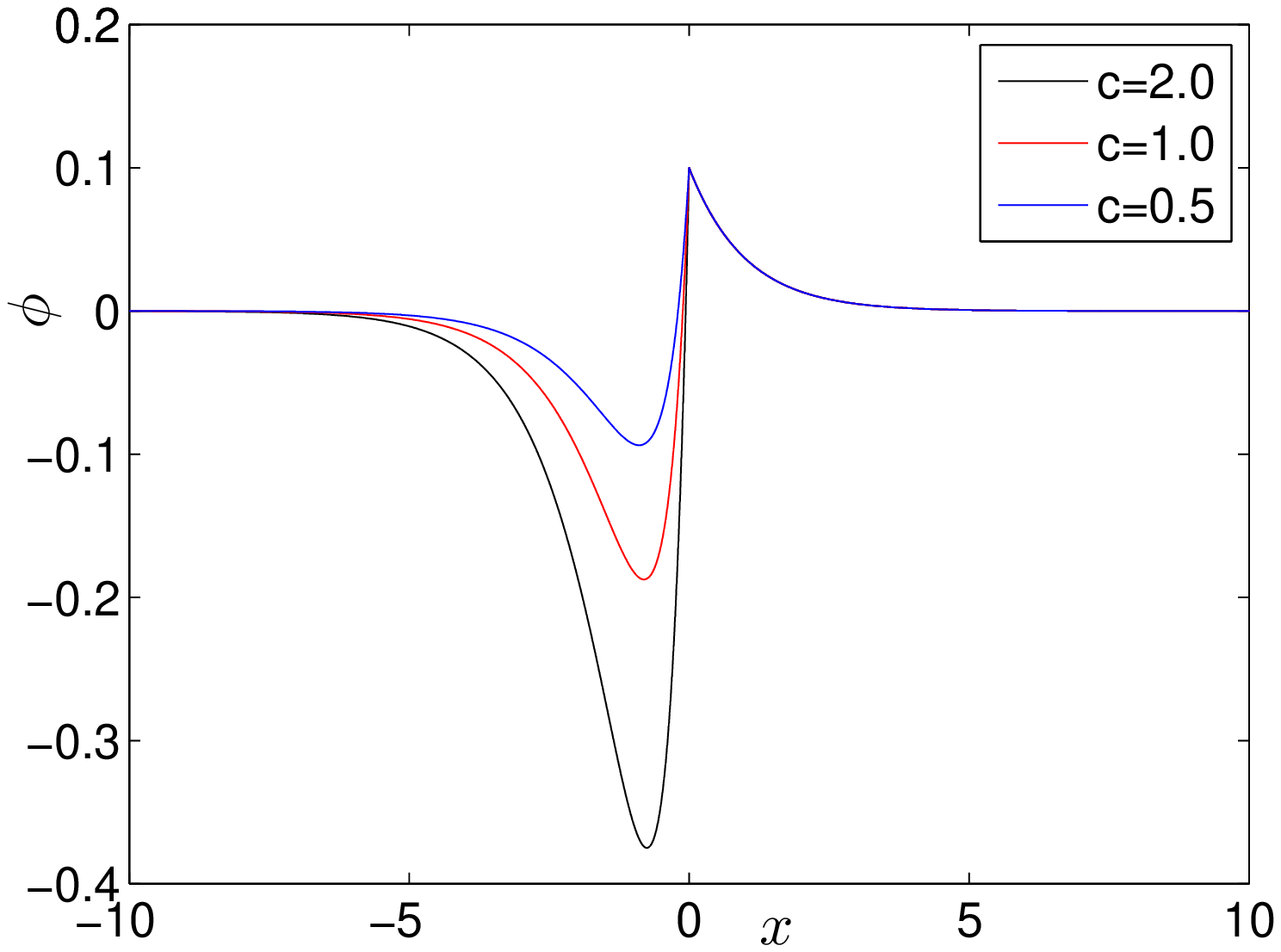}}
\subfigure[] {\epsfxsize=3 in \epsfysize=2.4 in\epsfbox{vuoto.eps}}
\end{center}
\caption{\label{eq2_Fig_homZero1}The system parameters $g=0$ and $c = 0.5, 1$, and $2$. The series solution $\phi(z)$ (as in \eqref{eq4_hom_orbit}, with $a_k$ and $b_k$ respectively given in \eqref{eq2_series_coeffk} and \eqref{eq2_series_coeffk_b}) for the homoclinic orbit to the point $(x_0=0,0)$, plotted as a function of $x$ at $t=0$,  (a) for negative values of $b_1$, and (b) for positive values of $b_1$.}
\end{figure}

Next, we discuss the behavior of solution for the particular case $g = 0$. When $g=0$, Eq. \eqref{eq2_sd_reg} has two fixed points $(0,0)$ which is a saddle (it corresponds to the equilibrium point $z_1$) and $\left(-\displaystyle\frac{1}{8} c,0\right)$, which is a center. Let us compute the series solution to the origin, i.e. $x_0=0$. Interestingly, we have noticed that all the series coefficients $a_k$ in the solution $\phi^+$ become zero for all values of $k$, independently on the choice of $a_1$. Therefore, only the zero solution can be mathematically obtained using this method when $z>0$.
Note that the zero solution resulting for $z>0$ is worth for future investigation as to whether it is the only possible homoclinic solution in that domain. It is also reminiscent of the phenomenon of \textit{Quenching} in nonlinear oscillators where the solutions go to zero in certain parts of the domain.
For $z\leq0$, $b_k$ is zero for $k > 2$. Therefore, to obtain $\phi^-(0)=0$, from Eq. \eqref{eq2_cont_b} we get $b_1+b_2=0$, where $b_2=\displaystyle\frac{4b_1^2}{3c}$. So, to satisfy the continuity as in Eq. \eqref{eq2_cont_b} the value of $b_1$ will be $b_1=\displaystyle\frac{3c}{4}=-b_2$. We also get another solution $b_1=0$, which obviously will not yield any series solution. However in present case for $g=0$, since the right side of the solution does not have any $a_1$ that satisfies Eq. \eqref{eq2_cont_a} we took $b_1$ so that $\phi^-(0)=\phi^+(0)$.  In Fig.\ref{eq2_Fig_homZero1}, $\phi (z)$ is plotted against $x$ for $c=0.5, 1$ and $2$.

The initial value was arbitrarily taken as $a_1=0.1$ and as mentioned before all other $a_k$ for $k>2$ are zero. Therefore, the initial value was $\phi^+(0)=0.1$. For a particular value of $c$, to maintain $\phi^-(0)=\phi^+(0)$, we obtain two different solutions of $b_1$. Choosing $c=0.5$, the solutions for $b_1$ are $b_1=-0.4570$ or $b_1=0.0821$; for $c=1.0$ the solutions are $b_1= -0.8393$ or $b_1=0.0894$; and for $c=2.0$ the solutions are $b_1= -1.5940$ or $b_1=0.0941$. For negative values of $b_1$ in Fig.\ref{eq2_Fig_homZero1}(a), the negative side is more susceptible to variation of $c$.

\begin{figure}
\begin{center}
\subfigure[] {\epsfxsize=3 in\epsfbox{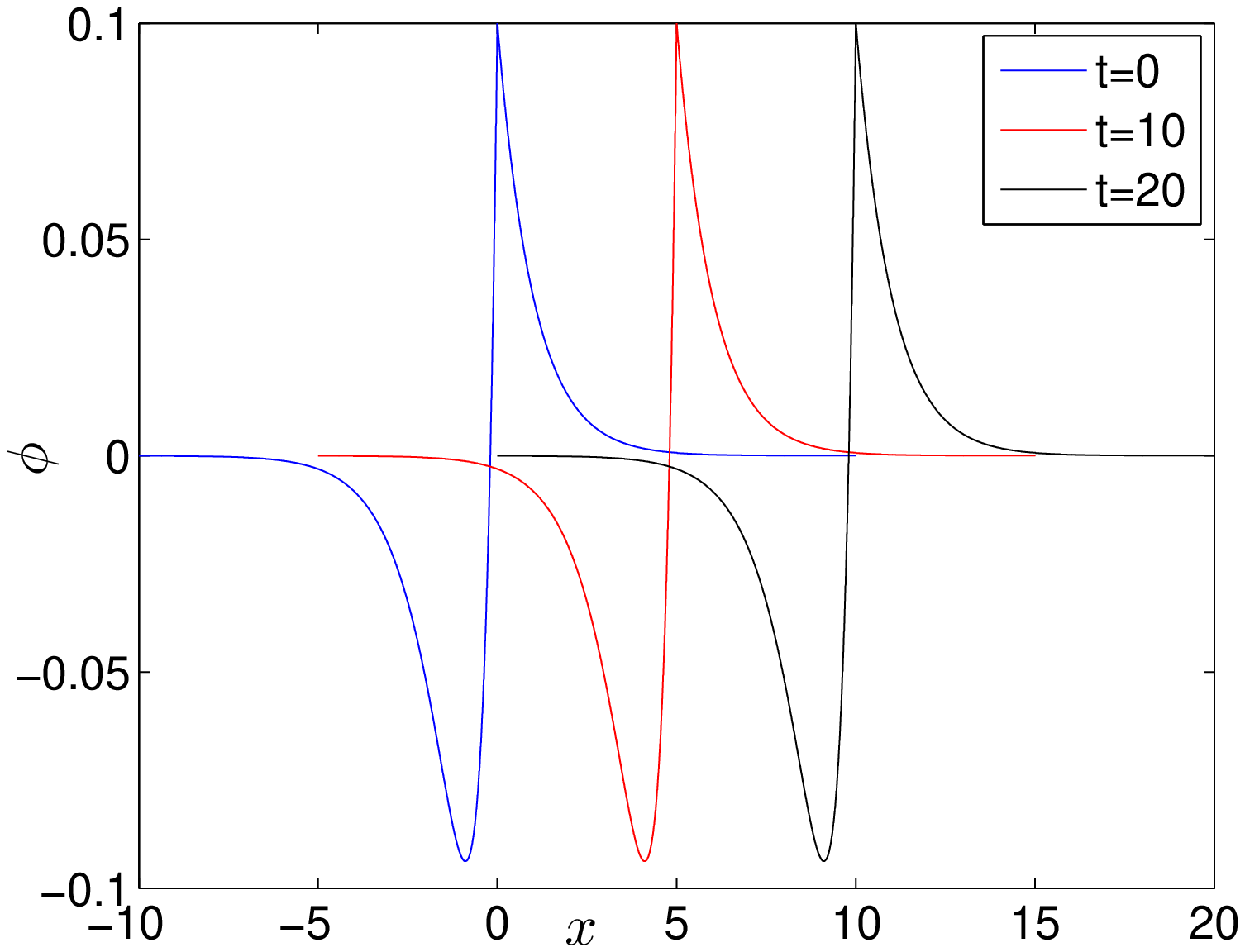}}
\subfigure[] {\epsfxsize=3 in \epsfbox{vuoto.eps}}
\end{center}
\caption{\label{eq2_Fig_homZero2} The system parameters are chosen as $g=0$ and $c = 0.5$. The series solution $\phi(z)$ (as in \eqref{eq4_hom_orbit}, with $a_k$ and $b_k$ respectively given in \eqref{eq2_series_coeffk} and \eqref{eq2_series_coeffk_b}) for the homoclinic orbit to the point $(x_0=0,0)$, plotted as a function of $x$ at different $t$ showing the wave nature of the solution for (a) $b_1=-0.4570$, and (b) $b_1=0.0821$.}
\end{figure}

 In both figures the solution $\phi$ is shown with respect to the variable $x$, for fixed $t = 0$. As seen from Fig.\ref{eq2_Fig_homZero1}, the plots are typical M-shaped wave in negative side, where the positive portion of $x$ is truncated. For positive values of $c$, with the change in $x$ the graph drops sharply before coming back to zero asymptotically and gradually (for negative values of $c$, the graph rises initially, then drops back to zero with change in $x$). The initial decrease in the graph is due to the contribution of the second term of the series solution, whereas with the change of $x$ the contribution of the first term dominates (see Fig.\ref{eq2_Fig_homZero2}).

From Fig.\ref{eq2_Fig_homZero1}, we can also see that for higher numerical value of $c$, the absolute value of $\phi$ is higher. This is due to the fact that larger $|c|$ yielded larger numerical value of $b_1$ (or $b_2$).
To demonstrate the traveling wave nature of the solution $\phi (z)$, Fig.\ref{eq2_Fig_homZero2} shows the plots of $\phi (z)$ versus $x$ at constant times $t=0, 10,$ and $20$. As mentioned before, for $c=0.5$ we have two solutions of $b_1$ to maintain continuity of $\phi^-(0)=\phi^+(0) =0.1$ which are $b_1=-0.4570$ or  $b_1=0.0821$. In Fig.\ref{eq2_Fig_homZero2}(a) the waves move to the positive direction of $x$ since $c$ is positive, whereas in Fig.\ref{eq2_Fig_homZero2}(b) the waves move to the left with the progression of time as $c$ is negative.

\subsection{Infinite Series for homoclinic orbits of Eq. \eqref{eq3_trav}}\label{subsec4_3}

In the particular case  $g=0$, Eq. \eqref{eq3_trav} can be rewritten as:
\begin{equation}\label{eq3_scomp}
(c-\phi^2+\phi'^2)(\phi^{''}-\phi)=0,
\end{equation}
where the prime $'$ indicates the derivative with respect to $z$ and it can be exactly integrated. The solutions are:
\begin{equation}\label{eq3_g0sol}
\begin{split}
\phi_1(z)&\,=k_1 e^z+k_2 e^{-z},\\
\phi_2(z)&\,=\frac{1}{4} e^{-z-k_3} (4c e^{2z}+e^{2k_4}),\\
\phi_3(z)&\,=\frac{1}{4} e^{-z-k_5} (4c e^{2z}+e^{2z+2k_6}),
\end{split}
\end{equation}
for all $k_i\in \mathbf{R}, i=1,\dots, 6$.

In the general case with $g$ different from zero, Eq. \eqref{eq3_trav} can be written as:
\begin{equation}\label{eq3_trav_s}
-c\phi+c\phi^{''}=\phi^2\phi^{''}+\phi\phi^{'2}-\phi^{'2}\phi^{''}-\phi^3-g.
\end{equation}
Let us consider a regular equilibrium point $(x_0,0)$ of Eq. \eqref{eq3_trav} (corresponding to an equilibrium $\bar{z}$ of \eqref{eq3_trav_s}) and assume to choose the parameters $c$ and $g$ in such a way that it is a saddle point and a homoclinic orbit to this equilibrium is given (see the stability analysis in Section \ref{subsec3.3}, see also Fig.\ref{eq3_Fig_phase}(b) and (c)).

Also in this case, we look for a series solution of the form given in \eqref{eq4_hom_orbit} and \eqref{eq4_hom_spsm},
where $\alpha<0$ and $\beta>0$ are undetermined constants and $a_k, b_k$, with $k\geq 1$ are, at the outset, arbitrary coefficients. Substituting the series \eqref{eq4_hom_spsm} for $\phi^+(z)$ into Eq. \eqref{eq3_trav_s}, we obtain the following expressions for each term:
\begin{eqnarray}\label{eq3_terms1}
\phi^{''}&=&\sum_{k=\,1}^{\infty}a_k(k\alpha)^2e^{k\alpha z},\\\label{eq3_terms2}
\phi^{2}\phi^{''}&=&\sum_{k=\,3}^{\infty}\sum_{j=\,2}^{k-1}\sum_{l=\,1}^{j-1}a_la_{j-l}a_{k-j}(k-j)^2\alpha^2e^{k\alpha z}+x_0^2\sum_{k=\,1}^{\infty}a_k (k\alpha)^2 e^{k\alpha z}
\\
\nonumber&\ &+
2x_0 \sum_{k=\,2}^{\infty}\sum_{j=\,1}^{k-1}(k-j)^2\alpha^2a_{k-j}a_j e^{k\alpha z},\\\label{eq3_terms3}
\phi\phi^{'2}&=&\sum_{k=\,3}^{\infty}\sum_{j=\,2}^{k-1}\sum_{l=\,1}^{j-1}a_la_{j-l}a_{k-j}(j-l)l\alpha^2e^{k\alpha z}
+x_0\sum_{k=\,2}^{\infty}\sum_{j=\,1}^{k-1}(k-j)j\alpha^2a_{k-j}a_j e^{k\alpha z},\\\label{eq3_terms4}
\phi^{'2}\phi^{''}&=&\sum_{k=\,3}^{\infty}\sum_{j=\,2}^{k-1}\sum_{l=\,1}^{j-1}a_la_{j-l}a_{k-j\,}l(j-l)(k-j)^2\alpha^4e^{k\alpha z},\\\label{eq3_terms5}
\phi^{3}&=&\sum_{k=\,3}^{\infty}\sum_{j=\,2}^{k-1}\sum_{l=\,1}^{j-1}a_la_{j-l}a_{k-j}e^{k\alpha z}
+3x_0\sum_{k=\,2}^{\infty}\sum_{j=\,1}^{k-1}a_{k-j}a_j e^{k\alpha z}\\\nonumber
&\ &+3x_0^2\sum_{k=\,1}^{\infty}a_k e^{k\alpha z}+x_0^3.
\end{eqnarray}
Using \eqref{eq3_terms1}-\eqref{eq3_terms5} into the Eq. \eqref{eq3_trav_s} we have:
\begin{equation}\label{eq3_series}
\begin{split}
-cx_0+x_0^3&\,+g+
\sum_{k=1}^{\infty}((k\alpha)^2 (c-x_0^2)+3x_0^2-c)a_k e^{k\alpha z}\\
+\sum_{k=\,3}^{\infty}&\,\sum_{j=\,2}^{k-1}\sum_{l=\,1}^{j-1}(-(k-j)^2\alpha^2-(j-l)l\alpha^2
+l(j-l)(k-j)^2\alpha^4+1)a_la_{j-l}a_{k-j}e^{k\alpha z}\\
&\,\qquad+\sum_{k=2}^{\infty}\sum_{i=1}^{k-1} (-2(k-j)^2\alpha^2-(k-j)j\alpha^2+3)x_0a_{k-i}a_i e^{k\alpha z}=0.
\end{split}
\end{equation}
As $x_0$ is an equilibrium for the Eq. \eqref{eq3_trav},  $-cx_0+x_0^3+g=0$. Comparing the coefficients of $e^{k\alpha z}$ for each $k$, one has for $k=1$:
\begin{equation}\label{eq3_series1}
(\alpha^2 (c-x^2_0)-c+3x_0^2)a_1=0.
\end{equation}
Assuming $a_1\neq0$ (otherwise $a_k=0$ for all $k>1$ by induction), results in the two possible values of $\alpha$:
\begin{equation}\label{eq3_series_eig}
\alpha_{1}= \sqrt{\frac{3x_0^2-c}{x_0^2-c}},\qquad\qquad \alpha_{2}=-\sqrt{\frac{3x_0^2-c}{x_0^2-c}}.
\end{equation}
We are dealing with the case when the equilibrium $x_0$ is a saddle (for the suitable choice of the parameter $c$ and $g$ see the details in Section \ref{subsec3.3}). In this case, as our series solution $\phi^+(z)$ needs to converge for $z >0$, we pick the negative root $\alpha=\alpha_2$. Thus we have:
\begin{equation}\label{eq3_series_2}
F(2\alpha_2)a_2=3x_0(1-\alpha^2_2)a_1^2,
\end{equation}
where  $F(k\alpha_2)=(k\alpha_2)^2 (c-x_0^2)+3x_0^2-c$ and the coefficient $a_2$ is easily obtained in terms of $a_1$:
\begin{equation}\label{eq3_series_coeff2}
a_2=\frac{3x_0(1-\alpha^2_2)}{F(2\alpha_2)}a_1^2.
\end{equation}
For $k=3$ we obtain:
\begin{equation}\label{eq3_series_coeff3}
a_3=\frac{2x_0(-7\alpha_2^2+3)a_1a_2+(\alpha_2-1)^2a_1^3}{F(3\alpha_2)}.
\end{equation}
Once substituted the formula \eqref{eq3_series_coeff2} into the Eq. \eqref{eq3_series_coeff3}, one obtains $a_3$  in terms of $a_1$. And so on for all $k$ the series coefficients $a_k$ can be iteratively computed in terms of $a_1$:
\begin{equation}\label{eq3_series_coeffk}
a_k=\varphi_k a_1^k,                                                                                  \end{equation}
where $\varphi_k, k>1$ are functions which can be obtained using Eqs. \eqref{eq3_series_coeff2}-\eqref{eq3_series_coeff3} and so on. The coefficients $\varphi_k$ depend on $\alpha_2$ and the constant coefficients of the Eq. \eqref{eq3_trav_s}.
The first part of the homoclinic orbit corresponding to $z>0$ has thus been determined in terms of $a_1$:         		 %
\begin{equation}\label{eq3_series_p}
\phi^+ (z)=x_0+a_1 e^{\alpha_2 z}+\sum_{k=2}^\infty \varphi_k a_1^k e^{k\alpha_2 z}.                                                 \end{equation}
Being the Eq. \eqref{eq3_trav_s} reversible, the series solution for $z < 0$ can be obtained using the intrinsic symmetry property of the equation simply defining:
\begin{equation}\label{eq3_series_m}
\phi^- (z)=x_0-a_1 e^{\alpha_2 z}-\sum_{k=2}^\infty \varphi_k a_1^k e^{k\alpha_2 z}.                                                 \end{equation}
As we want to construct a solution continuous at $z=0$, we impose:
\begin{equation}\label{eq3_cont}
x_0+a_1+\sum_{k=2}^\infty \varphi_k a_1^k=0.
\end{equation}
Hence we choose $a_1$ as the nontrivial solutions of the above polynomial Eq. \eqref{eq3_cont}.

In practice the Eq. \eqref{eq3_cont} is numerically solved and the corresponding series solutions are not unique.

\begin{figure}
\begin{center}
{\epsfxsize=3 in \epsfbox{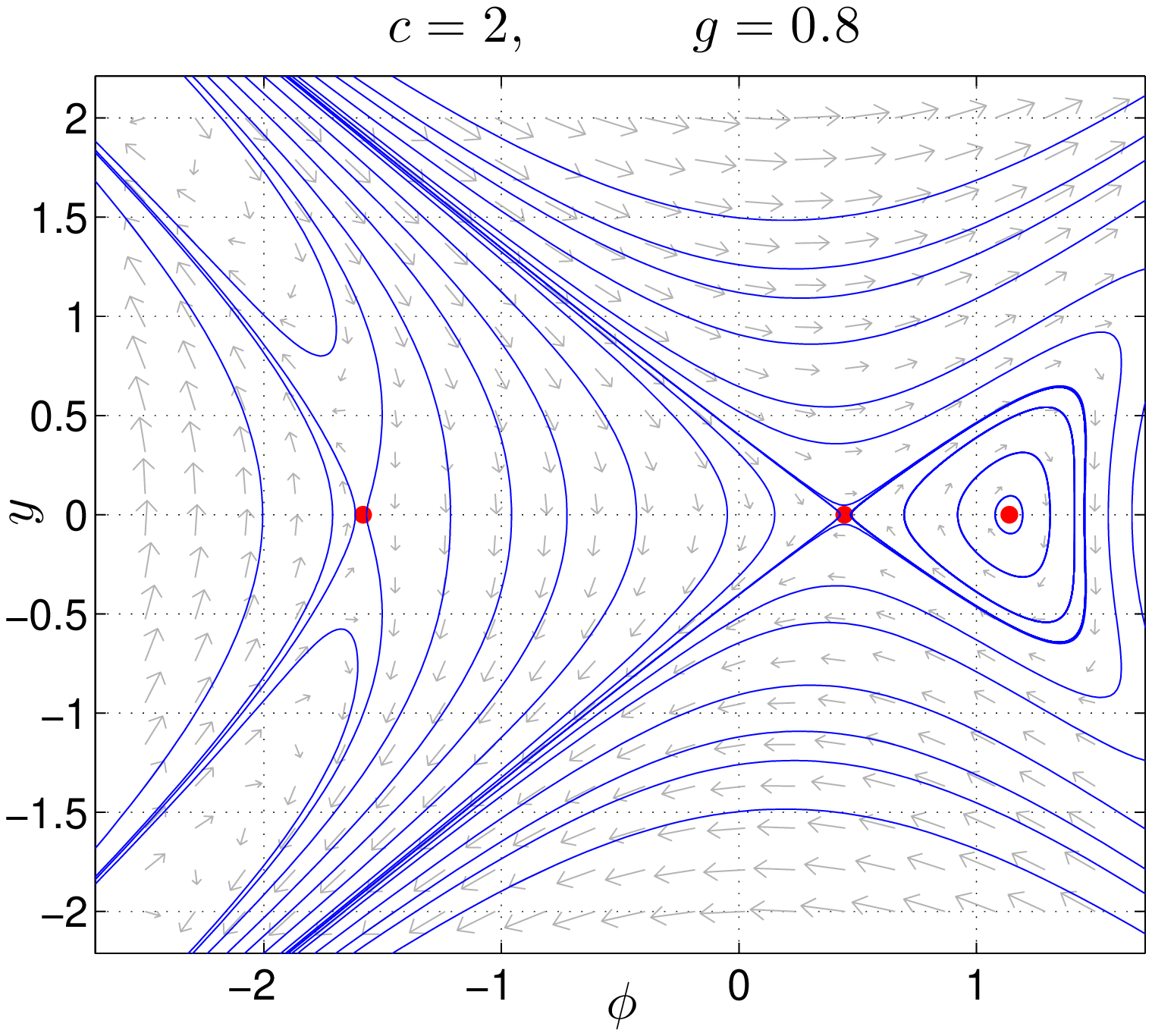}}
\end{center}
\caption{\label{Fig_eq3Hom1} Phase plane orbits of Eq. \eqref{eq3_sd_reg} when $c = 2$ and $g = 0.8$.}
\end{figure}

Let us now choose $c = 2$ and $g =0.8$. In this case the Eq. \eqref{eq3_trav} admits three real equilibrium points, two equilibria $\bar{z}_1=0.4436$  and $\bar{z}_2=-1.5828$ are saddles and the other $\bar{z}_3=1.1391$  is a center. We do not observe any homoclinic orbit at the saddle point $\bar{z}_2=-1.5828$ (corresponding to the regular equilibrium $(\bar{z}_2,0)$ of the system \eqref{eq3_sd}) in its phase portrait, see Fig.\ref{Fig_eq3Hom1}.  Thus we build homoclinic orbit to the saddle point $\bar{z}_1$ (corresponding to the regular equilibrium $(\bar{z}_1,0)$ of the system \eqref{eq3_sd}).

\begin{figure}
\begin{center}
\subfigure[] {\epsfxsize=3 in\epsfbox{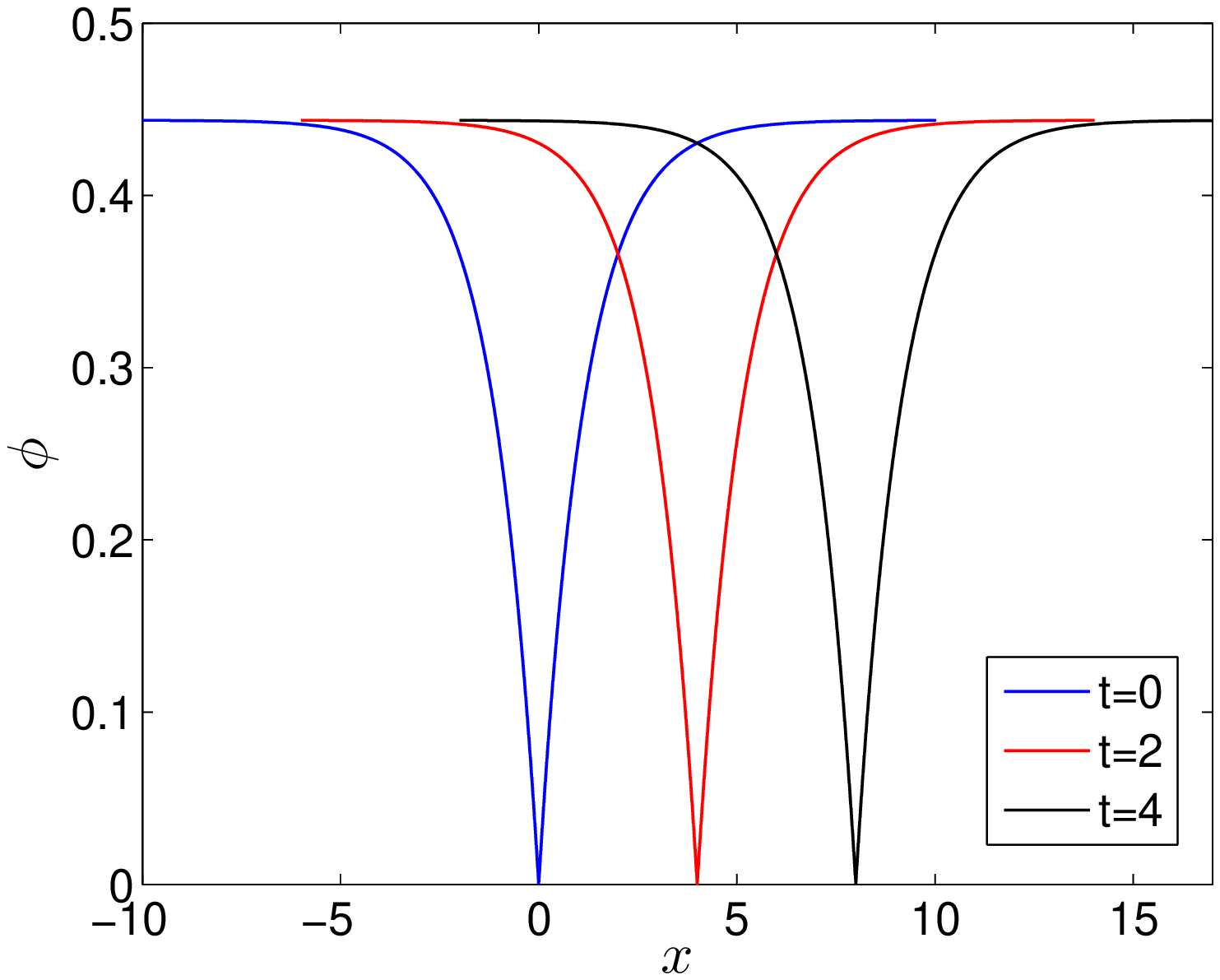}}
\subfigure[] {\epsfxsize=3 in \epsfbox{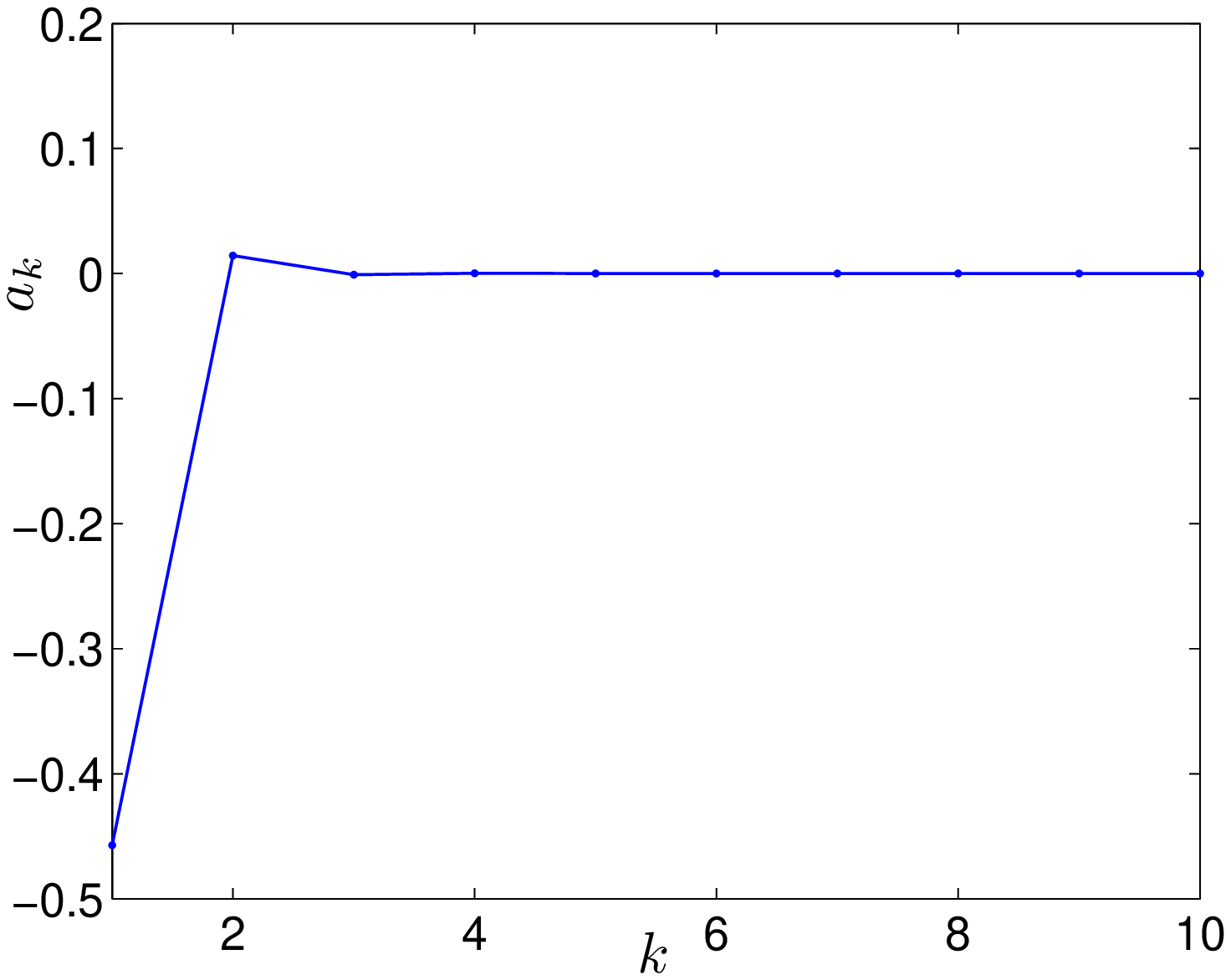}}
\end{center}
\caption{\label{eq3_Fig_hom1} The parameters are chosen as $c = 2$ and $g =0.8$. (a) The series solution $\phi(z)$ (as in \eqref{eq4_hom_orbit}, with $a_k$ given in \eqref{eq3_series_coeffk}) for the homoclinic orbit to the point $(x_0=-0.4436,0)$, plotted as a function of $x$ for different $t$. Here $a_1= -0.4572$ is the solution of the continuity condition \eqref{eq3_cont}  truncated at $M=25$. (b) Plot of $a_k$ in \eqref{eq3_series_coeffk} versus $k$, shows the series coefficients converge.}
\end{figure}

Truncating the series at $M = 25$, we find that the continuity condition \eqref{eq3_cont} is satisfied only for $a_1= -0.4572$. This choice of $a_1$ guarantees the continuity (see Fig.\ref{eq3_Fig_hom1}(a)) and the series coefficients $a_k$ also converge, see Fig.\ref{eq3_Fig_hom1}(b).

Let us now choose $c = 0.5$ and $g = 0.13$. In this case the Eq. \eqref{eq3_trav_s} admits three real equilibrium points, two equilibria $\bar{z}_1=-0.8124$  and $\bar{z}_2=0.3356$ are saddles and the other $\bar{z}_3=0.4768$  is a center.

\begin{figure}
\begin{center}
{\epsfxsize=3 in \epsfbox{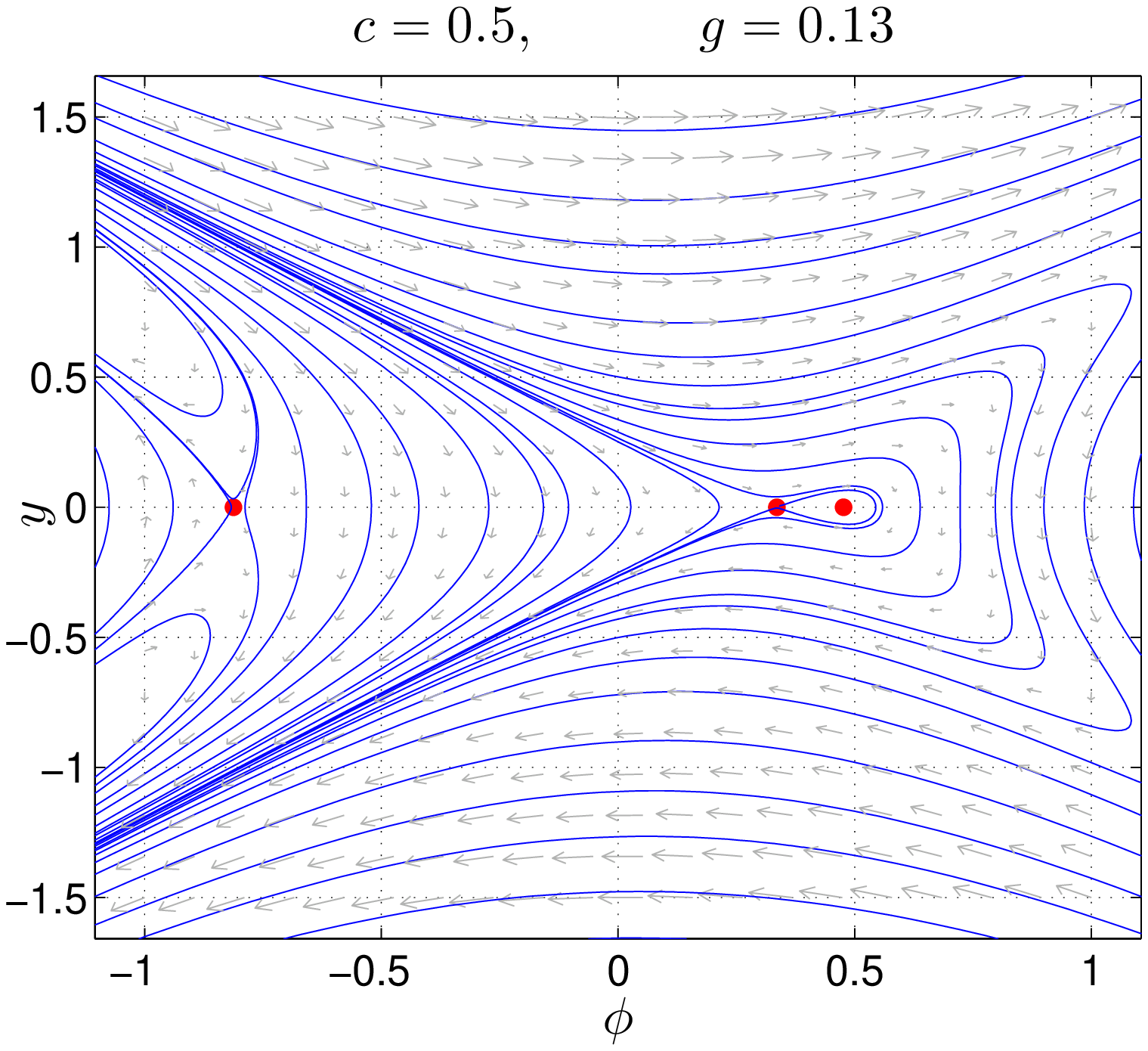}}
\end{center}
\caption{\label{Fig_eq3Hom2} Phase plane orbits of Eq. \eqref{eq3_sd_reg} when $c = 0.5$ and $g = 0.13$.}
\end{figure}
From Fig.\ref{Fig_eq3Hom2} we do not observe any homoclinic orbit at the saddle point $\bar{z}_1$, thus we construct the homoclinic orbit to the saddle $\bar{z}_2$. We find that the continuity condition \eqref{eq3_cont} does not admit any real solution. Therefore the solution will be discontinuous at the origin, or we have to choose an arbitrary $a_1$ and impose that in zero the solution does not take the value zero, but the common value of $\phi^+(0)$ and $\phi^-(0)$ for that $a_1$.

\begin{figure}
\begin{center}
\subfigure[] {\epsfxsize=3 in \epsfysize=2.4 in\epsfbox{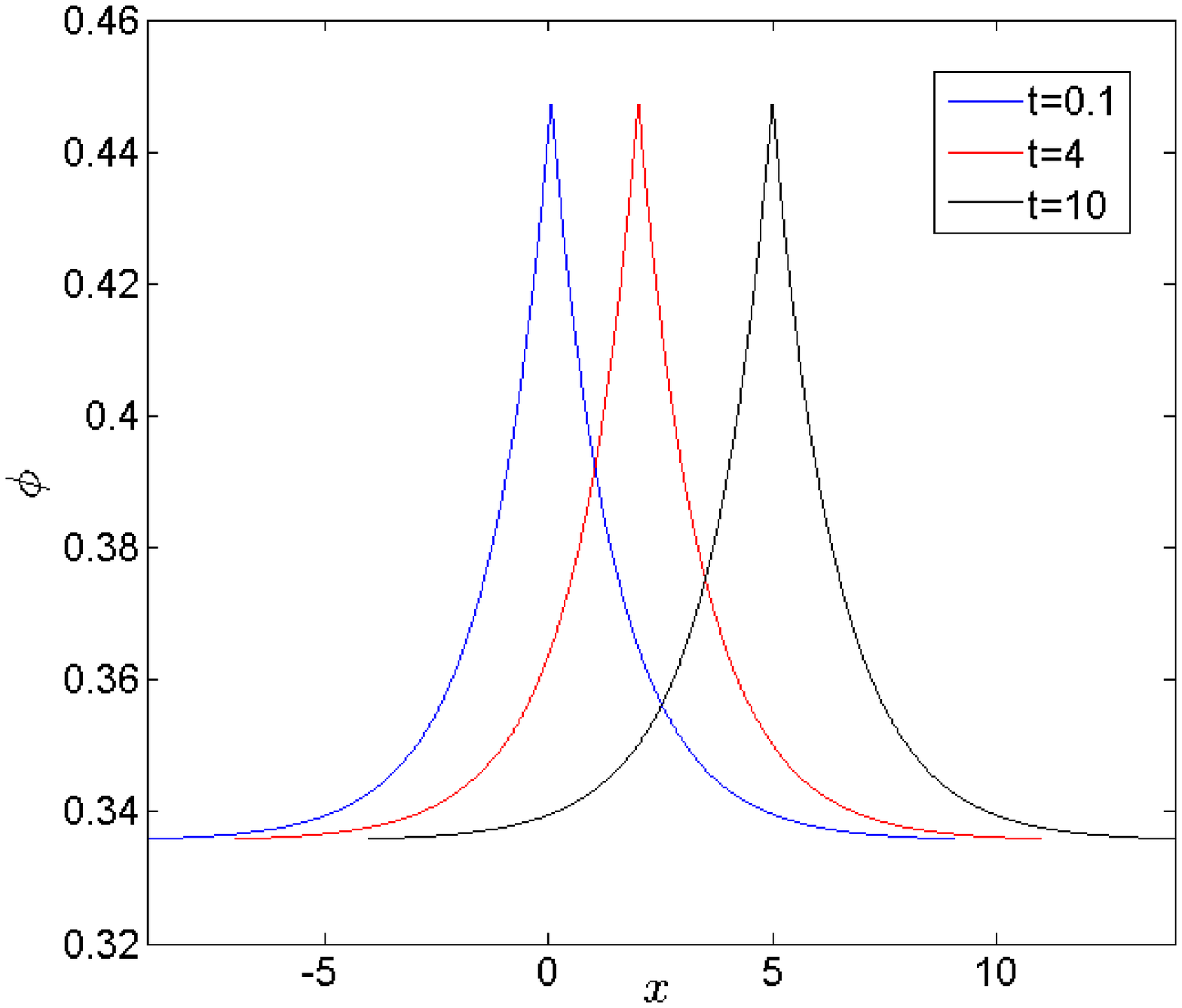}}
\subfigure[] {\epsfxsize=3 in \epsfbox{vuoto.eps}}
\end{center}
\caption{\label{eq3_Fig_hom2} The parameters are chosen as $c = 0.5$ and $g = 0.13$. (a) The series solution $\phi(z)$ (as in \eqref{eq4_hom_orbit}, with $a_k$ given in \eqref{eq3_series_coeffk}) for the homoclinic orbit to the point $(x_0=0.3356,0)$, plotted as a function of $x$ for different $t$. Here $a_1= 0.1$ is chosen to obtain the continuity at the origin. (b) Plot of $a_k$ in \eqref{eq3_series_coeffk} versus $k$, shows the series coefficients converge.}
\end{figure}
In Fig.\ref{eq3_Fig_hom2} we have chosen $a_1= 0.1$ and  $\phi(0)=0.44705$ and the traveling nature of the solution has been shown in Fig.\ref{eq3_Fig_hom2}(a). Moreover, for this choice of the parameters the series solution converges, as shown in the Fig.\ref{eq3_Fig_hom2}(b) where the $a_k$ rapidly goes to zero.

\begin{figure}
\begin{center}
{\epsfxsize=3 in \epsfbox{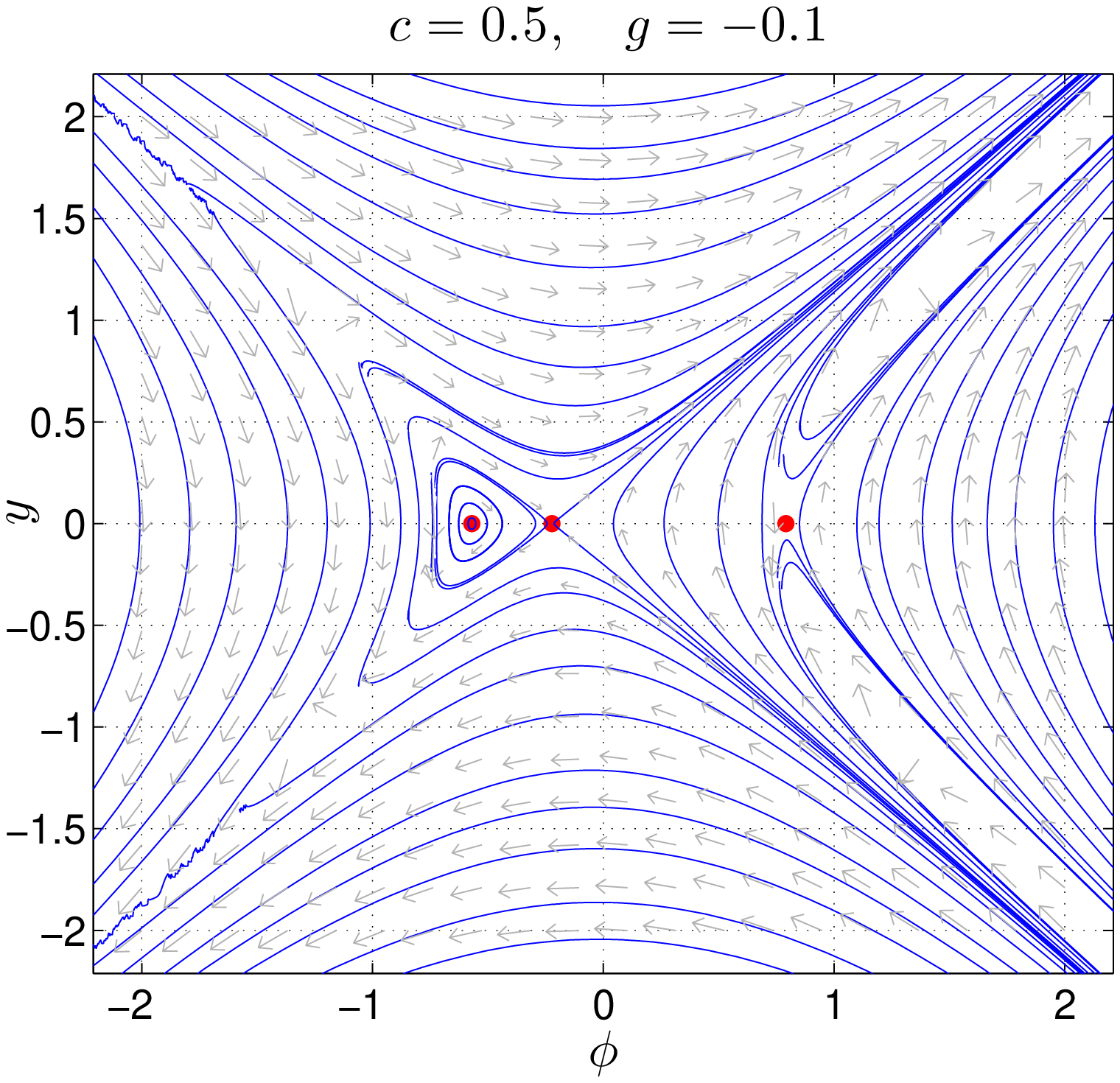}}
\end{center}
\caption{\label{Fig_eq3Hom3} Phase plane orbits of Eq. \eqref{eq3_sd_reg} when $c = 0.5$ and $g = -0.1$.}
\end{figure}

Let us now choose $c = 0.5$ and $g = -0.1$. In this case the Eq. \eqref{eq3_trav_s} admits three real equilibrium points, two equilibria $\bar{z}_1=0.7914$  and $\bar{z}_2=-0.2218$ are saddles and the other $\bar{z}_3=-0.5696$ is a center. From Fig\eqref{Fig_eq3Hom3}, we do not observe any homoclinic orbit at the saddle point $\bar{z}_1$, thus we construct the homoclinic orbit to the saddle $\bar{z}_2$.

\begin{figure}
\begin{center}
\subfigure[] {\epsfxsize=3 in\epsfbox{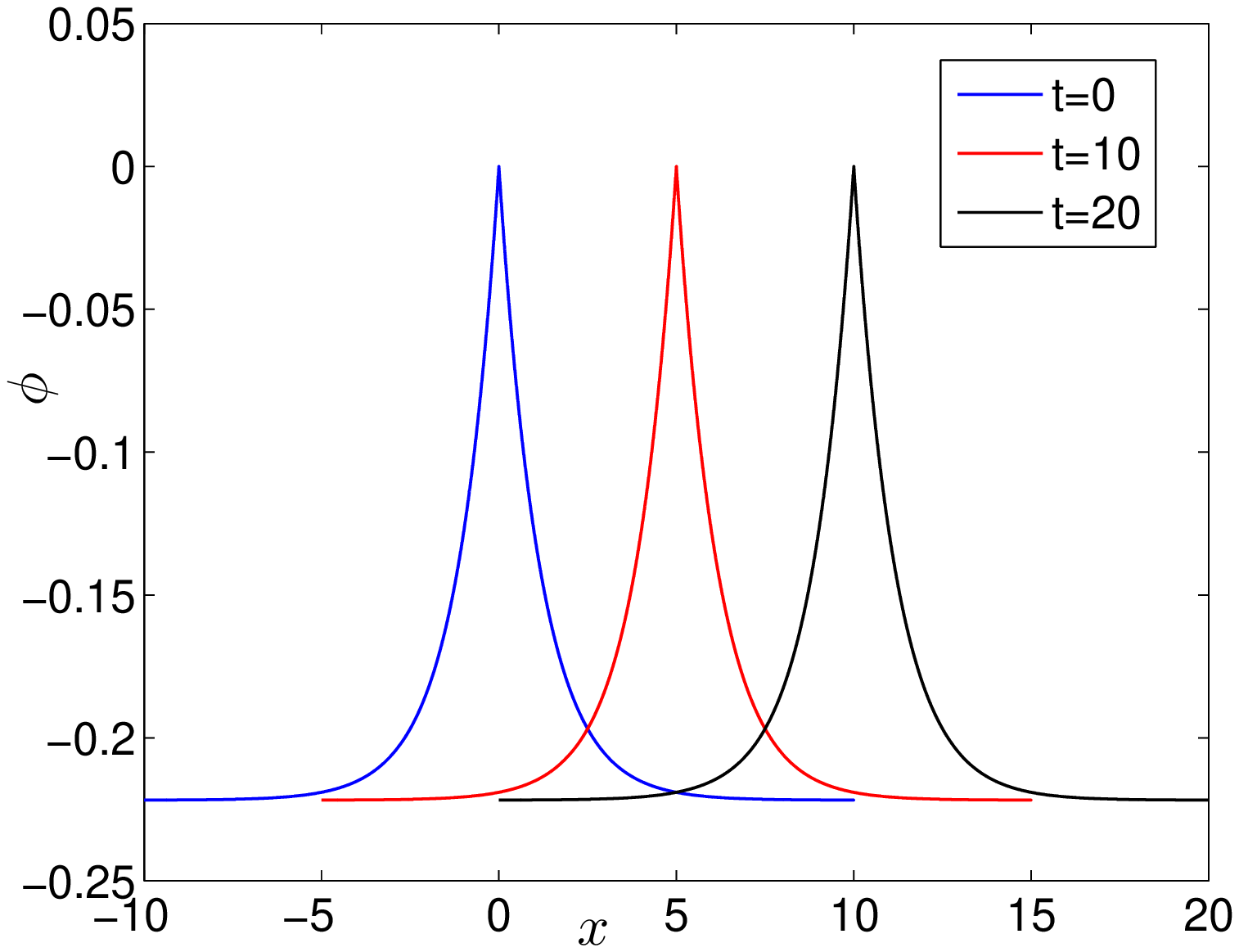}}
\subfigure[] {\epsfxsize=3 in \epsfbox{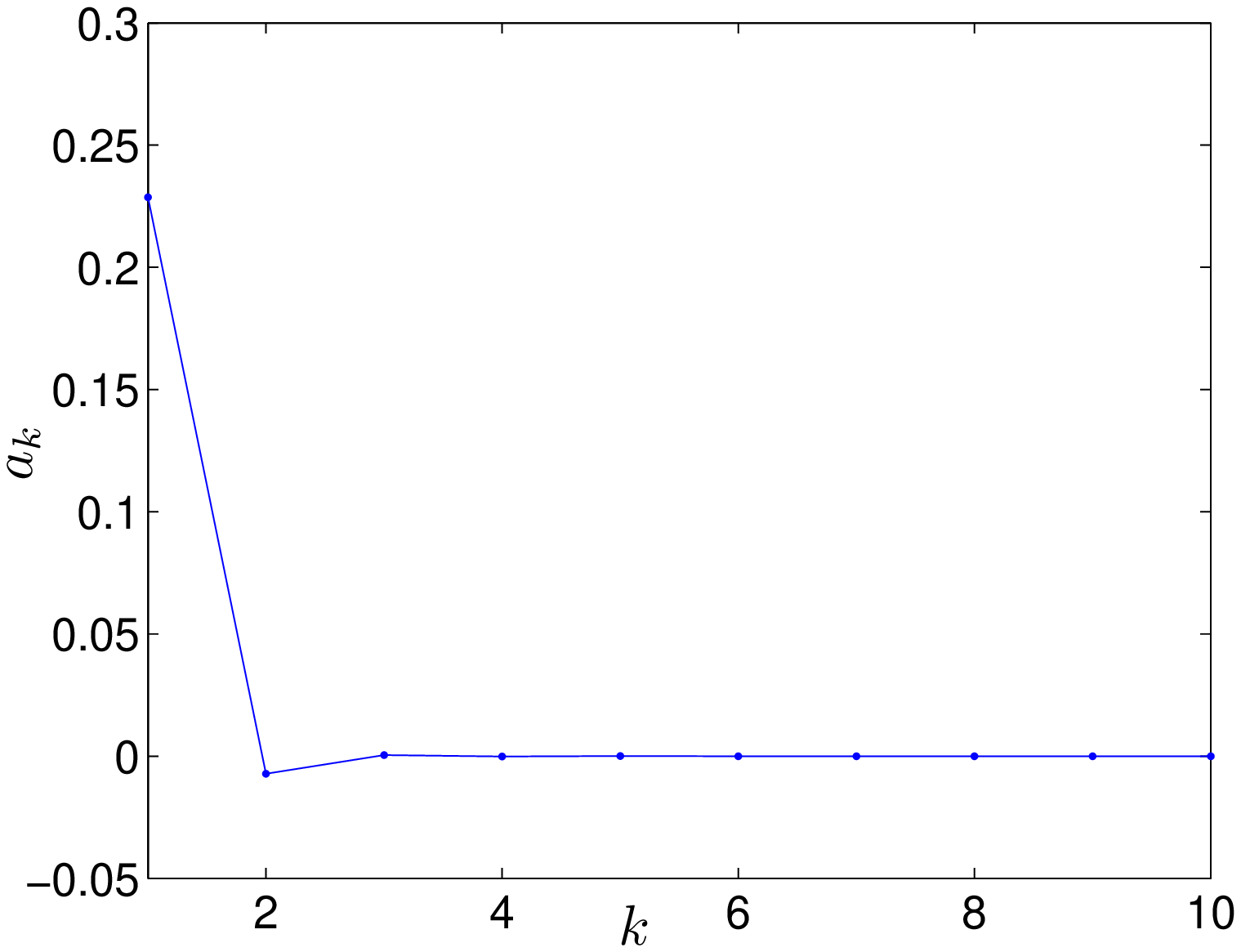}}
\end{center}
\caption{\label{eq3_Fig_hom3} The parameters are chosen as $c = 0.5$ and $g = -0.1$. (a) The series solution $\phi(z)$ (as in \eqref{eq4_hom_orbit}, with $a_k$ given in \eqref{eq3_series_coeffk}) for the homoclinic orbit to the point $(x_0=-0.2218,0)$. Here $a_1=0.2287$ is the solution of the continuity condition \eqref{eq3_cont} truncated at $M=10$. (b) Plot of $a_k$ versus $k$ showing the series coefficients converge.}
\end{figure}

We find that the continuity condition \eqref{eq3_cont} admits real solution for $a_1=0.2287$. The values of $a_k$ also converges to zero. In Fig.\ref{eq3_Fig_hom3} we have chosen $a_1= 0.2287$ and the traveling nature of the solution has been shown in Fig.\ref{eq3_Fig_hom3}(a). Moreover, for this choice of the parameters the series solution converges, as shown in the Fig.\ref{eq3_Fig_hom3}(b) where the $a_k$ rapidly goes to zero.

\section{Conclusions}

In this paper we have employed two recent analytical approaches to investigate the possible classes of traveling wave solutions of three members of a recently-derived integrable family of 27 generalized Camassa-Holm (GCH) NLPDEs.

A recent, novel application of phase-plane analysis is employed to analyze the singular traveling wave equations of these GCH NLPDEs, i.e. the possible non-smooth peakon and cuspon solutions. The first GCH equation \eqref{eq4} is found to support both solitary (peakon) and periodic (cuspon) cusp waves in different parameter regimes. The GCH equation \eqref{eq2} is found not to support singular traveling waves. The GCH NLPDE \eqref{eq3} is known to support four-segmented, non-smooth $M$-wave solutions \cite{DF10}, so we did not consider its singular solutions in this paper.

We also consider smooth, non-singular traveling waves of the three GCH NLPDEs. We use a recent technique to derive convergent multi-infinite series solutions for the homoclinic orbits of their traveling-wave equations, corresponding to pulse (kink or shock) solutions respectively of the original GCH NLPDEs.

We performed many numerical tests in different parameter regime to pinpoint real saddle equilibrium points of the corresponding GCH equations, as well as ensuring simultaneous convergence and continuity of the multi-infinite series solutions for the homoclinic orbits anchored by these saddle points. Unlike the majority of unaccelerated convergent series, high accuracy is attained with relatively few terms. We also show the traveling wave nature of these pulse and front solutions to the GCH NLPDEs.


\begin{thebibliography}{}
\bibitem{BC07} A. Bressan, A. Constantin, Global conservative solutions of the Camassa-Holm equation,
Arch. Rational Mech. 183(2) (2007) 215--239.
\bibitem{CH93}	R. Camassa, D. Holm, {An integrable shallow water equation with peaked solitons}, Phys. Rev. Lett. {71}(11) (1993) 1661--1664.
\bibitem{CHH94} R. Camassa, D. Holm, J. Hyman, {A new integrable shallow water equation}, Adv. Appl. Mech. {31} (1994) 1--33.
\bibitem{C98}	A. R. Champneys, {Homoclinic orbits in reversible systems and their application in mechanics, fluids and optics}, Physica D 112 (1998) 158--186.
\bibitem{CG13}	S. R. Choudhury, G. Gambino, {Convergent analytic solutions for homoclinic orbits in reversible and non-reversible systems},  Nonlinear Dynam. 73(3) (2013) 1769--1782, .
\bibitem{DHH02} A. Degasperis, D. D. Holm, A. N. W. Hone, {A new integrable equation with peakon solutions}, Theor. Math. Phys. 133(2) (2002) 1463--1474.
\bibitem{CGS12} G. M. Coclite, F. Gargano, V. Sciacca, Analytic solutions and singularity formation for the Peakon b-family equations. Acta Appl. Math. 122 (2012), 419–434.
\bibitem{DIS10} A. B. De Monvel, A. Its, D. Shepelsky,
Painlev\'{e}-type asymptotics for the Camassa-Holm equation,
SIAM J. Math. Anal. 42 (4) (2010) 1854--1873.
\bibitem{DKST09} A.B. De Monvel, A. Kostenko, D. Shepelsky, G. Teschlh, Long-time asymptotics for the Camassa-Holm equation, SIAM J. Math. Anal. 41(4) (2009) 1559--1588.
\bibitem{DLSS06} G. Della Rocca, M.C. Lombardo, M. Sammartino, V. Sciacca, Singularity tracking for Camassa-Holm and Prandtl's equations, Appl. Num. Math. 56(8) (2006) 1108--1122.
\bibitem{G94}	P. Glendinning, {Stability, Instability and Chaos}, Cambridge Univ. Press, Cambridge, 1994.
\bibitem{HR09} H. Holden, X. Raynaud, Dissipative solutions for the Camassa-Holm equation, Discrete Cont. Dyn. S. 24(4) (2009) 1047--1112.
\bibitem{HW08} A. N. W. Hone, J. P. Wang, {Integrable peakon equations with cubic nonlinearity}, J. Phys. A: Math. Theor. 41 (2008) 372002.
\bibitem{J03} R. S. Johnson, {On the solutions of the Camassa-Holm equation}, Proc. R. Soc. Lond. A {459}(2035) (2003) 1687--1708.
\bibitem{J02} R. S. Johnson, {Camassa-Holm, Korteweg-de Vries and related models for water waves}, J. Fluid Mech. 455 (2002) 63--82.
\bibitem{K95} Y. A. Kuznetsov, {Elements of Applied Bifurcation Theory}, Springer-Verlag, New York, 1995.
\bibitem{LD07}	J. Li, H. Dai, {On the study of singular nonlinear traveling wave equations: dynamical approach}, 2nd edition, Beijing, Science Press, 2007.
\bibitem{DF10} J. Li, {The dynamics of two classes of singular nonlinear travelling wave equations and loop solutions}, in: C. David, Z. Feng (Eds.), {Solitary waves in fluid media}, Bentham Science, Sharjah, 2010, pp. 123--201.
\bibitem{LSS05} M.C. Lombardo, M. Sammartino, V. Sciacca, A note on the analytic solutions of the Camassa-Holm equation, Comp. Rend. Math. 341(11) (2005) 659--664.
\bibitem{M10} T. Matsuo,
A Hamiltonian-conserving Galerkin scheme for the Camassa-Holm equation
J. Comp. Appl. Math. 234(4) (2010) 1258--1266.
\bibitem{MN02} A. V. Mikhailov, V. S. Novikov, {Perturbative symmetry approach}, J. Phys. A: Math. Gen. 35 (2002) 4775.
\bibitem{N09} V. Novikov, {Generalization of the Camassa-Holm equation}, J. Phys. A: Math. Theor. 42 (2009) 342002.
\bibitem{RTW10} J. Rong, S. Tang, W. Huang, Bifurcations of travelling wave solutions for a class of nonlinear fourth order variant of a generalized Camassa-Holm equation,
Commun. Nonlinear Sci. Numer. Simul., 15(11) (2010) 3402--3417.
\bibitem{W09}	X. Wang, {Si'lnikov chaos and Hopf bifurcation analysis of Rucklidge system}, Chaos Solitons Fractals 42 (2009) 2208--2217.
\bibitem{W06}A. M. Wazwaz, Peakons, kinks, compactons and solitary patterns solutions for a family of Camassa-Holm equations by using new hyperbolic schemes, Appl. Math. Comput. 182(1) (2006) 412--424
\bibitem{XWZ12} S. Xie, L. Wang, Y. Zhang, Explicit and implicit solutions of a generalized Camassa-Holm Kadomtsev-Petviashvili equation, Commun. Nonlinear Sci. Numer. Simul., 17(3) (2012) 1130--1141.
\bibitem{ZTW10}K. Zhang, S. Tang, Z. Wang, Bifurcation of travelling wave solutions for the generalized Camassa-Holm-KP equations, Commun. Nonlinear Sci. Numer. Simul., 15(3) (2010) 564--572.



\end{thebibliography}
\end{document}